\documentclass[times, twocolumn]{aastex7} 

\usepackage{array}
\usepackage{comment}
\usepackage{hyperref}
\usepackage{ifthen}
\usepackage{mathdots}
\usepackage{multirow}
\usepackage{pgfplots}
\usepackage{physics}
\usepackage{ragged2e}
\usepackage{tikz}

\usetikzlibrary{calc, 3d, perspective, decorations.markings, matrix}
\usepgfplotslibrary{colormaps}

\pgfplotsset{compat=1.18} 

\def\Hubble{\textit{Hubble}}
\def\Roman{\textit{Roman}}
\def\Romanc{\textit{Nancy Grace Roman}}

\newcommand{\vecbold}[1]{\boldsymbol{{\rm #1}}} 
\newcommand{\parest}[1]{\hat{ #1}} 
\newcommand{\data}{\vecbold{y}} 
\newcommand{\cube}{\vect{\D}} 
\newcommand{\cubetilde}{\vect{\vecbold{\tilde{D}}}} 
\newcommand{\model}{\vecbold{H}} 
\newcommand{\X}{\vecbold{X}} 
\newcommand{\cov}{\vecbold{\Sigma}} 
\newcommand{\dset}{\data} 
\newcommand{\weight}{\vecbold{W}} 
\newcommand{\transpose}{^\intercal} 
\newcommand{\eye}{\vecbold{I}} 
\newcommand{\res}{\vecbold{r}} 
\newcommand{\posterior}{p(\cube|\X,\data)} 
\newcommand{\likelihood}{p(\data|\X,\cube)} 
\newcommand{\prior}{p(\cube)} 
\newcommand{\scalelength}{\sigma_\text{\rm p}} 
\newcommand{\datacube}{\mathcal{D}}
\newcommand{\D}{\vecbold{D}} 
\newcommand{\G}{\vecbold{G}} 
\renewcommand{\L}{\vecbold{L}} 
\newcommand{\C}{\vecbold{C}} 
\newcommand{\Dp}{\vecbold{R}} 
\newcommand{\PSF}{\vecbold{P}} 
\newcommand{\spec}{\vecbold{S}} 
\newcommand{\coords}{\vecbold{\xi}} 
\newcommand{\wave}{\vecbold{\lambda}} 
\newcommand{\wavetilde}{\vecbold{\tilde{\lambda}}} 
\newcommand{\A}{\vecbold{A}} 
\newcommand{\F}{\vecbold{F}} 
\newcommand{\krl}{\vecbold{K}} 
\newcommand{\zero}{\vecbold{\emptyset}} 
\newcommand{\matricize}[3]{{\rm unvec}_{#2,#3}(\vect{#1})} 
\newcommand{\matricizend}[4]{{\rm unvec}_{#2,#3,#4}(\vect{#1})} 
\newcommand{\vectorize}[1]{{\rm vec}(#1)} 
\newcommand{\vect}[1]{\MakeLowercase{#1}} 
\newcommand{\floor}[1]{\left\lfloor #1 \right\rfloor}

\newcommand{\baylor}{\affil{Department of Physics, Baylor University, Waco, Texas 76798-7316}}
\newcommand{\gsfc}{\affil{NASA Goddard Space Flight Center, 8800 Greenbelt Rd, Greenbelt, MD 20771}}

\newcommand{\lbnl}{\affil{E.O. Lawrence Berkeley National Laboratory, 1 Cyclotron Rd., Berkeley, CA 94720}}
\newcommand{\nist}{\affil{National Institute of Standards and Technology, Remote Sensing Group, Sensor Science Division, 100 Bureau Drive, Gaithersburg, MD 20899}}

\newcommand{\stsci}{\affil{Space Telescope Science Institute, 3700 San Martin Drive, Baltimore, MD 21218}}

\newcommand{\ucb}{\affil{Department of Physics, University of California Berkeley, 366 LeConte Hall MC 7300, Berkeley, CA 94720}}
\newcommand{\uhawaii}{\affil{Department of Physics and Astronomy, University of Hawai`i at M{\=a}noa, Honolulu, Hawai`i 96822}}
\newcommand{\umbc}{\affil{University of Maryland Baltimore County, 1000 Hilltop Circle, Baltimore, MD 21250}}
\newcommand{\upenn}{\affil{Department of Physics and Astronomy, University of Pennsylvania, 209 South 33rd Street, Philadelphia, PA 19104}}

\begin{document}
\title{Three-dimensional scene reconstruction using \textit{Roman} slitless spectra}

\author[0000-0002-9888-2704]{Tri L. Astraatmadja}
\email{tastraatmadja@stsci.edu}
\stsci

\author[0000-0002-6652-9279]{Andrew S. Fruchter}
\email{fruchter@stsci.edu}
\stsci

\author[0000-0003-2823-360X]{Susana E. Deustua}
\email{susana.deustua@nist.gov}
\stsci
\nist

\author[0000-0003-1899-9791]{Helen Qu}
\email{helenqu@sas.upenn.edu}
\upenn

\author[0000-0003-2764-7093]{Masao Sako}
\email{masao@sas.upenn.edu}
\upenn

\author[0000-0003-0894-1588, gname='Russell', sname='Ryan', suffix='Jr']{Russell E. Ryan, Jr.}
\email{rryan@stsci.edu}
\stsci

\author[0000-0002-5317-7518]{Yannick Copin}
\email{y.copin@ipnl.in2p3.fr}
\affil{Universit\'{e} de Lyon, Universit\'{e} Claude Bernard Lyon 1, CNRS/IN2P3, IP2I Lyon, F-69622 Villeurbanne, France}

\author[0009-0009-1689-4874]{Greg Aldering}
\email{galdering@lbl.gov}
\lbnl

\author[0000-0002-0476-4206]{Rebekah A. Hounsell}
\email{rebekah.a.hounsell@nasa.gov}
\umbc
\gsfc

\author[0000-0001-5402-4647]{David Rubin}
\email{drubin@hawaii.edu}
\uhawaii

\author[0000-0002-1296-6887]{Llu\'{i}s Galbany}
\email{lluisgalbany@gmail.com}
\affil{Institute of Space Sciences (ICE, CSIC), Campus UAB, Carrer de Can Magrans, s/n, E-08193 Barcelona, Spain}
\affil{Institut d'Estudis Espacials de Catalunya (IEEC), E-08034 Barcelona, Spain}

\author[0000-0002-4436-4661]{Saul Perlmutter}
\email{saul@lbl.gov}
\lbnl
\ucb

\author[0000-0002-1873-8973]{Benjamin M. Rose}
\email{Ben_Rose@baylor.edu}
\baylor

\shorttitle{3d scene reconstruction using \textit{Roman} slitless spectra}
\shortauthors{Astraatmadja, Fruchter, Deustua et al.}

\begin{abstract}
The \Romanc{} Space Telescope will carry out a wide-field imaging and slitless spectroscopic survey of Type~Ia Supernovae to improve our understanding of dark energy.  Crucial to this endeavor is obtaining supernova spectra uncontaminated by light from their host galaxies. However, obtaining such spectra is made more difficult by the inherent problem in wide-field slitless spectroscopic surveys:  the blending of spectra of close objects. The spectrum of a supernova will blend with the host galaxy, even from regions distant from the supernova on the sky. If not properly removed, this contamination will introduce systematic bias when the supernova spectra are later used to determine intrinsic supernova parameters and to infer the parameters of dark energy. To address this problem we developed an algorithm that makes use of the spectroscopic observations of the host galaxy at all available observatory roll angles to reconstruct a three-dimensional (3d; 2d spatial, 1d spectral) representation of the underlying host galaxy that accurately matches the 2d slitless spectrum of the host galaxy when projected to an arbitrary rotation angle. We call this ``scene reconstruction''. The projection of the reconstructed scene can be subtracted from an observation of a supernova to remove the contamination from the underlying host. Using simulated \Roman{} data, we show that our method has extremely small systematic errors and significantly less random noise than if we subtracted a single perfectly aligned spectrum of the host obtained before or after the supernova was visible.
\end{abstract}

\section{Introduction}
\label{sec:intro}
\begin{figure*}
    \centering
    \includegraphics[width=\textwidth]{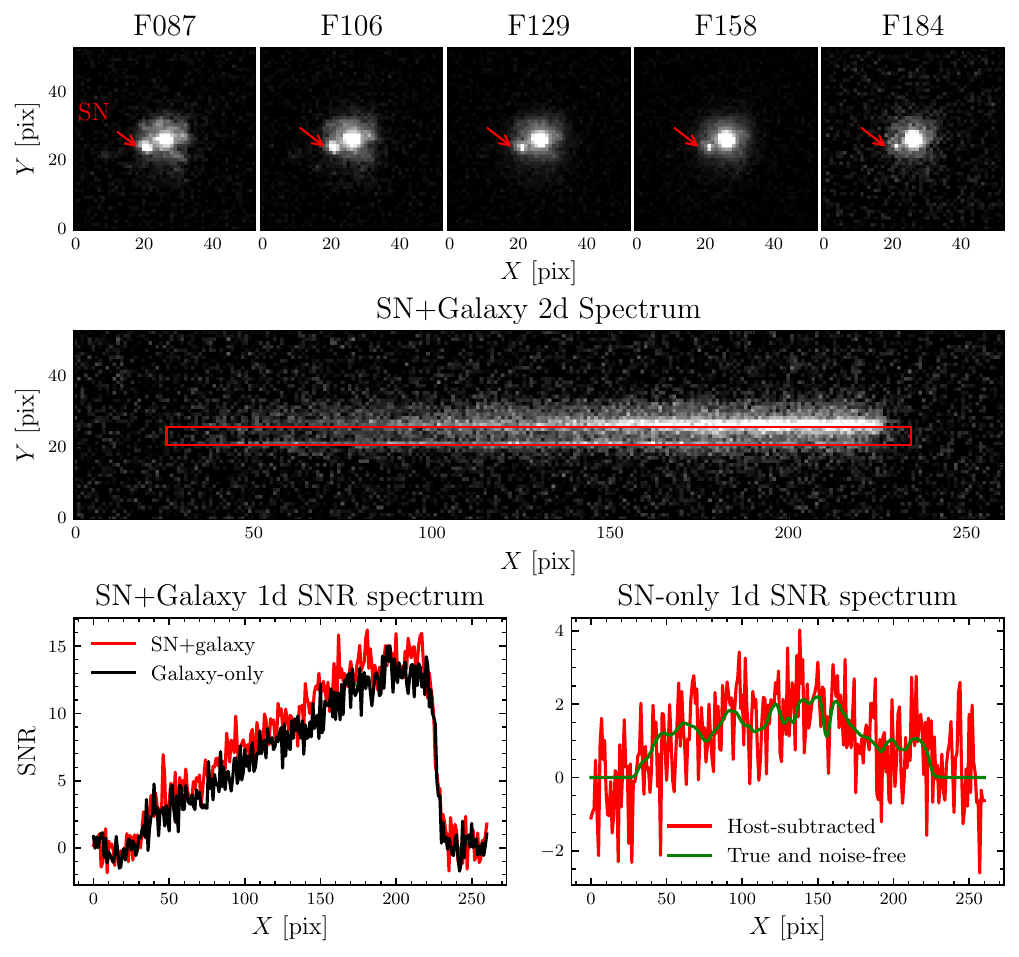}
    \caption{This illustration shows simulated observations of a supernova and its host galaxy, with images in five \Roman{} filters (top panels) and the corresponding spectral image taken using the slitless prism (middle panel). The location of the supernova relative to its host galaxy is marked by the red arrows. If we extract the 1d spectrum by concentrating on the boxed area marked in red, we will obtain the SN+galaxy 1d signal-to-noise-ratio (SNR) spectrum as shown in red on the bottom-left panels. If there is no supernova, the similarly-extracted galaxy-only 1d SNR spectrum in the same area is overplotted in black on this panel. Suppose there is no host galaxy, the observed supernova-only 1d SNR spectrum is shown in red at the bottom-right panel. For comparison, the noise-free supernova-only spectrum is overplotted as the green curve. Sky background, the dominant source of noise (see Section~\ref{sec:noise}), has been subtracted from these spectra. Note the different vertical axis ranges between the panels in the bottom row. The exposure times of these observations are 900\,s and 1800\,s for respectively the filter images and the spectral image. These are single images and not co-added images. To generate these data, for the host galaxy we use a datacube based on spectral energy distribution (SED) fitting of a galaxy from the VELA galaxy simulation \citep{sim19}, using the CIGALE \citep{boq19} code. The galaxy is at redshift $z=1.0$. We use \texttt{SNCosmo} \citep{sncosmo} to generate the injected supernova spectrum, which is an extended SALT2 spectrum \citep{guy10,bet14,hou18} and redshifted to the same $z$ as the host galaxy.}
    \label{fig:example}
\end{figure*}

\begin{figure}
    \centering
    \includegraphics[width=\columnwidth]{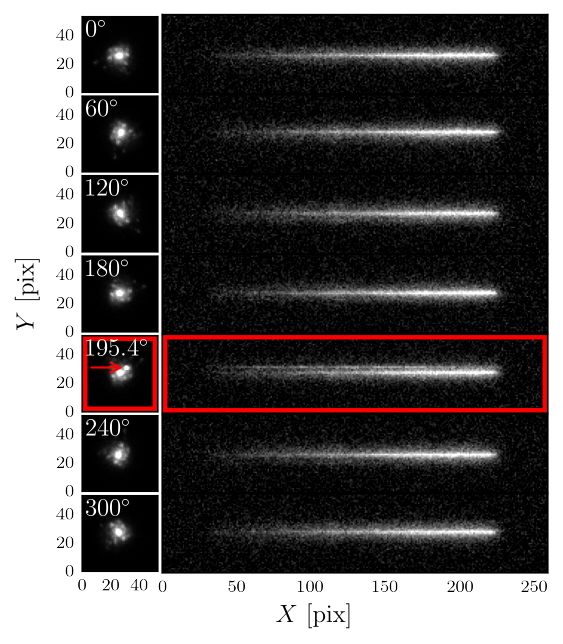}
    \caption{In this illustration, \Roman{} observed a galaxy at seven different roll angles, whose values are shown on the top left corners in the left column of each row. The images at each roll angle are shown on the left column, while the corresponding spectra are shown on the right column. At roll angle $\phi = 195.4^\circ$ (outlined in red), a supernova occurred (marked with red arrow). We set aside this spectrum and use the other roll angles to predict what the galaxy-only spectrum would look like at roll angle $\phi = 195.4^\circ$, in order to subtract the host-galaxy spectrum and obtain the supernova-only spectrum (In later demonstrations we will use more roll angles). This is the same simulated galaxy and supernova shown in Figure~\ref{fig:example}.}
    \label{fig:obs}
\end{figure}

\begin{figure*}
\centering
\def\azimuth{40}
\def\elevation{30}
\begin{tikzpicture}[3d view={\azimuth}{\elevation}, scale=1.0]
    \def\nplot{24} 
    \def\pull{6}   
    \def\stdwidth{3} 
    \def\imwidth{84pt} 
    \def\sep{-0.8} 
    \def\axlength{\stdwidth} 
    \pgfmathsetmacro{\lwidth}{2.0*\stdwidth} 
    \pgfmathsetmacro{\specwidth}{\lwidth} 
    \def\xxx{0.05} 
    \def\yyy{0.05} 
    \def\l0{0.00*\lwidth} 
    \pgfmathsetmacro{\dl}{\lwidth/(\nplot+1)}
    \def\lambdamin{7000}
    \def\lambdastep{500}
    \def\monimcol{blue}
    \def\loscolor{red}
    \def\imcol{green}
    \def\filterimagepos{1.9*\lwidth}
    \def\integralopacity{0.7}
    \def\integrallineopacity{1.0}
    \def\specpos{-1.40*\axlength}

    \def\losx{37}
    \def\losy{48}
    \def\cubewidth{100}
    \def\cubeheight{100}
    \pgfmathsetmacro{\xx}{\losx/\cubewidth*\axlength}
    \pgfmathsetmacro{\yy}{\losy/\cubeheight*\axlength}
    \def\dashfactor{1.29}
    \def\losstart{-6}

    \pgfkeys{/pgf/number format/.cd,fixed,set thousands separator={}, precision=0}
    
    \tikzset{->-/.style={decoration={markings,mark=at position #1 with {\arrow{>}}},postaction={decorate}}}
    \tikzset{-<-/.style={decoration={markings,mark=at position #1 with {\arrow{<}}},postaction={decorate}}}

    \begin{scope}[canvas is yz plane at x=\specpos]
        \fill[black] (0,0) rectangle (\specwidth cm,\imwidth);
        \node[anchor=south west, name=spec, inner sep=\sep, transform shape] at (0,0) {\includegraphics[height=\imwidth, width=\specwidth cm]{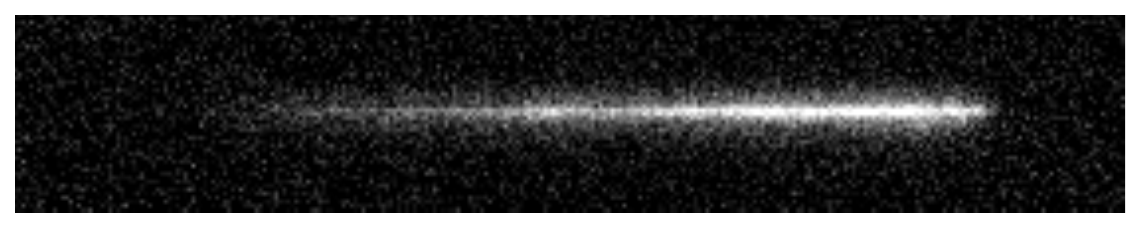}};
        \node[anchor=south, text width=288, align=center, black, transform shape] at (spec.north) {{\large 2d spectrum}};
    \end{scope}

    \draw[dashed, green, opacity=\integrallineopacity] (0,\lwidth,0) -- (spec.south east);
    \draw[dashed, green, opacity=\integrallineopacity] (0,\lwidth,\axlength) -- (spec.north east);
    \draw[dashed, green, opacity=\integrallineopacity] (0,0,0) -- (spec.south west);
    \draw[dashed, green, opacity=\integrallineopacity] (0,0,\axlength) -- (spec.north west);

    \def\lstart{0.60*\lwidth}
    \def\lend{0.80*\lwidth}
    
    \fill[green,opacity=\integralopacity] (0,\lstart,0)
    -- (0,\lstart,\axlength)
    -- (0,\lend,\axlength)
    -- (0,\lend,0) -- cycle;
    
    \fill[green,opacity=\integralopacity] (0,\lstart,0)
    -- (0,\lend,0)
    -- (\axlength,\lend,0)
    -- (\axlength,\lstart,0) -- cycle;

    \begin{scope}[canvas is xz plane at y=\filterimagepos]
        \fill[black] (0,0) rectangle (\axlength,\axlength);
        \node [name=imaging, anchor=south west, inner sep=0, transform shape] at (\xxx,\yyy) {\includegraphics[width=\imwidth]{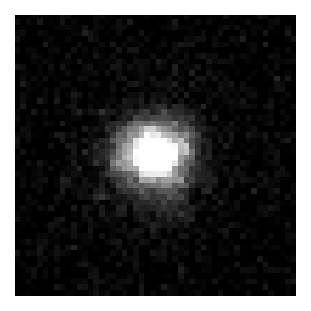}};
        \node[anchor=north, transform shape, white] at (imaging.north) {F158};
        \node[anchor=south, text width=144pt, align=center, black, transform shape] at (imaging.north) {Filter image};
    \end{scope}

    \draw[dashed, green, opacity=\integrallineopacity] (0,\lend,0) -- (imaging.south west);
    \draw[dashed, green, opacity=\integrallineopacity] (0,\lend,\axlength) -- (imaging.north west);
    
    \foreach \i in {\nplot,...,0} {
        \pgfmathsetmacro{\l}{\l0 + \i*\dl}
        \begin{scope}[canvas is xz plane at y=\l]
            \ifthenelse{\i=\pull}{
                \def\y{0.25+\yyy}
            }{\def\y{\yyy}}; 
            \node [name=im\i, anchor=south west, fill=white, inner sep=\sep, transform shape] at (\xxx,\y) {\includegraphics[width=\imwidth]{datacube\i.png}};
        \end{scope}
    }

    \draw[->, thick] (0,0,0) -- (\axlength,0,0) node[pos=0.5, below]{\Large $\xi$};                      
    \draw[->, thick] (\axlength,0.05,0) -- (\axlength,\lwidth,0) node[pos=0.5, below]{\Large $\lambda$}; 
    \draw[->, thick] (0,0,0) -- (0,0,\axlength) node[pos=0.5, left]{\Large $\eta$}; 

    \fill[green,opacity=\integralopacity] (0,\lstart,\axlength)
    -- (0,\lend,\axlength)
    -- (\axlength,\lend,\axlength)
    -- (\axlength,\lstart,\axlength) -- cycle;
    
    \fill[green,opacity=\integralopacity] (\axlength,\lstart,0)
    -- (\axlength,\lstart,\axlength)
    -- (\axlength,\lend,\axlength)
    -- (\axlength,\lend,0) -- cycle;

    \pgfmathsetmacro{\lambda}{\lambdamin + \pull * \lambdastep}
    \pgfmathsetmacro{\lpull}{\l0 + \pull * \dl}
    \begin{scope}[canvas is xz plane at y=\lpull]
        \node [name=impull, anchor=south west, fill=white, inner sep=\sep, transform shape] at (\xxx,7.0) {\includegraphics[width=\imwidth]{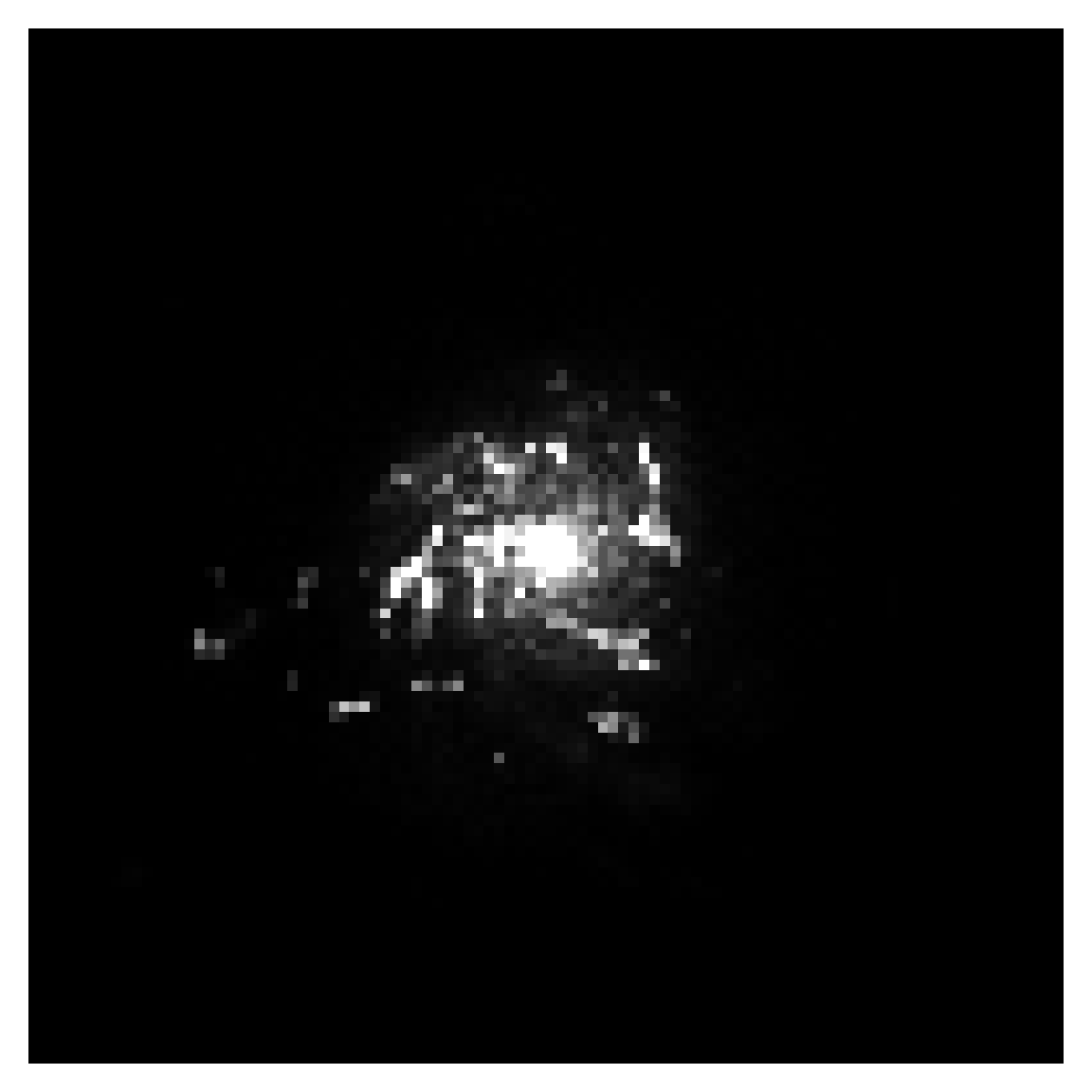}};
        \node[anchor=north, transform shape, white] at (impull.north) {\pgfmathprintnumber{\lambda}\,\AA};
        \draw[->, \monimcol, thick] (im\pull.north) to (impull.south);
        \node[anchor=south, text width=68pt, align=center, black, transform shape] at (impull.north) {Monochromatic image};
    \end{scope}

    \draw[dashed, green, opacity=\integrallineopacity] (\axlength,\lend,\axlength) -- (imaging.north east);
    \draw[dashed, green, opacity=\integrallineopacity] (\axlength,\lend,0) -- (imaging.south east);
    
    \begin{scope}[canvas is xy plane at z=0]
        \node[anchor=north west] at (1.1 * \axlength, -0.3\axlength) {{\LARGE Datacube}};
    \end{scope}

    \begin{scope}[canvas is xz plane at y=0]
        \fill[\loscolor]  (\xx,\yy) circle(1.5pt);
    \end{scope}
    \begin{scope}[canvas is yz plane at x=\xx]
        \node[anchor=center, name=spec1d] at (\losstart,\yy) {\includegraphics[width=3in]{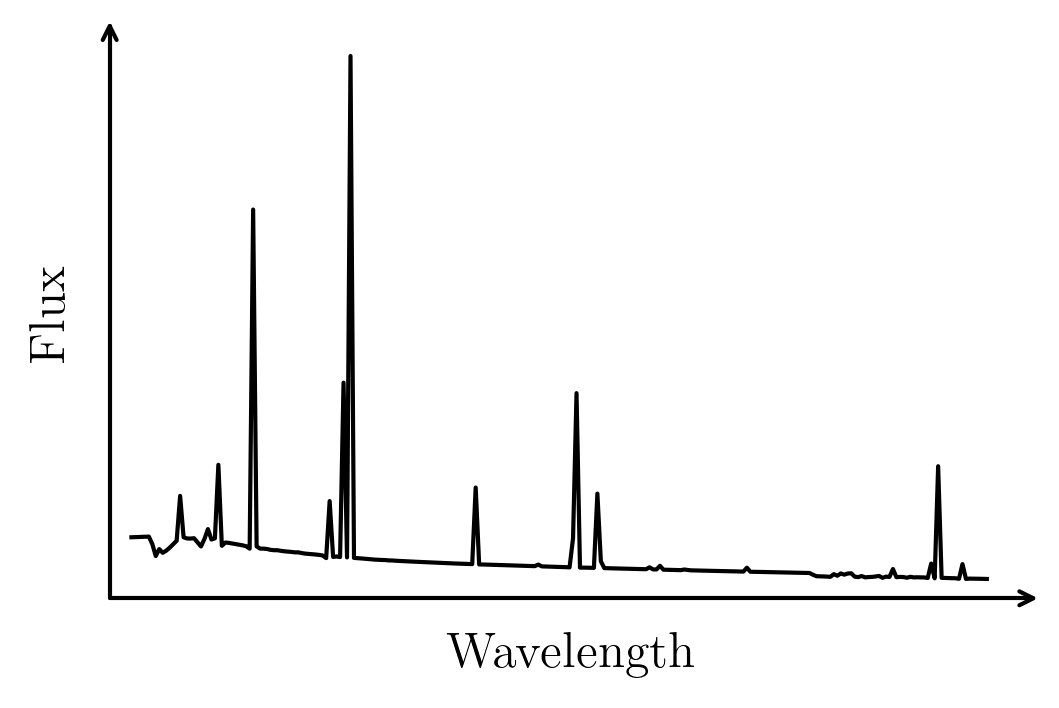}};
        \draw[->-=0.5, \loscolor, thick, transform shape ] (1.5*\lwidth, \yy) -- (\dashfactor*\lwidth, \yy);
        \draw[\loscolor, thick, dashed, opacity=0.7] (\dashfactor*\lwidth, \yy) -- (0, \yy);
        \draw[->,\loscolor, thick, transform shape] (0, \yy) -- (\losstart, \yy) node[name=text, pos=0.6]{};
        \node[anchor=center, align=center, black, text width=96pt, transform shape] at (text) {{\Large 1d spectrum at $(\xi,\eta)$}};
    \end{scope}
\end{tikzpicture}
\caption{An illustration of a datacube and its various products through projections. A datacube is a flux mapping of a scene of the sky in two spatial axes $(\xi,\eta)$ and the wavelength axis $\lambda$. If we slice the datacube at a constant wavelength, each slice is a monochromatic image at wavelength $\lambda$ (blue arrow). If we look through a particular spatial coordinate $(\xi,\eta)$ along the wavelength axis (red arrow), we will obtain a 1d spectrum. In imaging mode, we convolve the datacube with the corresponding PSF and the passband of a particular filter, and integrate it along the wavelength range of the filter (highlighted as the green box), obtaining a filter image as recorded by the detector. In slitless spectroscopy mode, the prism is put in the optical path rather than the filter. The datacube is convolved with the PSF and the prism passband then integrated along the prism dispersion curve (see Section~\ref{sec:datasim} for details) to obtain a 2d spectral image. Note that the horizontal axis of the 2d spectral image is actually not an exact mapping of the wavelength axis, but rather it is a mapping of both the spatial axis $\xi$ and wavelength axis $\lambda$ (and in some cases also $\eta$) through the dispersion curve. See Section~\ref{sec:prism}, Section~\ref{sec:Dp}, and Figure~\ref{fig:disp} for details. Illustration adapted from \cite{pon16}, with the same datacube used in Figure~\ref{fig:example}.}
\label{fig:datacube}
\end{figure*}

Type~Ia supernovae (SNe~Ia) are cosmological probes with a long and well-established history as standardizable candles for measuring the geometry of the universe (see \citealt{bra92,bra98,fre10} for review). Analysis of SNe~Ia observations resulted in the discovery of the accelerating expansion of the universe \citep[e.g.][]{rie98,sch98,per99}, attributed to a hitherto unknown force dubbed dark energy. Subsequent observations of the baryon acoustic peak in the correlation function of SDSS\footnote{Sloan Digital Sky Survey, \url{https://www.sdss.org/}} galaxies by \citealt{eis05} and the fluctuations in the cosmic microwave background by The \citealt{planck15} confirmed the original SNe result. This led to the concordance $\Lambda$CDM model: a universe that is geometrically flat, dominated by dark energy and cold dark matter (see \citealt{frie08} for review). Despite these advances, the nature of dark energy is still unknown. 


One of the primary goals of the \Romanc{} Space Telescope is to elucidate the nature of dark energy by carrying out a survey for Type~Ia supernovae. \Roman{} is a NASA flagship mission expected to be launched in late 2026. It has the same collecting area as the \Hubble{} Space Telescope (HST), but \Roman{}'s Wide Field Instrument (WFI) has a field-of-view (FoV) of 0.281\,square-degrees, approximately 100~times that of \Hubble{}'s Wide Field Camera 3 (WFC3) in the infrared. 
\Roman{}'s WFI consists of 18 $4\text{k pixel}\times 4\text{k pixel}$ Hawaii-4RG (H4RG) HgCdTe photodiode arrays, eight imaging filters that cover the wavelength range from 4800\,\AA{} to 23000\,\AA{}, a prism for slitless spectroscopy between  7500\,\AA{} and 18000\,\AA{}, and a grism sensitive between 10000\,\AA{} and 19300\,\AA{} \citep{ake19}. \Roman{} is thus a wide-field survey instrument.

The \Roman{} supernovae survey will be carried out as part of the core community High-Latitude Time Domain Survey (HLTDS).  Its high-cadence (every 5 days) will acquire imaging and spectroscopic data in  a 20 square-degree  area to a limiting magnitude of $25.4$\,AB--$26.7$\,AB magnitude \citep{ros21}. 
Data will be obtained at multiple roll angles of the observatory.  The details of the survey are yet to be finalized but some survey strategies have been presented by \cite{hou18, ros21, rub22}. \Roman{} is expected to discover tens of thousands of supernovae and obtain spectrophotometry for several thousand of them \citep{ros21}. SNe~Ia spectroscopy will be valuable for i)~confirming and refining photometric redshifts; ii)~improving the classification and subclassification of SNe~Ia \citep{fol13}, therefore expanding our understanding of the physics of the supernovae; iii)~improving the standardization of SNe~Ia light curves through twins embedding \citep[e.g.][]{fak15,boo21a,boo21b}; and iv)~expanding the available SNe spectral library particularly in the near infrared \citep{pie18,rub22}.

\subsection{Challenges of slitless spectroscopy}

One of the issues with slitless spectroscopy is the contamination of the target spectrum by nearby sources in the field. This makes the subtraction of the host galaxy from the supernova spectrum especially challenging because the SN~Ia spectrum will be blended with light from many regions in the host galaxy, not just the underlying light. An illustration of this problem is shown in Figure~\ref{fig:example} with simulated \Roman{} data.

Suppose a supernova occurs in a galaxy and we observe it in 6 \Roman{} filters (top panels of Figure~\ref{fig:example}) as well as with the prism (middle panel of Figure~\ref{fig:example}). If we extract the one-dimensional (1d) spectrum within the narrow red rectangle drawn in the middle panel, the resulting 1d spectrum is shown as the red line in the bottom-left panel of Figure~\ref{fig:example}. If there is no supernova, then the galaxy-only 1d spectrum of the same region is shown as the black line. Even though we have excluded a large part of the galaxy and concentrate on a small area around the supernova, the supernova spectrum is dominated by the galaxy spectrum.

It is therefore necessary to subtract the host galaxy spectrum from the blended spectrum to obtain the clean SN-only spectrum, to accurately measure the SN type and luminosity distance. This will in turn provide a more accurate determination of the dark energy parameters.

Determining the host galaxy spectrum underlying the SN~Ia is a variation on the problem of decontamination of blended spectra.  For example, \texttt{LINEAR} \citep{rya18} was originally developed to deblend  HST WFC3's infrared grism observations.  In that implementation, direct image positions of potential sources are used as the starting point for determining the transformation to spectral images at all wavelengths via the inversion of a large, sparse systems of linear equations, in order to extract the 1d spectrum integrated over a pre-defined area in the direct image.
\cite{jos22} recently applied \texttt{LINEAR} to simulated \Roman{} supernova spectra, and extracted the 1d spectra to recover the supernova parameters e.g. supernova redshift, phase, and line-of-sight dust extinction. They found that most failures to recover these parameters are due to their inability to fully subtract the underlying host spectrum. In this paper we describe the development of a host galaxy subtraction algorithm for \Roman{} slitless spectrographic data that is sufficiently accurate that removal of the underlying host spectra should not be a significant error in the measurement of cosmological parameters.


The algorithm makes use of a novel feature of the \Roman\ HLTDS in which observations of a host galaxy are obtained at numerous observatory roll angles. This can break the degeneracy between spatial and spectral information to reconstruct the underlying 3d scene of the host galaxy. We illustrate this in Figure~\ref{fig:obs} where we show spectra of a host galaxy observed at 7 different roll angles, one of which, at $195.4^\circ$ (outlined in red) has a SN at peak brightness. We set aside the SN+galaxy spectrum and use the others with no supernova contamination to reconstruct the plausible underlying scene that can reproduce the observed spectra. We then use the reconstructed scene to predict what the galaxy-only spectrum would look like at the roll angle where the supernova appears, allowing us to subtract the host-galaxy spectrum and obtain the supernova-only spectrum.


During the course of the \Roman{} surveys, the observatory roll angle could either follow the natural roll (changing by $\sim$1\,deg/day) or a set pattern approximating the natural roll. This is a detail yet to be finalized, but in any case the availability of observations of the same galaxy at many roll angles is a unique characteristic of the \Roman{} HLTDS, making our approach possible.  

\subsection{Overview of our approach}
In order to demonstrate our approach, we simulate host galaxy spectra at different observatory roll angles and use the simulated spectra to reconstruct the datacube, that is a  representation of the host galaxy flux densities in three-dimensions (3d): 2 spatial axes, the positions on the sky, $(\xi,\eta)$ and one spectral axis for wavelength $\lambda$ (see Figure~\ref{fig:datacube} for illustration). We inject a template supernova spectrum into the scene, so that the set of simulated spectroscopy includes one spectrum of the SN at peak plus host galaxy and the remainder are host galaxy alone, as shown previously in Figure~\ref{fig:obs}.


For simplicity, we create the model datacube of the galaxy assuming that each pixel in the galaxy has the same spectral energy distribution scaled with radius as a S\'{e}rsic power law \citep{ser63} with index $n=1$, semi-major axis of 3\,kpc, semi-minor axis of 1.5\,kpc, and integrated absolute luminosity of $-21$\,mag in the SDSS $z$-band. As a template for this simple galaxy we use the spectrum of IC~4553 from the \cite{bro14} spectral atlas. We use \texttt{Pandeia} \citep{pon16} to generate the datacube specified as such. The resulting datacube consists of monochromatic images uniformly spaced by 10\,\AA{} in the restframe between 3500\,\AA{} to 23500\,\AA. The spacing is lower than the spectral sampling of the prism at its bluest wavelength (see Figure~\ref{fig:specsamp}) and is sufficient for interpolation purposes.

The next step is to take the model datacube and use it to create simulated images by passing it through a model of the \Roman{} optical and detector systems.  This is described in more detail in Section~\ref{sec:datasim}. 
These transformations are linear in nature and, as we shall see in Section~\ref{sec:forward}, can be combined into a single operator, $\model$, that acts upon the datacube $\vect{\cube}$ to \textit{project} it into spectrum $\vect{\spec}$:
\begin{equation}
    \vect{\spec} = \model\cube.
    \label{eq:projection}
\end{equation}
Here both $\vect{\cube}$ (a 3d matrix) and $\vect{\spec}$ (a 2d matrix) are \textit{vectorized}, i.e. transformed into one-dimensional vectors. The projection of a datacube into a spectrum is illustrated in Figure~\ref{fig:datacube}.

To accurately map all the various operations that transform a datacube into spectrum, the datacube must sample the scene at a higher spatial and spectral resolution than does \Roman{}. In practice, we will thus have more pixels in our model datacube than there are pixels in all the combined dispersed images of the scene observed by \Roman{}. The reconstruction is then an \textit{ill-posed problem}, which has no unique solution (Section~\ref{sec:rec}). We approach this problem in a probabilistic manner, in which we use Bayes' Theorem to introduce a prior probability to constrain the solution within physical plausibility (Section~\ref{sec:rec}). To construct such a plausible prior, we use \Roman{} images of the same scene taken with multiple filters. We employ cross-validation techniques to optimize the parameters of the prior probability, with the aim that the reconstructed datacube can be used to accurately predict spectra at new rotation angles. Because this is our main goal to reconstruct the underlying datacube, the physical plausibility of the reconstructed datacube is not really important to achieve this goal.\footnote{See Figure~2 in \cite{rub21}, which shows a reconstructed high-resolution galaxy model shows various unphysical features that are suppressed when the model is convolved with the PSF and the detector pixel.}

Our scene reconstruction approach is similar to those of \cite{out20}, who also build a forward model of galaxy datacube projection into a 2d spectrum, but with an assumed light distribution and internal kinematic model of the galaxy as part of their forward model. The inferred parameters are then the parameters of the galaxy model, rather than the underlying datacube itself. Similarly, \cite{nev24} build a forward model of a point source projection into a 2d spectrum for ground-based observations, with the aim of inferring atmospheric transmission parameters from slitless spectroscopic observations. We generalize these earlier works by not making any assumption about the datacube that we want to reconstruct. The approach is thus applicable to reconstruct arbitrary scenes from slitless spectroscopy data. This is similar to \cite{bon11} who build a forward model of galaxy datacube projection into Integral Field Spectrograph (IFS) data and use regularized fitting to deconvolve the data and reconstruct the underlying datacube.



\subsection{Mathematical notations}
\label{sec:math}
In this paper we denote non-scalar entities such as matrices and vectors as upright (non-italicized) boldfaced characters. Two- or higher-dimensional matrices are denoted by bold upper case ($\A$), while vectors and one-dimensional column matrices are denoted by bold lower case ($\vecbold{a}$). Because of this upper/lower case differentiation between matrices and vectors, individual elements of matrices and vectors will still be written out with their corresponding cases, i.e. $A_{ij}$ for an element of a matrix and $a_{i}$ for an element of a vector. The transpose of a matrix is denoted as $\A\transpose$. We will also vectorize matrices, viz. converting two or higher-dimensional matrices of dimension $N_1\times N_2 \times\cdots\times N_d$ into column vectors of dimension $N_1 N_2\cdots N_d\times 1$. This operation will be denoted as $\vect{\A} = \vectorize{\A}$. The inverse operation of vectorization is matricization, denoted as $\A = \matricize{\A}{M}{N}$ for a matricization of an $MN\times 1$ vector $\vect{\A}$ into an $M\times N$ matrix $\A$. We denote the vectorized matrices as the same character but in lower case to reflect the conversion into vectors. The detail of these operations, which includes how we define them and how to transform between indices, is given in Appendix~\ref{app:vec}. In Appendix~\ref{app:acronyms} we provide a table of mathematical symbols used in this paper as well as a list of acronyms and a glossary of key terminologies used in this paper.

\section{Two-dimensional spectrum formation}
\label{sec:datasim}


We do not yet have \Roman{} data with which to test our method for subtracting the host contribution to a supernova spectrum, so we must create our own simulated data. In this section, we show how we create synthetic \Roman{} slitless spectroscopic observations using the known properties of the \Roman{} optical system and detector.


A scene is represented by a datacube $\datacube$, a mapping of flux densities at spatial coordinates $(\xi,\eta)$ and wavelength $\lambda$, i.e.
\begin{equation}
    \datacube\equiv\datacube(\lambda,\xi,\eta),
\end{equation}
with, for instance, the photon flux density in photons/s/cm$^{2}$/\AA{}. This is the photon flux density before it enters the \Roman{} optical system, and thus is free of instrumental effects such as convolution with PSF or filter bandpasses. The datacube already includes any astrophysical effects before it reaches the telescope (e.g. absorption effects by interstellar matter), and is normalized to a specific magnitude in a photometric band. For now we assume that the datacube has also been geometrically transformed such that it aligns to the observatory roll angle and dither (we discuss how to do this in Section~\ref{sec:Dp}).

An illustration of a datacube and its various projections is shown in Figure~\ref{fig:datacube}. A datacube can be projected into data space according to the observing mode. In imaging mode,  the datacube is convolved with the corresponding PSF and the passband of a particular filter, then integrated along the wavelength range of the filter. In slitless spectroscopy mode---the main subject of this paper---in addition to the PSF and spectroscopic passband convolution, the integration is performed along a slightly curved line called the trace, which is a projection of the coordinate system $(\lambda,\xi,\eta)$ in the datacube into the data space coordinates. We will discuss obtaining the spectrum from the trace in Sections~\ref{sec:specform} and \ref{sec:prism}.

In addition to representing the scene as a datacube and modelling its projection into spectra, the other ingredients are the properties of the \Roman{} optical system, namely the PSF, passband, and the prism dispersion curve. We describe how these are modelled in this Section. Later in Section~\ref{sec:forward} we will also show that the projection from datacube into a spectrum is linear in nature and can be expressed as matrix operations.

\subsection{Spectrum formation}
\label{sec:specform}

A beam of light from a source at position $(\xi,\eta)$ in the datacube $\datacube{}$ passes through the \Roman{} optical system, is dispersed by the prism, and reaches the detector on the focal plane, forming the observed spectrum. As the beam of light passes through the optical system, it is blurred by convolution with the PSF $\mathcal{P}(\lambda,\xi,\eta)$:
\begin{equation}
    \label{eq:conv}
    (\datacube\star\mathcal{P})(\lambda,\xi,\eta) = \iint\datacube(\lambda,\xi-\xi',\eta-\eta')\mathcal{P}(\lambda,\xi,\eta)\,\dd\xi'\dd\eta'.
\end{equation}
The light is then dispersed by the prism on the focal plane, where it is recorded by the detector.

Let $(\kappa,\mu)$ denote the \textit{continuous} two-dimensional coordinates in the detector space. Suppose that the datacube spatial coordinates $(\xi,\eta)$ are already transformed such that they are at the same scale and orientation as the detector space. The transformation from detector coordinates $(\kappa,\mu)$ back to datacube space is then
\begin{equation}
    \begin{aligned}
        \xi(\lambda,\kappa) &= \kappa - \kappa_\lambda(\lambda;\lambda_0,\kappa_0),\\
        \eta(\mu)           &= \mu,
    \end{aligned}
    \label{eq:traceline}
\end{equation}
where $\kappa_\lambda$ is the \textit{prism dispersion curve}, i.e. a function that maps the relation between wavelength $\lambda$ and its location in data space given the coordinate $\kappa_0$ of a reference wavelength $\lambda_0$. The dispersion curve $\kappa_\lambda$ is derived using the prism spectral sampling curve $\dd\lambda/\dd\kappa$ and will be discussed further in Section~\ref{sec:prism}. 

For many instruments, the direction of dispersion, or the spectral trace, is a line close to parallel to one axis of the detector (usually the horizontal or $x$-axis), with a very slight curvature \citep{rya18}. For simplicity, we assume that the trace of the \Roman{} slitless mode is in the horizontal direction only, with no dispersion in the vertical axis, hence $\eta = \mu$. Thus, for the purposes of this discussion, we ignore the small curvature that will need to be accounted for in actual practice. We adopt the FITS\footnote{Flexible Image Transport System, \url{https://fits.gsfc.nasa.gov/}} convention of placing the centres of pixels at integer values of the continuous coordinates system $\kappa$ and the size of each pixel is one unit.

Let $\mathcal{S}$ denote the formed two-dimensional spectral image in this continuous detector space. The observed brightness at $(\kappa,\mu)$ is then \citep{LL:AB-009, pon16}
\begin{equation}
   \label{eq:spectra}
   \mathcal{S}(\kappa,\mu) = \tau\int A_\text{\rm eff}(\datacube\star\mathcal{P})\left(\lambda,\xi(\lambda,\kappa),\eta(\mu)\right)\,\dd\lambda.
\end{equation}
The quantities in this Equation are listed in Table \ref{tab:im_params}, and $\datacube\star\mathcal{P}$ is the convolution of the datacube $\datacube$ with PSF $\mathcal{P}$ previously described in Equation~\ref{eq:conv}.  

The spectrum $\mathcal{S}$ represents a continuous version of the observed spectral image before it is sampled by the discrete array of pixels that comprise the detector. The data that we are actually going to observe is $\spec$, the pixelated version of $\mathcal{S}$, i.e. the convolution of $\mathcal{S}$ with the pixels and sampled at the centers of the pixels (cf. \citealt{LL:AB-009, LL:LL-089}). Each pixel measures the observed count:
\begin{equation}
   \label{eq:specsamp1}
   S_{km} = \iint\mathcal{S}(\kappa,\mu)\Pi(\kappa-k,\mu-m)\dd{\kappa}\dd{\mu}\,+\,\epsilon_S(k,m),
\end{equation}
where $k$ and $m$ are integers representing the \textit{discrete} two-dimensional coordinates in the detector space and are equivalent to the column and row indices of the detector array, $\Pi(x,y)$ is the pixel kernel function which has zero support outside the boundaries of the pixel $(k,m)$, here taken to be
\begin{equation}
\Pi(x,y) = \Pi_1(x)\Pi_1(y),
\label{eq:pixkern}
\end{equation}
where 
\begin{equation}
\Pi_1(a) = 
\begin{cases}
   1 & \text{\rm for }|a| < 0.5,\\
   0 & \text{\rm elsewhere},
\end{cases} 
\label{eq:smearing}
\end{equation}
and finally $\epsilon_S(k,m)$ is the noise that arise from the measurement process, consisting of photon and detector noise. These depend on the adopted noise model, which will be discussed in Section~\ref{sec:noise}.

\begin{deluxetable}{lp{0.75\columnwidth}}
\tablecaption{Parameters used in Equation~\ref{eq:spectra}.}
\label{tab:im_params}
\tablehead{\colhead{Term} & \colhead{Description}}
\startdata
$\mathcal{S}$                   & Continuous two-dimensional spectrum\\
$\tau$                          & Integration time in second\\
$A_\text{\rm eff}(\lambda)$     & Telescope effective area in m$^2$, which includes mirror transmissivities, passband, and detector quantum efficiency\\
$\datacube(\lambda,\xi,\eta)$   & Photon flux density at wavelength $\lambda$ and spatial coordinate $(\xi,\eta)$, in photons/s/m$^{2}$/\AA\\
$\mathcal{P}(\lambda,\xi,\eta)$ & Point spread function (PSF)\\
\enddata
\end{deluxetable}

\subsection{Model of the optical system and detector}
\label{sec:model}

Having described in the previous subsection how a 2d spectrum is formed, we now discuss the parameters needed to simulate it.

\subsubsection{The effective area}
\label{sec:area}
\begin{figure}
    \centering
    \includegraphics[width=\columnwidth]{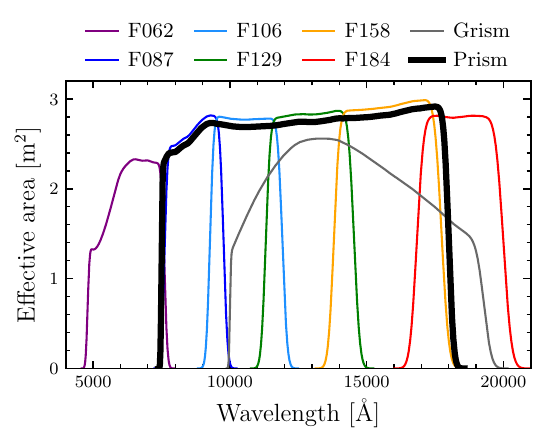}
    \caption{The effective area of the Wide Field Instrument (WFI) filters, grism, and prism. Not shown are the F213 filter and the very wide F146 filter.}
    \label{fig:area}
\end{figure}

The effective area of an optical system is the equivalent collecting area needed in a system of 100\% efficiency to obtain the same sensitivity as that of the system under discussion. In Figure~\ref{fig:area}, we show the effective area of \Roman{} using the individual WFI filters, grism, and prism. The data used to make this figure are available at the \Roman{} web page of the Goddard Space Flight Center (GSFC)\footnote{\url{https://roman.gsfc.nasa.gov/science/RRI/Roman_effarea_20210614.xlsx}}. The effective area already includes all transmissivities of the optical system, the area of the primary mirror and the quantum efficiency of the detector. 

\subsubsection{The prism}
\label{sec:prism}
\begin{figure}
    \centering
    \includegraphics[width=\columnwidth]{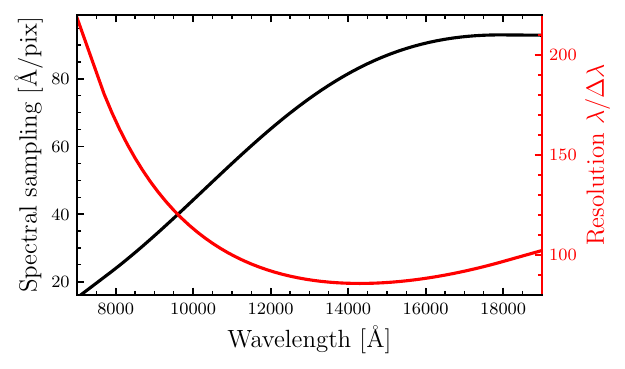}
    \caption{The spectral sampling (black line) at the center of the field of view (FoV) of the WFI and the two-pixel resolution (red line) of the prism. The axis on the left corresponds to the black line, while the axis on the right (in red) corresponds to the red line. Note that although the spectral sampling changes by a factor of $\sim$4 across the prism bandpass, the resolution changes by only a factor $\sim$2 across the same bandpass range.}
    \label{fig:specsamp}
\end{figure}

\begin{figure}
    \centering
    \includegraphics[width=\columnwidth]{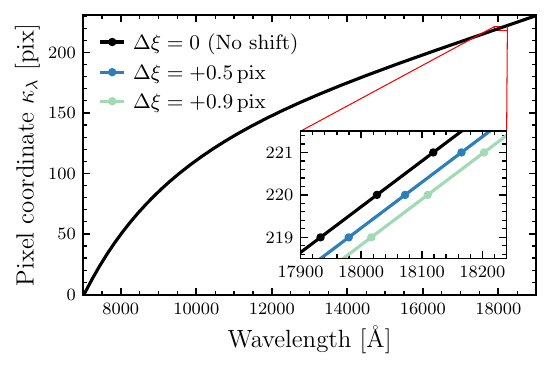}
    \caption{The dispersion curve $\kappa_\lambda$ as a function of wavelength $\lambda$. By dithering the telescope, we sample different sub-pixel phases of the spectrum. In the inset plot, the shift in wavelength is shown for two different sub-pixel dithers, 0.5\,pix (light blue line) and 0.9\,pix (light green line).}
    \label{fig:disp}
\end{figure}

\begin{figure*}
    \centering
    \includegraphics[width=\textwidth]{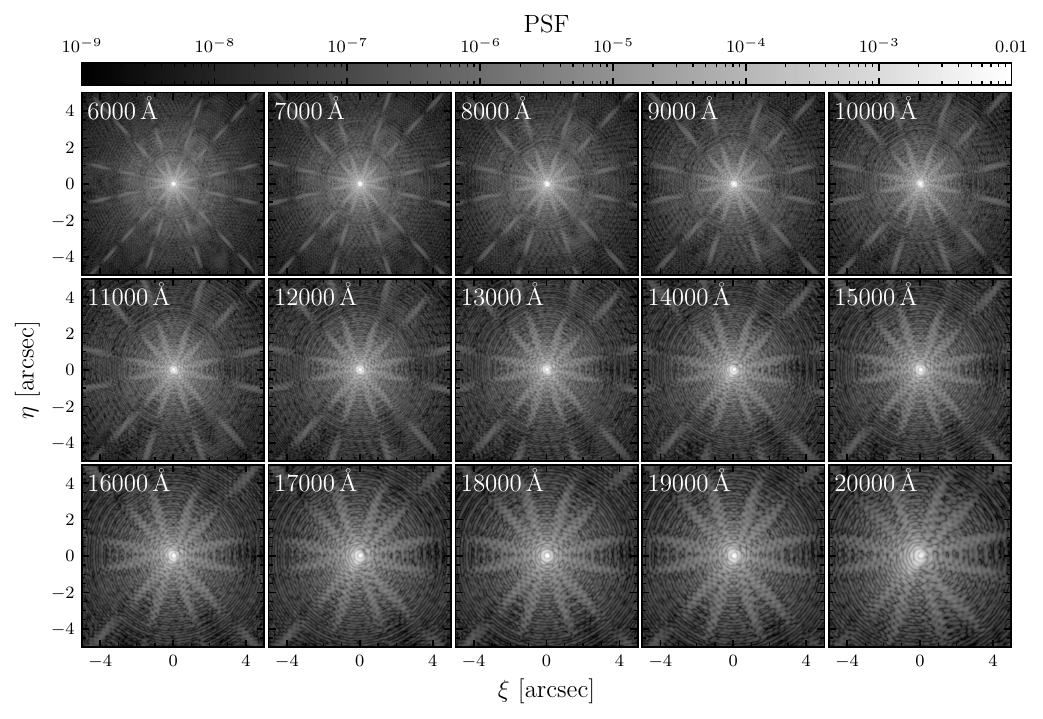}
    \caption{The \Roman{} point spread function (PSF) for monochromatic light with wavelengths as shown in the upper left corner of each panel, and for the central point of the WFI01 detector (See Figure~\ref{fig:wfi} for the location of the detectors on the focal plane). In each panel, the image covers a field of view of $10\,\text{\rm arcsec}\times 10\,\text{\rm arcsec}$. The PSF model is created using \texttt{STPSF} \citep{per11, per12, per14} using the \Roman{} optical model.}
    \label{fig:psf}
\end{figure*}

\begin{figure}
    \centering
    \includegraphics[width=\columnwidth]{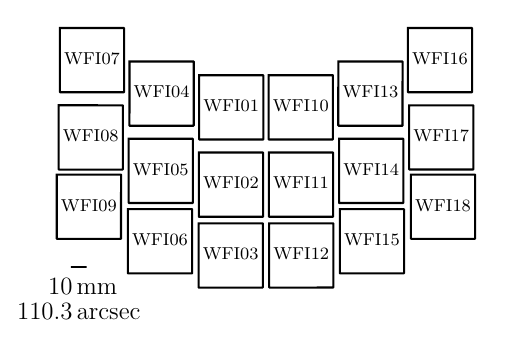}
    \caption{The arrangement of the 18 H4RG detectors that comprise the WFI. The detectors are labeled WFI01 through WFI18. The scale bar at the bottom left corner indicates the physical size and field of view of the detectors.}
    \label{fig:wfi}
\end{figure}

The spectral sampling of the prism at various points on the detector array is available at the \Roman{} web page at GSFC\footnote{\url{https://roman.gsfc.nasa.gov/science/2020-01/GRISM_PRISM_Dispersion_190510.xlsx}}. Figure~\ref{fig:specsamp} shows the current spectral sampling (black line) at the center of the field of view (FoV) of the WFI and resolution (red line) of the prism as of this writing.

We can use the spectral sampling curve to calculate the pixel coordinate $\kappa_\lambda$ as a function of wavelength $\lambda$ given the reference wavelength $\lambda_0$ and reference coordinate $\kappa_0$, by writing
\begin{equation}
    \frac{\dd\lambda}{\dd\kappa} = g(\lambda)\,\text{\rm \AA\,pix}^{-1},
    \label{eq:specsamp3}
\end{equation}
where $g(\lambda)$ is the spectral sampling curve shown by the black line in Figure~\ref{fig:specsamp}. Rearranging Equation~\ref{eq:specsamp3} and integrating both sides, we get
\begin{equation}
   \kappa_\lambda(\lambda;\lambda_0,\kappa_0) = \kappa_0 + \int^{\lambda}_{\lambda_0}\,\frac{\dd{\lambda}}{g(\lambda)}.
   \label{eq:dispcurve}
\end{equation}

Using Equation~\ref{eq:dispcurve}, it is then straightforward to calculate where a beam of light with a certain wavelength $\lambda$ will fall on the focal plane. In principle the reference wavelength $\lambda_0$ and its location on the data space $\kappa_0$ can be obtained from jointly analyzing the direct image and spectrum of a source with well-known spectral properties (cf. \citealt{kun09a, kun09b}). In this work we set $\lambda_0 = 7380$\,\AA{} at $\kappa_0 = -0.5$\,pix. These values are such that $7380$\,\AA{}, the bluest edge of the prism effective area, is located at the leftmost edge of the formed 2d spectrum image.

The black line in Figure~\ref{fig:disp} shows the dispersion curve $\kappa_\lambda$ plotted as a function of wavelength $\lambda$ using the abovementioned values of $\lambda_0$ and $\kappa_0$. If we dither the telescope, i.e. slightly shifting the pointing of the telescope, we sample different sub-pixel phases of a higher-resolution ``true'' spectrum\footnote{Just like sub-pixel dithering of images can be thought as sampling different phases of a higher-resolution ``true'' image, sub-pixel dithering of spectral images means we are sampling different sub-pixel phase of a higher-resolution spectrum.}. In the inset plot of Figure~\ref{fig:disp} we show how sub-pixel dithering $\Delta\xi$ change the spectral sampling. We show the shifts in wavelengths at the same pixel coordinates, if we shift the telescope by 0.5\,pix (light blue line) and 0.9\,pix (light green line). Because of the non-uniform spectral sampling, the shift in wavelength will vary.

\subsubsection{The point spread function (PSF)}
\label{sec:PSF}
The expected PSF of the \textit{Roman} WFI can be estimated if we know the pupil function and the coefficients of the Zernike polynomials that aberrate the PSF. These coefficients have been determined for each chip of the WFI, for 5 key positions (center, top left, top right, bottom left, and bottom right) on each detector, and for wavelengths spaced between 7500\,\AA{} and 18000\,\AA{}. Using these coefficients, we can construct a PSF for any wavelength, for any chip, and at any location on the chip by interpolating the coefficients in the location and wavelength space. This is what the package \texttt{STPSF} \citep{per11,per12,per14} does for us using \Roman{} optical model. In Figure~\ref{fig:psf} we show the PSF for the central point of the WFI01 detector, one of the 18 detectors that comprise the WFI. The arrangement of the detectors is shown in Figure~\ref{fig:wfi}.

We can see from Equation~\ref{eq:spectra} and Figure~\ref{fig:psf} that although simple ray tracing can tell us that the light with wavelength $\lambda$ will fall at position $(\kappa(\lambda), \mu)$, the PSF convolution spreads the light into the neighboring pixels as well. Thus at a particular position $(\kappa(\lambda),\mu)$, the light from all wavelengths in the WFI prism wavelength range actually contributes to the total flux at that position, although the majority of the contribution will come from the light with wavelength $\lambda$ \citep{LL:AB-005}. This situation is made more complex in the case of extended objects, where neighboring celestial points will have overlapping spectra with differing wavelengths projected into a single pixel.  In standard slit spectroscopy the width of the slit and the PSF produce a small wavelength overlap or blurring, but in slitless spectroscopy it is a combination of the full width of the object (convolved with the PSF) and the length of the dispersed spectrum that determine the extent of the overlap. 

\subsection{The noise model}
\label{sec:noise}

Observed spectra will have noise.   We simulate noisy spectra by first generating a noise-free spectrum as specified in the previous Sections, and then for each integrated count $S(k,m)$ at pixel coordinate $(k,m)$ we add the noise
\begin{equation}
S_\text{\rm noisy}(k,m) = S_\text{\rm noise-free}(k,m) + r_G(0|1)\sigma_S(k,m),
\label{eq:noise1}
\end{equation}
where $r_G(0|1)$ is a random number drawn from a standard normal distribution (the mean is zero, and 1 unit of standard deviation), and $\sigma_S(k,m)$ is the standard deviation of the count measurement at $(k,m)$.

The measurement variance of a single observation is the quadrature summation of the variances of the flux measurement, the measured sky background $n_\text{\rm bg}$\footnote{The measured sky background $n_\text{\rm bg}$ should be multiplied by two when the data is sky-subtracted}, as well as the total detector noise $r$ at $(k,m)$:
\begin{equation}
   \sigma_S(k,m) = \sqrt{S_\text{\rm noise-free}(k,m) + n_\text{\rm bg} + r^2}.
   \label{eq:noise2}
\end{equation}
The total detector noise $r$ can include (but is not limited to) the readout noise (RON), dark noise, and analog-to-digital converter (ADC) noise \citep{rau07, rau15, mos20}. We model the noise as uncorrelated white noise and do not (yet) include $1/f$ pink noise that characterizes the slope in the detector power spectrum or the correlations caused by interpixel capacitance and charge spreading. This will be the subject of future publications.

Here we are assume that the sky background is uniform in spectral data space $(k,m)$ even though the sky background flux is not uniform in wavelength. This is a valid assumption in slitless spectroscopy, because background light coming from all points in the sky will be dispersed on top of each other. In a particular pixel, the background light is effectively white noise, being a summation of background light at all wavelengths dispersed from directions along the trace line.

\subsection{Spectrum formation as a linear forward model}
\label{sec:forward}

\begin{figure}
    \centering
    \begin{tikzpicture}
    \matrix (C) [draw, xshift=8em, yshift=6.0em, fill=white]{
    \node{$D_{N_\lambda11}$}; & \node{$D_{N_\lambda12}$}; & \node{$\cdots$}; & \node{$D_{N_\lambda1J}$};\\
    \node{$D_{N_\lambda21}$}; & \node{$D_{N_\lambda21}$}; & \node{$\cdots$}; & \node{$D_{N_\lambda2J}$};\\
    \node{$\vdots$};  & \node{$\vdots$};  & \node{$\ddots$}; & \node{$\vdots$};\\
    \node{$D_{N_\lambda I1}$}; & \node{$D_{N_\lambda I2}$}; & \node{$\cdots$}; & \node{$D_{N_\lambda IJ}$};\\
    };
  
    \matrix[draw, xshift=3.5em, yshift=2em, fill=white]{
    \node{$D_{211}$}; & \node{$D_{212}$}; & \node{$\cdots$}; & \node{$D_{21J}$};\\
    \node{$D_{221}$}; & \node{$D_{222}$}; & \node{$\cdots$}; & \node{$D_{22J}$};\\
    \node{$\vdots$};  & \node{$\vdots$};  & \node{$\ddots$}; & \node{$\vdots$};\\
    \node{$D_{2I1}$}; & \node{$D_{2I2}$}; & \node{$\cdots$}; & \node{$D_{2IJ}$};\\
    };
    
    \matrix (A) [draw, fill=white]{
    \node(a){$D_{111}$}; & \node{$D_{112}$}; & \node{$\cdots$}; & \node(b){$D_{11J}$};\\
    \node(c){$D_{121}$}; & \node{$D_{122}$}; & \node{$\cdots$}; & \node(d){$D_{12J}$};\\
    \node{$\vdots$};  & \node{$\vdots$};  & \node{$\ddots$}; & \node{$\vdots$};\\
    \node{$D_{1I1}$}; & \node{$D_{1I2}$}; & \node{$\cdots$}; & \node{$D_{1IJ}$};\\
    };

    \draw [->,red,opacity=0.2, line width=5pt] (a.west) -- (b.east) -- (c.west) -- (d.east);
    \draw [->,blue,opacity=0.2, line width=5pt] (A.north west)++(0,2em) -- +(7.0em,6em);

    \draw (-10pt, 75pt) node {\Huge{$\iddots$}};
    \draw (100pt, 10pt) node {\Huge{$\iddots$}};
    \draw (-90pt, 30pt) node {\Huge{$\D = $}};

\end{tikzpicture}
    \caption{Direction of 3d datacube vectorization. An $N_\lambda\times I\times J$ datacube $\D$ can be thought of as $N_\lambda$ layers of $I\times J$ matrices $\D_l$, each layer representing a monochromatic image at wavelength $\lambda_l$. Vectorization of $\D$ starts from the first layer, i.e. $\D_1$, shown here as the frontmost layer, then continues to the next layer etc. (blue arrow) until the last layer, $\D_{N_\lambda}$. In a given layer, the direction of vectorization is shown by the red arrow on the frontmost layer. We start from the first element (the top left element) then to the right along the same row, down to the start of the next row, until the last element in the last row of the layer. The result of vectorizing $\D$ is shown in Equation~\ref{eq:vect}.}
    \label{fig:vect}
\end{figure}
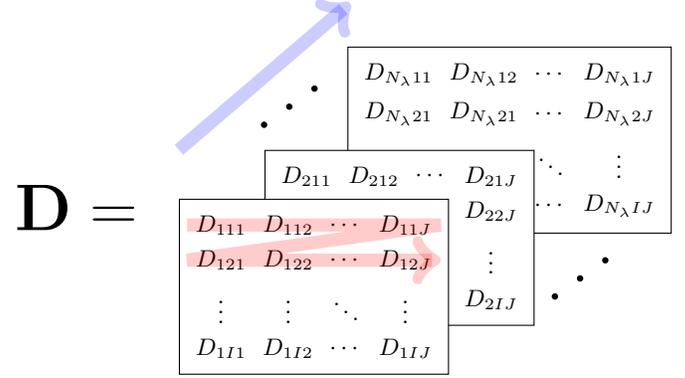

The formation of spectra as described in Equations~\ref{eq:spectra}--\ref{eq:specsamp1} are linear operations in nature and these can be expressed in terms of a projector matrix that operates on the datacube to transform it into a two-dimensional spectrum. We will build this projector matrix by rewriting the operations described in the previous Section in terms of matrices and combine them together into a single projector matrix.

It might already be obvious to readers, but before we go on we would like to point out that although in the previous Section the datacube $\datacube$ is presented as a continuous function, in real applications it is \textit{always digitized}, viz. binned into \textit{discrete three-dimensional grid} of flux densities, denoted in this paper as $\D$. It is indexed by a predefined, discrete, and uniform $I\times J$ grid of celestial coordinate $\coords = \{(\xi_i,\eta_j)\}_{i,j=1}^{I,J}$ and $N_\lambda$ grid of wavelength $\wave = \{\lambda_l\}_{l=1}^{N_\lambda}$. The number of elements in $\D$ is thus $P = N_\lambda\times I\times J$. A slice of monochromatic image at wavelength $\lambda_l$, denoted as $\D_l$, is then a matrix with dimension $I\times J$.  

In order to conveniently express Equations~\ref{eq:spectra}--\ref{eq:specsamp1} as linear algebraic operations, we \textit{vectorize} the datacube $\D$, i.e. convert $\D$ into a $P\times 1$ column vector:
\begin{equation}
\begin{split}
    \cube & = \vectorize{\D}\\
          & = \left(D_{111}, \cdots, D_{1IJ}, D_{211}, \cdots, D_{N_\lambda 00}, \cdots, D_{N_\lambda IJ}\right)\transpose
\end{split}
    \label{eq:vect}
\end{equation}
The direction of the vectorization is detailed in Figure~\ref{fig:vect}. We vectorize $\D$ layer-by-layer, i.e. first $\D_1$, then $\D_2$ etc. until $\D_{N_\lambda}$. For a given layer, we start from the top left element, i.e. $D_{l11}$, then move horizontally along the same row. In a similar manner, the spectrum $\spec$ with dimension $K\times M$ is converted into a $KM\times 1$ column vector:
\begin{equation}
\begin{split}
    \vect{\spec} & = \vectorize{\spec}\\
                 & = \left(S_{11}, \cdots, S_{1M}, S_{21}, \cdots, S_{2M}, \cdots, S_{K1}, \cdots, S_{KM}\right)\transpose
\end{split}
\end{equation}

With $\D$ and $\spec$ vectorized, we can then express Equations~\ref{eq:spectra}--\ref{eq:specsamp1} in terms of a matrix operation
\begin{equation}
\vect{\spec} = \model\cube,
\label{eq:shd}
\end{equation}
where $\model{}$ is the operator we intend to build in this Section, which projects the (vectorized) datacube $\cube$ into (vectorized) spectrum $\vect{\spec}$. The output vector $\vect{\spec}$ can be returned to matrix form, i.e.
\begin{equation}
\spec = \matricize{\spec}{K}{M},
\end{equation}
to properly display the 2d spectral image. As we shall see shortly, $\model$ is a large, though sparse, matrix; however, many sparse matrix implementations can handle at most two-dimensional matrices. This is the second reason for our vectorizing $\D$ and $\spec$. For more details about how vectorization and matricization (the operation that transform a vector back into matrix form) are defined, as well as how to transform between vector and matrix indices, see Appendix~\ref{app:vec}.

The operator $\model{}$ consists of three operations:
\begin{equation}
    \model{} = \Dp\times\C\times\L,
    \label{eq:H}
\end{equation}
where $\L$ is a $P\times P$ matrix that performs the interpolation of fluxes at the wavelengths of a particular dither of the observation (using the datacube fluxes at common wavelength grid $\wave$), $\C$ is a $P\times P$ matrix that performs convolution of the datacube with the PSF $\PSF$, and $\Dp$ is an $N\times P$ operator that performs shift, rotation, dispersion, and pixel sampling.

In order to accurately map all the various transformations in Equation~\ref{eq:H}, the datacube $\D$ must then sample the scene at a resolution higher than the \Roman{} pixel scale. In this Section we illustrate our methods using a datacube $50\,\text{\rm pix}\times 50\,\text{\rm pix}$ in size, oversampled by a factor of 2, making the dimension of an image slice $\D_l$ in the datacube $I\times J = 100\,\text{\rm sub-pixel}\times 100\,\text{\rm sub-pixel}$. 

The spectra in the datacube are uniformly sampled at the exact center of each pixel of the detector, which means that the wavelengths are sampled in non-uniform spacing following the dispersion curve $\kappa$ shown in Figure~\ref{fig:disp}. This corresponds to pixel coordinates $\kappa_\lambda = \left\{0,\right.$ $1$, $2,\ldots,204$,$\left.205\right\}$. As mentioned in Section~\ref{sec:prism}, this is calculated using Equation~\ref{eq:dispcurve}, with $\lambda_0 = 7380$\,\AA{} and $\kappa_0 = -0.5$\,pix. The wavelengths are sampled such that they are at the native \Roman{} pixel resolution (i.e., oversampled by a factor of 1) because, as we shall see later in this Section, we can use linear interpolation to interpolate the fluxes at wavelengths that fall at sub-pixel coordinates, e.g. at pixel coordinates $\kappa_\lambda = \left\{-0.25,\right.$$0.25$, $0.75$, $1.25$, $1.75$, $2.25, \ldots, 203.75$, $204.25$, $204.75$, $\left.205.25\right\}$, which twice oversamples the wavelength space.

At the native \Roman{} pixel resolution, there are $N_\lambda = 206$ wavelengths ranging from $7380$\,\AA{} to $18700$\,\AA{}. The total number of elements on the grid $\D$ is then $P = 206\times 100\times 100 = 2\,060\,000$.    


We will now discuss each of the operators in Equation~\ref{eq:H} from right to left.

\subsubsection{Linear interpolation matrix $\L$}
\label{sec:interp}
The wavelength grid $\wave$ that we defined previously in Section~\ref{sec:forward} is calculated for a single dither position. However, as discussed in Section~\ref{sec:prism}, if the telescope is dithered, we sample different sub-pixel phases of the ``true'' higher-resolution spectrum. Consequently, for each dithered 2d spectral image we will have a different set of wavelengths that fall on each pixel on that image. Multiple spectra observed with multiple dithers will then have their own grid of wavelengths. We will need to connect this grid of wavelengths at various dithers, denoted as $\wavetilde = \{\tilde{\lambda}_{\tilde{l}}\}_{\tilde{l}=1}^{N_{\tilde{\lambda}}}$, to the common grid of wavelengths $\wave$ that we previously defined.

To do this, we use linear interpolation to calculate the flux density at wavelength grid $\wavetilde$. Suppose a given wavelength $\tilde{\lambda}_{\tilde{l}}$ is bracketed by the interval $(\lambda_{l[\tilde{l}]}, \lambda_{l[\tilde{l}]+1})$ in the common
wavelength grid (here $l[\tilde{l}]$ denotes the index of the wavelength of the closest common grid to $\tilde{\lambda}_{\tilde{l}}$ but not larger). The fluxes of the datacube slice $\tilde{\D}_{\tilde{l}}$ are then the weighted summation of the fluxes at $\D_{l[\tilde{l}]}$ and $\D_{(l[\tilde{l}]+1)}$:
\begin{equation}
   \tilde{\D}_{\tilde{l}} = w_{\tilde{l},1}\D_{l[\tilde{l}]} + w_{\tilde{l},2}\D_{(l[\tilde{l}]+1)},
   \label{eq:d_interp}
\end{equation}
where
\begin{equation}
    \begin{aligned}
        w_{\tilde{l},1} &= \frac{\lambda_{(l[\tilde{l}]+1)} - \tilde{\lambda}_{\tilde{l}}}{\lambda_{(l[\tilde{l}]+1)} - \lambda_{l[\tilde{l}]}},\\
        w_{\tilde{l},2} &= 1 - w_{\tilde{l},1}.
        \label{eq:w_interp}
    \end{aligned}
\end{equation}
The weights $w_{\tilde{l},1}$ and $w_{\tilde{l},2}$ are then the normalized distances between a given wavelength $\tilde{\lambda}_{\tilde{l}}$ and its bracketing wavelengths $(\lambda_{l[\tilde{l}]}, \lambda_{(l[\tilde{l}]+1)})$.

With $\tilde{\D}$ and $\D$ vectorized, we can express Equation~\ref{eq:d_interp} as the matrix operation
\begin{equation}
    \cubetilde = \L\cube,
    \label{eq:dld}
\end{equation}
such that the operator $\L$ acts upon $\cube$ to generate $\cubetilde$ using linear interpolation as prescribed above.

The vectorized datacube $\cubetilde$ is a vertical concatenation of $N_{\tilde{\lambda}}$ vectorized monochromatic images $\cubetilde_{\tilde{l}}:$
\begin{equation}
    \cubetilde = 
    \begin{pmatrix}
        \cubetilde_{1}\\
        \cubetilde_{2}\\
        \vdots\\
        \cubetilde_{N_{\tilde{\lambda}}},
    \end{pmatrix}.
\end{equation}
Each monochromatic image $\cubetilde_{\tilde{l}}$ in turn is the weighted summation of the fluxes in the bracketing monochromatic image in the common grid wavelength $\wave$:
\begin{equation}
    \cubetilde = 
    \begin{pmatrix}
        w_{1,1}\cube_{l[1]} + w_{1,2}\cube_{(l[1]+1)}\\
        w_{2,1}\cube_{l[2]} + w_{2,2}\cube_{(l[2]+1)}\\
        \vdots\\
        w_{N_{\tilde{\lambda}},1}\cube_{l[N_{\tilde{\lambda}}]} + w_{N_{\tilde{\lambda}},2}\cube_{(l[N_{\tilde{\lambda}}]+1)}
    \end{pmatrix}.
    \label{eq:interp}
\end{equation}
The right hand side of Equation~\ref{eq:interp} above can be rearranged such that it is equivalent to the right hand side of Equation~\ref{eq:dld} if
\begin{equation}
\L =
\begin{pmatrix}
   w_{1,1}\eye & w_{1,2}\eye & \zero       & \cdots & \zero  & \zero\\
   \zero       & w_{2,1}\eye & w_{2,2}\eye & \cdots & \zero  & \zero\\
   \vdots      & \vdots      & \vdots      & \ddots & \vdots & \vdots\\
   \zero       & \zero       & \zero       & \cdots & w_{N_{\tilde{\lambda}-1},2}\eye & \zero\\
   \zero       & \zero       & \zero       & \cdots & w_{N_{\tilde{\lambda}},1}\eye & w_{N_{\tilde{\lambda}},2}\eye
\end{pmatrix}.
\end{equation}
Here $\eye$ is the identity matrix, and $\zero$ is a zero-filled block matrix with appropriate shape. As we can see, the matrix is very sparse, with most of the blocks being either zeroes or identity matrices.

\begin{figure}
    \centering
    \includegraphics[height=0.90\textheight]{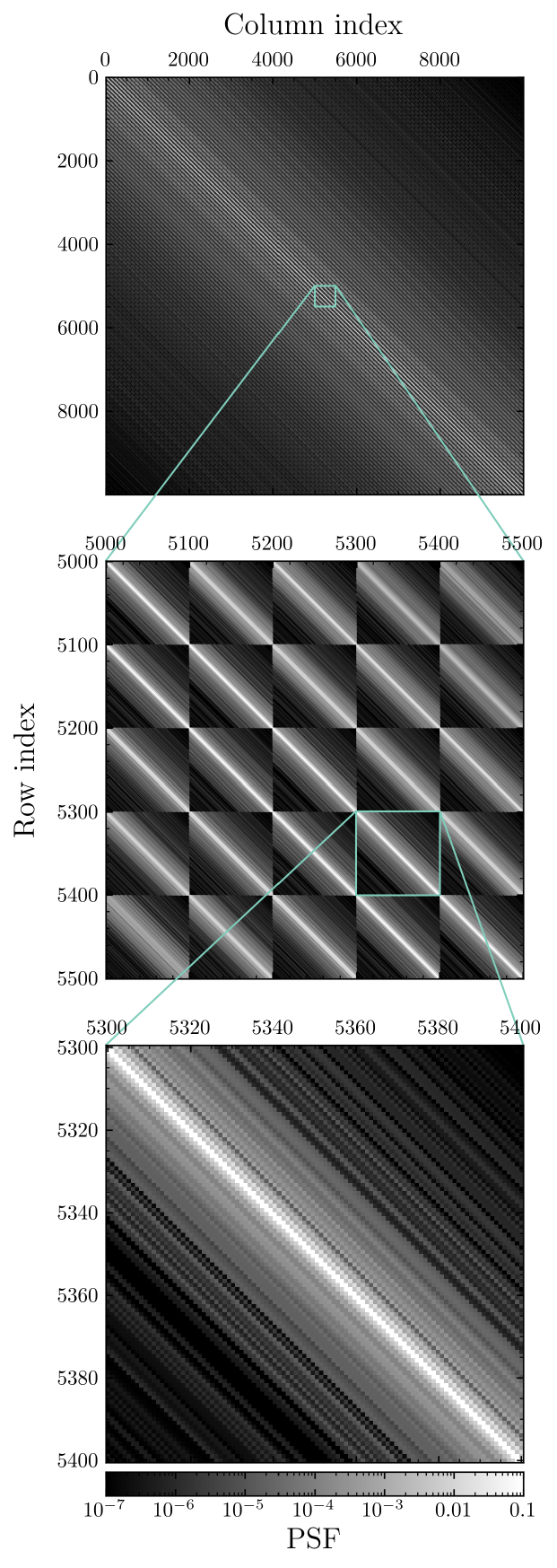}
    \caption{An example of a convolution operator $\C_l$, which convolves a vectorized image $\cube_l$ with the corresponding PSF $\PSF_l$, here shown in three different levels of detail. The top panel shows the whole matrix, the middle panel shows $5\times 5$ blocks around the main diagonal blocks, and the bottom panel shows a single (main diagonal) block. Here the size of the image $\D_l$ is $100\times 100$, thus the size of the corresponding vectorized image $\cube_l$ is $10000\times 1$.}
    \label{fig:c}
\end{figure}

\begin{figure}
    \centering
    \includegraphics[width=\columnwidth]{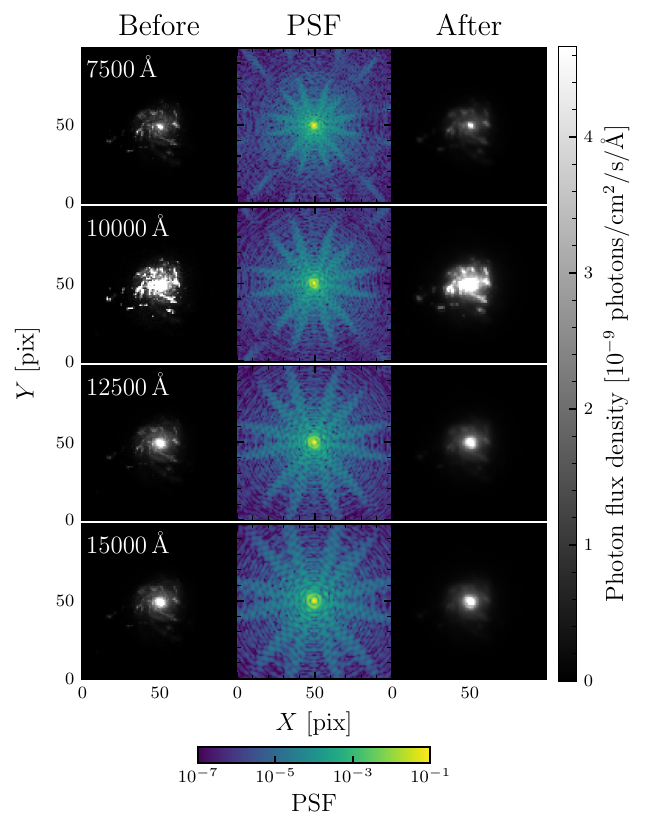}
    \caption{Examples of monochromatic images before (left column) and after (right column) the corresponding convolution operator $\C_{\lambda}$ is applied to the images, with each corresponding PSF shown in the middle column. Due to the decreasing resolution with wavelength, redder images are blurred more than the bluer images. The integrated spectrum of this galaxy is shown in the upper panel of Figure~\ref{fig:integratedflux_priorspec}. The before and after images are shown in linear scale, while the PSFs are shown in logarithmic scale (see the corresponding color bars at the right and bottom of the plot).}
    \label{fig:convolution}
\end{figure}

\subsubsection{The convolution operator $\C{}$}
\label{sec:C}


The matrix $\C$ is a $P\times P$ operator that convolves the vectorized datacube $\cube$ with the PSF $\PSF$ for all wavelength slices by matrix multiplication. $\C$ is a block diagonal matrix, with the main-diagonal blocks consisting of the convolution operator for each wavelength slice $\lambda_l$:
\begin{equation}
    \C{} = 
    \begin{pmatrix}
    \C_1   & \zero  & \cdots & \zero\\
    \zero  & \C_2   & \cdots & \zero\\
    \vdots & \vdots & \ddots & \vdots\\
    \zero  & \zero  & \cdots & \C_{N_\lambda}\\
    \end{pmatrix}.
\end{equation}
The submatrix $\C_l$ is a doubly-blocked Toeplitz matrix, i.e. a matrix with constant block diagonals (both main diagonal and off-diagonals) and the blocks themselves also have constant elements on the main diagonal and off-diagonals (this is described in more detail in Appendix~\ref{app:conv}). Figure~\ref{fig:c} shows a selected submatrix $\C_l$ in progressively finer detail. We can see in the middle panel of Figure~\ref{fig:c} that the submatrices are constant in diagonal, and the elements of the submatrices are themselves constant along the diagonal (cf. Equation~\ref{eq:bca}).

In Figure~\ref{fig:convolution} we show the datacube for selected wavelength slices before and after the vectorized images are multiplied by the corresponding convolution operator $\C_l$. Images in redder wavelengths are more blurred than those in bluer wavelengths due to the wider full-width at half-maximum (FWHM) of the PSF at redder wavelengths (see the middle column of Figure~\ref{fig:convolution}).

The submatrix $\C_l$ shown in Figure~\ref{fig:c} is dense, i.e. all elements in this submatrix are non-zero, because in this example the size of the PSF $\PSF_l$ is twice the size of the image $\D_l$, making all pixels in the image participate in the convolution at a particular sub-pixel $(i,j)$.

A fully dense $\C_l$ makes the convolution a bottleneck in computing the forward model $\model$, and will increase the computation time by $\mathcal{O}(n^2)$, where $n$ is the width of the image $\D_l$. The density of $\C_l$ will further affect the density of $\model$ and the computing time for using $\model$ to project the datacube into spectrum $\vect{\spec}$ will also increase. This creates a trade-off between the speed with which we can create the reconstruction and the accuracy of the reconstruction. We discuss this further in Appendix~\ref{app:psf}. 

When we convolve an interpolated datacube $\cubetilde$, we use the PSF $\PSF$ as a function of the wavelength grid $\wavetilde$. The convolved datacube $\cubetilde'$ in terms of the input datacube $\cube$ is then
\begin{equation}
    \begin{aligned}
        \cubetilde' =&~\C\L\cube\\
        =&~
        \begin{pmatrix}
            w_{1,1}\C_1 & w_{1,2}\C_1 & \zero       & \cdots & \zero  & \zero\\
            \zero       & w_{2,1}\C_2 & w_{2,2}\C_2 & \cdots & \zero  & \zero\\
            \vdots      & \vdots      & \vdots      & \ddots & \vdots & \vdots\\
            \zero       & \zero       & \zero       & \cdots & w_{N_{\tilde{\lambda}},1}\C_{N_{\tilde{\lambda}}} & w_{N_{\tilde{\lambda}},2}\C_{N_{\tilde{\lambda}}}
        \end{pmatrix}\\
        &~\times
        \begin{pmatrix}
            \cube_1\\
            \cube_2\\
            \vdots\\
            \cube_{N_\lambda}
        \end{pmatrix}.
    \end{aligned}
\end{equation}


\subsubsection{Geometric transformation, dispersion, and pixel sampling matrix $\Dp{}$}
\label{sec:Dp}

The operator $\Dp{}$ disperses the convolved (vectorized) datacube $\cube' = \C\cube$ onto the detector and pixelizes it to the pixel scale of the detector. The operator also simultaneously aligns the datacube to the detector by shifting and rotating it according to the dither and observatory roll angle.

To do all of these, we use the variable-pixel linear reconstruction algorithm commonly known as Drizzle \citep{fru02}, because it conserves the total flux of an object after transformation.

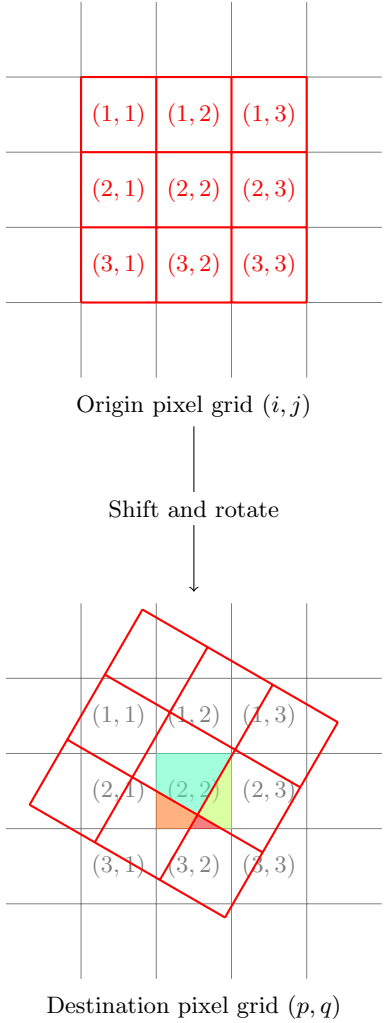
\begin{figure}
    \centering
    \begin{tikzpicture}
        \pgfplotsset{colormap/jet}
        
        \def\nx{3}   
        \def\ny{\nx} 
        
        \pgfmathsetmacro{\nxx}{\nx+2} 
        \pgfmathsetmacro{\nyy}{\ny+2} 
        
        \def\shiftlength{2}
        
        \pgfmathsetmacro{\xtext}{0.5 * \nxx}
        \pgfmathsetmacro{\ytext}{0.5 * \nyy}
        
        \pgfmathsetmacro{\frameshifty}{\nxx + \shiftlength + 1}
        
        \def\rotation{-30}  
        \def\xshift{-0.3cm} 
        \def\yshift{0.25cm} 
        \def\xrot{2.5}      
        \def\yrot{2.5}      
        \def\fillopacity{0.5}
        
        \pgfkeys{/pgf/number format/.cd,fixed,set thousands separator={}, precision=0}
        
        \draw[gray, very thin] (0.001,0.001) grid +(\nxx-0.002,\nyy-0.002);
        \draw[red, thick] (1,1) grid +(\nx,\ny);
        \foreach \x in {1,...,\nx}{
            \foreach \y in {1,...,\ny}{
                \pgfmathsetmacro{\i}{\ny - \y + 1}
                \pgfmathsetmacro{\j}{\x}
                \node at (\x+0.5,\y+0.5) [red] {$(\pgfmathprintnumber{\i},\pgfmathprintnumber{\j})$};
            }
        }
        
        \node[black, below, name=origintext] at (\xtext, -0.1){Origin pixel grid $(i,j)$};
        
        \draw[->] (origintext.south) -- +(0,-1.1*\shiftlength) node[pos=0.5, text width=72pt, fill=white, align=center]{Shift and rotate};
        
        \begin{scope}[yshift=-\frameshifty cm]
            \draw[gray, very thin] (0.001,0.001) grid +(\nxx-0.002,\nyy-0.002);
        
            \foreach \x in {1,...,\nx}{
            \foreach \y in {1,...,\ny}{
                \pgfmathsetmacro{\i}{\ny - \y + 1}
                \pgfmathsetmacro{\j}{\x}
                \node at (\x+0.5,\y+0.5) [gray] {$(\pgfmathprintnumber{\i},\pgfmathprintnumber{\j})$};
            }
        }
            
            \begin{scope}
                \clip (2,2) rectangle +(1,1);
                \foreach \x in {1,...,\nx}{
                    \foreach \y in {1,...,\ny}{
                        \pgfmathsetmacro{\i}{\ny - \y}
                        \pgfmathsetmacro{\j}{\x - 1}
                        \pgfmathsetmacro{\k}{\nx * \i + \j} 
                        \pgfmathtruncatemacro{\colidx}{\k*1000/\nx/\ny}
                        \fill[/pgfplots/color of colormap=\colidx of jet, opacity=\fillopacity, rotate around={\rotation:(\xrot,\yrot)}, xshift=\xshift, yshift=\yshift] (\x,\y) rectangle +(1,1);
                    }
                }
            \end{scope}
            
            \begin{scope}[rotate around={\rotation:(\xrot,\yrot)}, xshift=\xshift, yshift=\yshift] 
                \draw[red, thick] (1,1) grid +(\nx,\ny);
            \end{scope}
            
            \node[black, below] at (\xtext, -0.1){Destination pixel grid $(p,q)$};
        \end{scope}
    \end{tikzpicture}
    \caption{An illustration of how Drizzle works. The origin pixel grid (drawn as the red $3\times 3$ grid) is shifted and rotated onto the destination pixel grid. The flux in a given origin pixel is then redistributed into the overlapping destination grid. In the example above, the middle pixel in the destination grid receives a fraction of the fluxes from each of four partially overlapping origin pixels, each colored accordingly. These are pixels $(2,2)$, $(2,3)$, $(3,2)$, and $(3,3)$ in the origin grid.}
    \label{fig:drizzle}
\end{figure}

Drizzle redistributes the fluxes in the origin pixel grid onto the destination grid by calculating the fractions of the overlap area between the origin and the destination pixel grid. We illustrate this in Figure~\ref{fig:drizzle}. 

Suppose that a $3\times 3$ grid of pixels is rotated by $\phi$ and shifted by $(\Delta x, \Delta y)$ respectively in the $x$- and $y$-axis direction. A pixel coordinate $(x, y)$ in the origin frame of reference is mapped into the destination coordinate $(x', y')$ by a linear transform:
\begin{equation}
    \begin{pmatrix}
        x' - x'_c\\
        y' - y'_c
    \end{pmatrix}
    =
    \varsigma
    \begin{pmatrix}
    \cos\phi & -\sin\phi\\
    \sin\phi &  \cos\phi
    \end{pmatrix}
    \begin{pmatrix}
        x - x_{c}\\
        y - y_{c}
    \end{pmatrix}
    +
    \varsigma
    \begin{pmatrix}
    \Delta x\\
    \Delta y
    \end{pmatrix}.
    \label{eq:trans}
\end{equation}
where $(x'_{c}, y'_{c})$ and $(x_c, y_c)$ respectively are the centers of the original and transformed images, and $\varsigma$ is the pixel oversampling factor. In Figure~\ref{fig:drizzle} we set $\varsigma = 1$, i.e. the destination pixel has the same size as the origin pixel.

\begin{figure}
    \centering
    \includegraphics[width=\columnwidth]{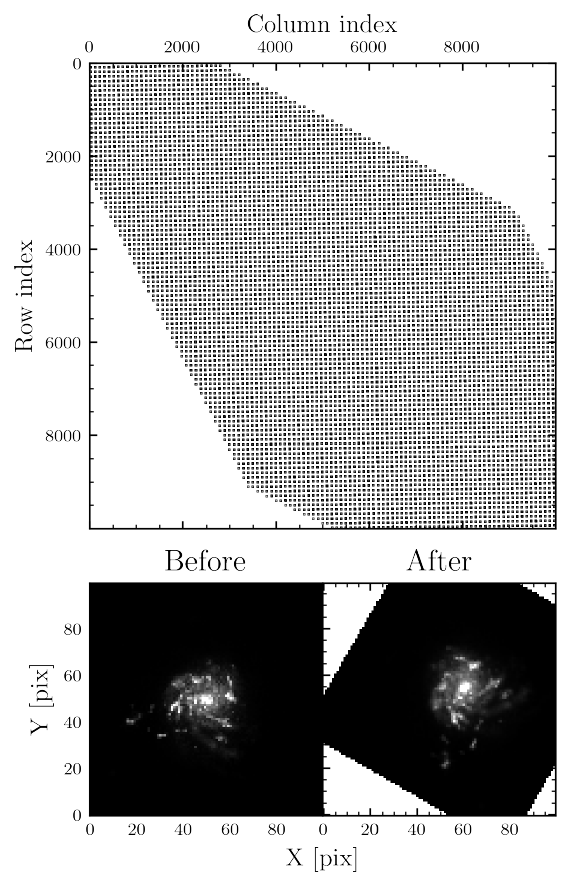}
    \caption{The sparsity structure of an example geometric transformation matrix $\G$ (top panel) and example of the transformed image after $\G$ is applied (bottom panels). In this example $\G$ rotates a vectorized image $\cube_l$ by $\phi = 60^\circ$ and shifts by $\vecbold{\xi} = (10\,\text{\rm pix},5\,\text{\rm pix})$. The matrix $\G$ maps the flux redistribution from the origin pixels (column indices) to the destination pixels (row indices). Here the size of the image is $100\times 100$, making the size of the corresponding vectorized image equal $10000\times 1$; $\G$ becomes a $10000\times 10000$ matrix. This image is oversampled by a factor of 2, making the actual image size in \Roman{} pixel scale $50\times 50$ (but this transformation keeps the oversampling factor. The pixelization into \Roman{} pixel scale is illustrated in Figures~\ref{fig:downsample}--\ref{fig:downsample2}).}
    \label{fig:G}
\end{figure}

\begin{figure}
    \centering
    \begin{tikzpicture}
        \pgfplotsset{colormap/jet}
        
        \def\nx{4}   
        \def\ny{\nx} 

        \def\step{1.0}
        \def\stepd{2.0}
            
        \pgfmathsetmacro{\nxx}{\nx+2} 
        \pgfmathsetmacro{\nyy}{\ny+2} 
        
        \pgfmathsetmacro{\nxxx}{3} 
        \pgfmathsetmacro{\nyyy}{3} 
            
        \def\shiftlength{2}
        
        \pgfmathsetmacro{\xtext}{0.5 * \nxx}
        \pgfmathsetmacro{\ytext}{0.5 * \nyy}
        
        \pgfmathsetmacro{\frameshiftx}{0}
        \pgfmathsetmacro{\frameshifty}{-\stepd*\nyyy - \shiftlength - 1}
        
        \def\rotation{-30}  
        \def\xshift{-0.3cm} 
        \def\yshift{0.25cm} 
        \def\xrot{2.5}      
        \def\yrot{2.5}      
        \def\fillopacity{0.5}
        
        \pgfkeys{/pgf/number format/.cd,fixed,set thousands separator={}, precision=0}
        
        \draw[gray, very thin, step=\step] (0.001,0.001) grid +(\nxx-0.002,\nyy-0.002);
        \draw[red, thick] (\step,\step) grid +(\nx,\ny);
        
        \node[black, below, name=origintext] at (\xtext, -0.1){Origin pixel grid $(i,j)$};
        
        \draw[->] (origintext.south) -- +(0,-1.1*\shiftlength) node[pos=0.5, anchor=center, text width=84pt, fill=white, align=center]{Shift, rotate, and pixelize};
        
        \begin{scope}[xshift=\frameshiftx cm, yshift=\frameshifty cm]
            \draw[gray, very thin, step=\stepd] (0.001,0.001) grid +(\stepd*\nxxx-0.002,\stepd*\nyyy-0.002);
        
            \begin{scope}
                \clip (2,2) rectangle +(\stepd,\stepd);
                \foreach \x in {1,...,\nx}{
                    \foreach \y in {1,...,\ny}{
                        \pgfmathsetmacro{\i}{\ny - \y}
                        \pgfmathsetmacro{\j}{\x - 1}
                        \pgfmathsetmacro{\k}{\nx * \i + \j} 
                        \pgfmathtruncatemacro{\colidx}{\k*1000/\nx/\ny}
                        \fill[/pgfplots/color of colormap=\colidx of jet, opacity=\fillopacity, rotate around={\rotation:(\xrot,\yrot)}, xshift=\xshift, yshift=\yshift] (\x,\y) rectangle +(\step,\step);
                    }
                }
            \end{scope}
            
            \draw[red, thick, rotate around={\rotation:(\xrot,\yrot)}, xshift=\xshift, yshift=\yshift] (\step,\step) grid +(\nx,\ny);
            
            \node[black, below] at (\xtext, -0.1){Destination pixel grid $(p,q)$};
        \end{scope}
    \end{tikzpicture}
    \caption{How pixelizing works with Drizzle. Here the origin pixel (red $4\times 4$ grid) is shifted and rotated onto the destination pixel grid, now twice the area of the origin grid. The middle pixel now receives a fraction of the fluxes from 12 partially overlapping origin pixels (each colored accordingly).}
    \label{fig:downsample}
\end{figure}
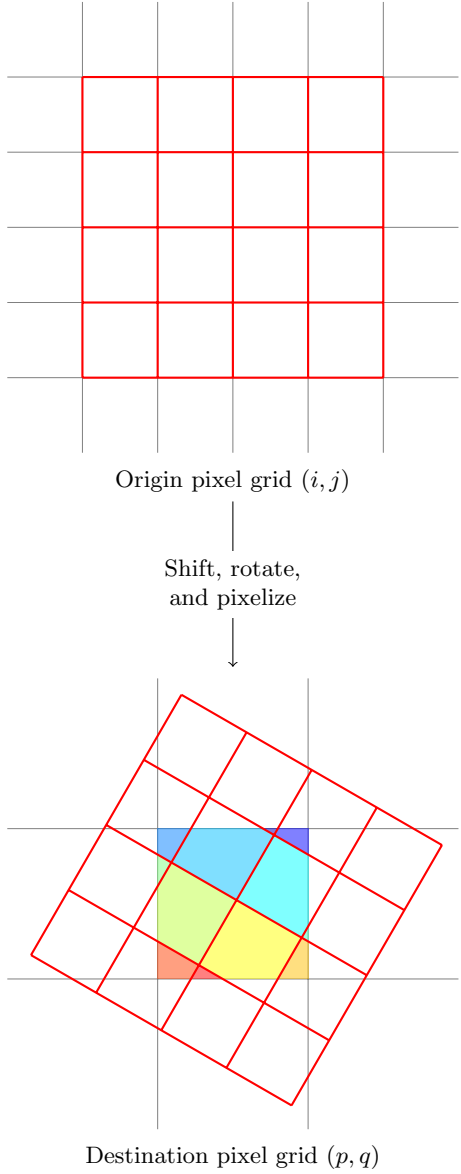

\begin{figure}
    \centering
    \includegraphics[width=\columnwidth]{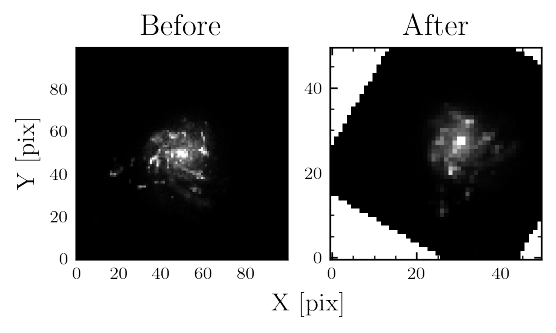}
    \caption{An example monochromatic image $\D_l$ before (left panel) and after (right panel) shift, rotation, and downsampling by a factor of 2 is applied. The shift and rotation is the same as in Figure~\ref{fig:G}, i.e. $\phi = 60^\circ$ and $\vecbold{\xi} = (10\,\text{pix},5\,\text{pix})$.}
    \label{fig:downsample2}
\end{figure}

The fluxes in the origin pixel grid are then distributed into the overlapping pixels in the destination grid:
\begin{equation}
    f'_{pq} = \sum_{i,j\in p,q} a_{pqij}f_{ij},
    \label{eq:drizzle}
\end{equation}
where $0 \leq a_{pqij} \leq 1$ is the fractional overlap of origin pixel $(i,j)$ in destination pixel $(p,q)$. Here the summation is performed over all origin pixels $(i,j)$ that overlap with the destination pixel $(p,q)$. For example in Figure~\ref{fig:drizzle}, after the shift and rotation is performed, the middle destination pixel, $(p,q) = (2,2)$, partially overlaps with four pixels in the origin grid. It then receives a fraction of the fluxes in each of these four pixels, hence $a_{pqij} > 0$ for these four overlapping pixels and zero otherwise.

We can express Equation~\ref{eq:drizzle} as a matrix operating on the vectorized array of fluxes that comprise the origin image, producing a vectorized array as the destination image. In the example shown in Figure~\ref{fig:drizzle}, we can denote the origin grid as $\F$ and index its pixels accordingly:
\begin{equation}
\F =
\begin{pmatrix}
    f_{11} & f_{12} & f_{13}\\
    f_{21} & f_{22} & f_{23}\\
    f_{31} & f_{32} & f_{33}
\end{pmatrix},
\end{equation}
and similarly with the destination grid, denoted as $\F'$. With both $\F$ and $\F'$ vectorized into respectively $\vect{\F}$ and $\vect{\F'}$, by inspecting each destination pixel $(p,q)$ and identifying which origin pixels $(i,j)$ have non-zero overlap with the given destination pixel $(p,q)$, we can form the operator $\G$ that performs the transformation given in Figure~\ref{fig:drizzle}, i.e. $\vect{\F'} = \G\,\vect{\F}$:
\begin{equation}
    \begin{aligned}
    \begin{pmatrix}
        f'_{1}\\
        f'_{2}\\
        f'_{3}\\
        f'_{4}\\
        f'_{5}\\
        f'_{6}\\
        f'_{7}\\
        f'_{8}\\
        f'_{9}
    \end{pmatrix}
    = &
    \begin{pmatrix}
        a_{11} & 0      & 0      & a_{14} & a_{15} & 0      & a_{17} & 0      & 0\\
        a_{21} & a_{22} & 0      & a_{24} & a_{25} & 0      & 0      & 0      & 0\\
        0      & a_{32} & a_{33} & 0      & a_{35} & a_{36} & 0      & 0      & 0\\
        0      & 0      & 0      & a_{44} & a_{45} & 0      & a_{47} & a_{48} & 0\\
        0      & 0      & 0      & 0      & a_{55} & a_{56} & 0      & a_{58} & a_{59}\\
        0      & 0      & a_{63} & 0      & a_{65} & a_{66} & 0      & 0      & 0\\
        0      & 0      & 0      & 0      & 0      & 0      & a_{77} & a_{78} & 0\\
        0      & 0      & 0      & 0      & 0      & a_{86} & 0      & a_{88} & a_{89}\\
        0      & 0      & 0      & 0      & 0      & a_{96} & 0      & 0      & a_{99}
    \end{pmatrix}\\
    &\times
    \begin{pmatrix}
        f_{1} & f_{2} & f_{3} & f_{4} & f_{5} & f_{6} & f_{7} & f_{8} & f_{9}
    \end{pmatrix}\transpose.
    \end{aligned}
    \label{eq:drizzle2}
\end{equation}
Here the pixel indices $(p,q)$ and $(i,j)$ are transformed into their respective vector indices $r$ and $k$ using Equation~\ref{eq:mnp}. 


Generally $\G$ is thus an $IJ \times IJ$ matrix that applies a prescribed geometric transformation to a vectorized $I\times J$ monochromatic image at wavelength slice $\lambda_l$, $\cube_l = \vectorize{\D_l}$, producing the transformed image $\cube'_l$ as output, i.e. $\cube'_l = \G\cube_l$. In the top panel of Figure~\ref{fig:G} we show the sparsity structure of $\G$ which transforms a $100\,\text{pix}\times 100\,\text{pix}$ image into an image with similar size, making $\G$ a $10000\times 10000$ matrix. As \cite{fru02} noted, very few origin pixels overlap a given destination pixel, resulting in sparse $\G$ matrix. 

The bottom panels of Figure~\ref{fig:G} compare an example galaxy image before and after transformation. Due to rotation and shift, some of the origin pixels are transformed out of the image boundaries, while pixels from outside boundaries get transformed into the image boundaries. Care should be taken in determining the image size to be large enough to ensure that all significantly bright pixels stay inside the datacube boundaries.

To downsample an image such that it simulates the integration of light over one pixel, we set e.g. $\varsigma = 1/2$ in Equation~\ref{eq:trans} to downsample the image by a factor of 2. We illustrate this in Figure~\ref{fig:downsample}, in which we shift and rotate a $4\times 4$ grid onto a destination grid of pixels, in which the pixels are twice as large as the pixels in the origin grid. Because of the larger size of the destination pixels, a single destination pixel receives a fraction of fluxes from more overlapping origin pixels. In Figure~\ref{fig:downsample2} we show the comparison before (left panel) and after (right panel) a transformation that includes downsampling is applied. 

\begin{figure}
    \centering
    \includegraphics[width=\columnwidth]{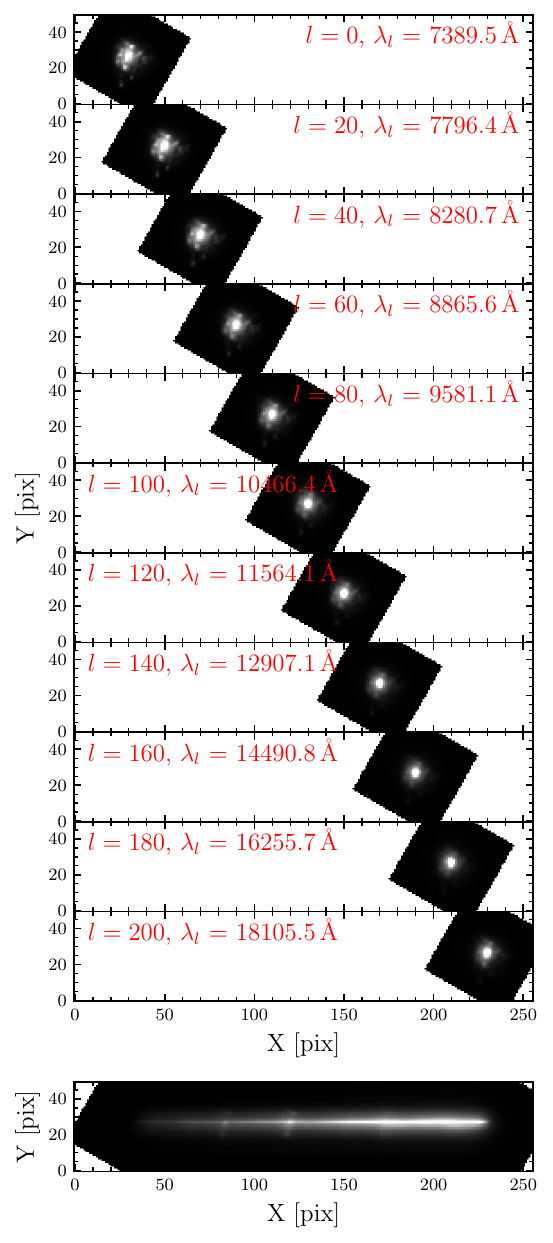}
    \caption{Drizzling the datacube onto the detector plane to form a 2d spectrum image. Each layer of monochromatic image $\D_l$, shown here for 11 selected wavelength, is shifted and rotated according to the observatory dither and roll angle, pixelized, then dropped along the trace. The drop point is determined using Equation~\ref{eq:traceline}. The bottom panel shows the 2d spectrum image formed from the summation of all the transformed mages.}
    \label{fig:specform}
\end{figure}

Prism dispersion of the datacube to form a 2d spectrum image is then a matter of Drizzling the convolved monochromatic images $\D'$ onto the focal plane, with each slice $\D'_{l}$ with wavelength $\lambda_{l}$ appropriately shifted along the trace line (cf. \citealt[Sect.~2.2]{nev24}). This shift is determined using Equation~\ref{eq:traceline}, which calculates where a beam of light will fall on the focal plane. We illustrate this in Figure~\ref{fig:specform}, in which we show as examples 11 convolved monochromatic images that have been geometrically transformed, downsampled, and appropriately shifted along the trace\footnote{The monochromatic images are convolved with the corresponding PSFs that have been rotated---using Drizzle---to opposite direction of the observatory roll angle, such that the PSF will align with the detector when the convolved datacube is later rotated and shifted along the trace.}. Each monochromatic image $\D'_{l}$ will then have its own Drizzle matrix, ${\Dp}_{l}$, all with the same geometric transformation and downsampling parameters but different shift along the trace depending on the wavelength $\lambda_l$ of $\D'_{l}$. The matrix ${\Dp}_{l}$ is then applied to the corresponding (vectorized) image $\cube'_{l} = \vectorize{\D'_{l}}$:
\begin{equation}
    \vect{\spec}_{l} = \tau A_\text{eff}(\lambda_{l})\left.\frac{\dd\lambda}{\dd\kappa}\right|_{\lambda_l}{\Dp}_{l}\cube'_{l},
\end{equation}
to form the layer of the 2d spectrum image for wavelength $\lambda_{l}$ only. Here $\tau$ is the integration time, $\dd\lambda/\dd\kappa$ is the spectral sampling (black curve in Figure~\ref{fig:specsamp}), and $A_\text{eff}$ is the passband of the prism (shown as the thick black curve in Figure~\ref{fig:area}). We show $\vect{\spec}_{l}$ in Figure~\ref{fig:specform} for 11 selected wavelengths. As discussed previously, the grid of wavelengths $\wave$ are uniform in pixel coordinates (but non-uniform in wavelength, due to the non-uniformity of the spectral sampling as shown in Figure~\ref{fig:specsamp}), thus they are Drizzled onto the detector plane at uniform distance from each other.

The 2d spectrum image is then formed by summing all these Drizzled monochromatic images over the wavelength grid $\wave = \{\lambda_{l}\}_{l=1}^{N_\lambda}$:
\begin{equation}
    \vect{\spec} = \tau \sum_{l=1}^{N_{\lambda}}A_\text{eff}(\lambda_{l})\left.\frac{\dd\lambda}{\dd\kappa}\right|_{\lambda_l}{\Dp}_{l}\C_l\cube_{l}.
    \label{eq:specsum}
\end{equation}
The result of this summation is shown at the bottom panel of Figure~\ref{fig:specform}. Note that in Equation~\ref{eq:specsum} we substitute $\cube'_l$ with $\C_l\cube_l$.

Expressed in terms of block matrices, the forward model operator $\model$ that acts upon the input datacube $\cube$ to transform it into spectrum $\vect{\spec}$ is then of the form
\begin{equation}
\model = 
\begin{pmatrix}
    \model_1 & \model_2 & \cdots & \model_{N_\lambda-1} & \model_{N_\lambda}
\end{pmatrix},
\label{eq:Hblock}
\end{equation}
which is a horizontal concatenation of the blocks $\model_l$, where
\begin{equation}
    \model_l = \tau A_\text{eff}(\lambda_l)\left.\frac{\dd\lambda}{\dd\kappa}\right|_{\lambda_l}{\Dp}_{l}\C_l.
    \label{eq:Hblock2}
\end{equation}
Each block matrix $\model_l$ is then an $N \times IJ$ matrix, where $N = K \times M$ is the size of the 2d spectrum image. The matrix $\model$ is then an $N\times P$ matrix. Although very large, $\model$ is a sparse matrix. The sparsity arises because each geometric transform matrix ${\Dp}_{l}$ only drizzles the datacube onto a small part of the 2d spectrum image.

When we include interpolation, we substitute $\cube_l$ in Equation~\ref{eq:specsum} with $\cubetilde_l$ in Equation~\ref{eq:d_interp} to obtain
\begin{equation}
    \vect{\spec} = \tau\sum_{\tilde{l}=1}^{N_{\tilde{\lambda}}}A_\text{eff}(\tilde{\lambda}_{\tilde{l}})\left.\frac{\dd\lambda}{\dd\kappa}\right|_{\tilde{\lambda}_{\tilde{l}}}{\Dp}_{\tilde{l}}\C_{\tilde{l}}\left(w_{{\tilde{l}},1}\cube_{l[\tilde{l}]} + w_{{\tilde{l}},2}\cube_{(l[\tilde{l}]+1)}\right).
    \label{eq:specsum2}
\end{equation}

The forward model operator $\model$ that includes linear interpolation will have the same form as in Equation~\ref{eq:Hblock}, but now each block $\model_l$ will include the weighted summations involving the relevant image $\cube_l$. Let $c_{\tilde{l},1}$ and $c_{\tilde{l},2}$ be 
\begin{equation}
\begin{aligned}
    c_{\tilde{l},1} &= \tau A_\text{eff}(\tilde{\lambda}_{\tilde{l}})\left.\frac{\dd\lambda}{\dd\kappa}\right|_{\tilde{\lambda}_{\tilde{l}}}w_{\tilde{l},1},\\
    c_{\tilde{l},2} &= \tau A_\text{eff}(\tilde{\lambda}_{\tilde{l}})\left.\frac{\dd\lambda}{\dd\kappa}\right|_{\tilde{\lambda}_{\tilde{l}}}w_{\tilde{l},2},
\end{aligned}
\end{equation}
which are the scalars that depend on the interpolated wavelength $\tilde{\lambda}_{\tilde{l}}$, weighted respectively by $w_{\tilde{l},1}$ and $w_{\tilde{l},2}$ as defined by Equation~\ref{eq:w_interp}. The blocks $\model_l$ are then
\begin{equation}
\begin{aligned}
\model_1 & = c_{1,1}{\Dp}_1\C_1,\\
\model_2 & = c_{1,2}{\Dp}_1\C_1 + c_{2,1}{\Dp}_2\C_2,\\
\model_3 & = c_{2,2}{\Dp}_2\C_2 + c_{3,1}{\Dp}_3\C_3,\\
         & \vdots\\
\model_{N_\lambda-1} & = c_{N_{\tilde{\lambda}}-1,2}{\Dp}_{N_{\tilde{\lambda}}-1}\C_{N_{\tilde{\lambda}}-1} + c_{N_{\tilde{\lambda}},1}{\Dp}_{N_{\tilde{\lambda}}}\C_{N_{\tilde{\lambda}}},\\
\model_{N_\lambda} & = c_{N_{\tilde{\lambda}},2}{\Dp}_{N_{\tilde{\lambda}}}\C_{N_{\tilde{\lambda}}}.
\end{aligned}
\end{equation}

\begin{figure}
    \includegraphics[width=\columnwidth]{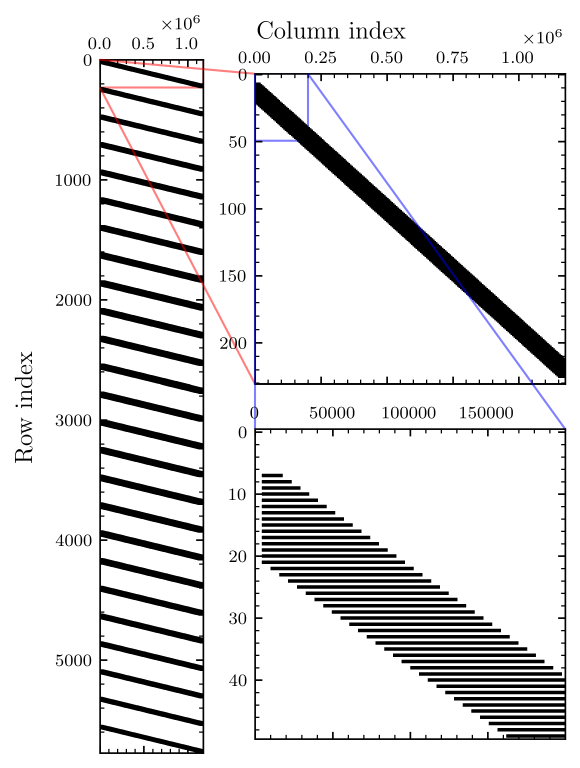}
    \caption{The sparsity structure of the dispersion matrix $\model$ that projects a vectorized datacube $\vect{\cube}$ into vectorized spectrum $\vect{\spec}$ given the prism passband, dispersion curve, and integration time. The matrix is shown here in three level of details. The left panel shows the whole matrix. On the top right panel the sparsity structure is shown for a segment of $\model$ that projects the datacube into columns of constant row index $p$ of the 2d spectrum image, and on the bottom right a small part of the segment above is shown. Note the different range of the vertical and horizontal axes.}
    \label{fig:H}
\end{figure}

Figure~\ref{fig:H} shows the sparsity structure of $\model$. Overall $\model$ consists of repeating blocks that span the whole columns (i.e. block rows), where each block projects the datacube into pixels along the horizontal axis in the data space. On the top right panel at Figure~\ref{fig:H} we show such block. Here we see the manifestation of Equations~\ref{eq:Hblock}--\ref{eq:Hblock2}, in which a given pixel in the 2d spectrum image is the summation of certain parts of the datacube (in the spatial-wavelength space). Which parts of the datacube contribute to a given pixel, and at what fraction, can be identified using the dispersion curve and other components, as manifest in Equation~\ref{eq:Hblock2}.

\section{Reconstructing the datacube}
\label{sec:rec}

Reconstructing a datacube $\D$ given that we know the forward model $\model$ and its projection into spectrum $\spec$, is an \textit{inverse problem}. From a physical standpoint, an inverse problem is the reverse of the cause and effect sequence: Finding the underlying cause given the observed consequences. This is the opposite of a \textit{direct problem}, in which we compute the consequence of a given cause \citep{ber22}.

When $\D$ is projected into $\spec$ using $\model$, some degree of information is lost because the spectral axis is projected into the spatial axis by means of the dispersion curve. As we have seen in the previous Section, in spectral space, the axis along the dispersion direction is a combination of one spatial axis and the spectral axis of the datacube space. The most tangible manifestation of this loss of information is the fact that the resulting number of pixels in $\spec$ is much less than the number of elements $P$ in the datacube $\D$. Even though we observe the scene multiple times at various orientations, the total number of available pixels $N$ in all the spectra remains less than the number parameters $P$ we have to infer. 

To illustrate this, let $\data$ be the vertical concatenation of $N_\text{\rm s}$ vectorized spectra:
\begin{equation}
    \data = 
    \begin{pmatrix}
        \vect{\spec}_1\\
        \vect{\spec}_2\\
        \vdots\\
        \vect{\spec}_{N_\text{\rm s}}\\
    \end{pmatrix},
\end{equation}
and similarly let $\X$ be the vertical concatenation of the corresponding $N_\text{\rm s}$ forward models $\model_i$ that transform the vectorized datacube $\vect{\D}$ into vectorized spectra $\vect{\spec}_i$:
\begin{equation}
    \X = 
    \begin{pmatrix}
        \model_1\\
        \model_2\\
        \vdots\\
        \model_{N_\text{\rm s}}
    \end{pmatrix}.
\end{equation}
The size of $\data$ is $N\times 1$ where $N = N_\text{\rm s}\times K\times M$ is the total number of pixels in the concatenated 2d spectral images, and the size of $\X$ is $N\times P$ where $P = N_\lambda \times I \times J$ is the size of the datacube $\D$ we want to reconstruct. With $P\gg N$, if we invert $\X$ to reconstruct $\D$, i.e. $\cube = \vectorize{\D} = \X^{-1}\data$, there will be no unique solution. This is further made worse with noise---which is ever present in realistic observations. A lack of unique solution makes this an \textit{ill-posed problem}.

An approach to ``cure'' this ill-posedness is to search for an approximate solution that not only can reproduce the observed data within their uncertainties but is also physically plausible \citep{ber22}. Under these constraints we approach the problem in a probabilistic manner using Bayesian theorem. Using this framework, the presence of observational uncertainties is addressed using the likelihood function, and prior information is used to constrain the solution within what is physically plausible.

Starting from Bayes theorem $\posterior\propto\likelihood\prior$---in which $\posterior$ is the probability that the inferred datacube $\cube$ is true given that the data $\data$ and forward model $\X$ are true, $\likelihood$ is the likelihood of observing the data given the datacube and forward model, and $\prior$ is the prior probability of the inferred datacube---we can derive an objective function
\begin{equation}
    Q(\cube) = \res\transpose\weight\res + \alpha(\cube- \cube_\text{\rm p})\transpose(\cube - \cube_\text{\rm p}),
\end{equation}
where $\res = \data - \X\cube$ is the vector of residuals between the observed and predicted brightnesses in all the pixels in the 2d spectral images, $\weight = \cov^{-1}$ is the inverse covariance matrix of the data and for simplicity we assume no correlation between pixels, thus consisting only of the inverse variances of the pixels on the main diagonal and zero elsewhere:
\begin{equation}
    \weight = 
    \begin{pmatrix}
        1/\sigma^2_1 &              &        & \\
                     & 1/\sigma^2_2 &        & \\
                     &              & \ddots & \\
                     &              &        & 1/\sigma^2_{N}\\
    \end{pmatrix},
\end{equation}
$\cube_\text{\rm p}$ is the prior on the datacube, and $\alpha$ is what we call the regularization parameter. A full derivation is given in Appendix~\ref{app:bayesridge}, in which we can see that $Q$ is the inverse of $\ln\posterior$.

The objective function $Q$ encapsulates our approach to the ill-posed inverse problem by measuring how close the predicted data $\X\cube$ is to the observed data $\data$ (within the observational uncertainties) and how close $\cube$ is to a prior datacube $\cube_\text{\rm p}$. The regularization parameter $\alpha$ determines how much $\cube_\text{\rm p}$ is emphasized.

To infer the restored datacube $\parest{\cube}$ that is not only physically plausible but can also accurately predict the spectra at new rotation angles, it is crucial to properly assign a prior datacube $\cube_\text{\rm p}$ that is close to the true datacube we want to infer, and to determine the appropriate value of the regularization parameter $\alpha$. We discuss these respectively in Section~\ref{sec:prior} and Sections~\ref{sec:alpha}--\ref{sec:alphacv}.

For fixed values of $\cube_\text{\rm p}$ and $\alpha$, we can infer the restored datacube $\parest{\cube}$ by seeking $\cube$ that minimizes $Q$, i.e.
\begin{equation}
    \parest{\cube} = \underset{\cube}{\arg\min}\ Q(\cube),
    \label{eq:objfunc}
\end{equation}
which is equivalent to finding the \textit{maximum a posteriori} (MAP) estimate of $\cube$ (see Appendix~\ref{app:bayesridge} for details).

A direct solution to $\parest{\cube}$ can be analytically found by finding $\parest{\cube}$ that solves 
\begin{equation}
    \left.\frac{\partial Q}{\partial\cube}\right|_{\cube=\hat{\cube}} = 0.
\end{equation}
We then solve the matrix equation $\vecbold{A}\hat{\cube} = \vecbold{b}$ for $\hat{\cube}$, where $\vecbold{A}$ is
\begin{equation}
    \vecbold{A} = \X\transpose\weight\X + \alpha\eye.
\end{equation}
Here $\eye$ stands for the identity matrix. The right hand side is
\begin{equation}
    \vecbold{b} = \X\transpose\weight\data + \alpha\cube_\text{\rm p}.
\end{equation}
However, we can see that $\vecbold{A}$ is a $P\times P$ matrix which, although sparse and symmetric, is very large and will take a very long time to invert, with no guarantee that the inverse\footnote{Directly inverting $\A$ to obtain $\hat{\cube} = \A^{-1}\vecbold{b}$ is never recommended. Instead, one should directly solve the linear equation $\A\hat{\cube} = \vecbold{b}$ using appropriate decomposition technique based on the shape and properties of $\A$. In this case, since $\A$ is a square positive-definite matrix, we can then use Cholesky Decomposition to find $\hat{\cube}$.} is also sparse. Even if we do not directly invert $\vecbold{A}$ and solve the equation using certain decomposition techniques (e.g. Cholesky decomposition), for very large $P$---which we expect to deal with---this can still takes a very long time to solve and requires a large amount of memory to contain $\vecbold{A}$.

A more feasible method to find $\hat{\cube}$ is to use iterative minimization techniques, in which we repeatedly guess $\cube$, calculate the direction of descent $\nabla Q$ at each trial $\cube$ so we know where to go, and stop when we have arrived at the minimum $Q$. At the $k$-th iteration, the successive update $\cube_{k+1}$ can be calculated by
\begin{equation}
    \cube_{k+1} = \cube_{k} - \left[\nabla^2 Q\left(\cube_{k}\right)\right]^{-1}\nabla Q\left(\cube_k\right),
\end{equation}
where $\nabla Q\left(\cube_k\right)$ is the gradient of $Q$ at $\cube_k$:
\begin{equation}
    \begin{split}
        \nabla Q\left(\cube_k\right) &= \left(\left.\frac{\partial Q}{\partial\cube}\right|_{\cube=\cube_k}\right)\transpose \\
         &= -\X\transpose\weight\res_k + \alpha\left(\cube_k-\cube_\text{\rm p}\right),
    \end{split}
\end{equation}
and $\nabla^2Q\left(\cube_k\right)$ is the Hessian matrix, a $P\times P$ matrix of second-order derivatives of $Q$:
\begin{equation}
    \begin{split}
        \nabla^2Q\left(\cube_k\right) & =
        \begin{pmatrix}
            \frac{\partial^2 Q}{\partial\cube_1^2} & \frac{\partial^2 Q}{\partial\cube_1\partial\cube_2} & \cdots & \frac{\partial^2 Q}{\partial\cube_1\partial\cube_{P}}\\
            \frac{\partial^2 Q}{\partial\cube_2\cube_1} & \frac{\partial^2 Q}{\partial\cube_2^2} & \cdots & \frac{\partial^2 Q}{\partial\cube_2\partial\cube_{P}}\\
            \vdots & \vdots & \ddots & \vdots\\
            \frac{\partial^2 Q}{\partial\cube_{P}\cube_1} & \frac{\partial^2 Q}{\partial\cube_{P}\cube_2} & \cdots & \frac{\partial^2 Q}{\partial\cube_{P}^2}
        \end{pmatrix}\\
        & = \X\transpose\weight\X + \alpha\eye.
    \end{split}
\end{equation}
The Hessian $\nabla^2Q\left(\cube_k\right)$ describes the curvature of $Q$ at $\cube_k$ and is used to calculate the step size along the direction of descent $\nabla Q$.

An advantage of employing the iterative solution is that the gradient $\nabla Q\left(\cube_k\right)$ is a $P \times 1$ column vector and $Q$ is a scalar, neither require a lot of memory to store. Furthermore, both $Q$ and $\nabla Q\left(\cube_k\right)$ are (weighted) summations that can be done in parallel if the matrices $\X$, $\data$, and $\weight$ are split into block rows, further reducing computing time and memory usage.

We still have to calculate the Hessian though, and invert it (the Hessian is actually the $\A$ matrix, which we do not want to explicitly calculate). We address this by employing the limited memory Broyden–Fletcher–Goldfarb–Shanno (L-BFGS, \citealt{byr95, noc06}) algorithm as implemented by \texttt{SciPy} \citep{scipy} to minimize $Q$. L-BFGS is a quasi-Newton method, which does not explicitly calculate the inverse Hessian, but rather it uses a few vectors to approximate and successively update it at each iteration. It is therefore the suitable algorithm to address the problem of solving for a large number of parameters $P$.

\begin{figure}
   \centering
   \includegraphics[width=\columnwidth]{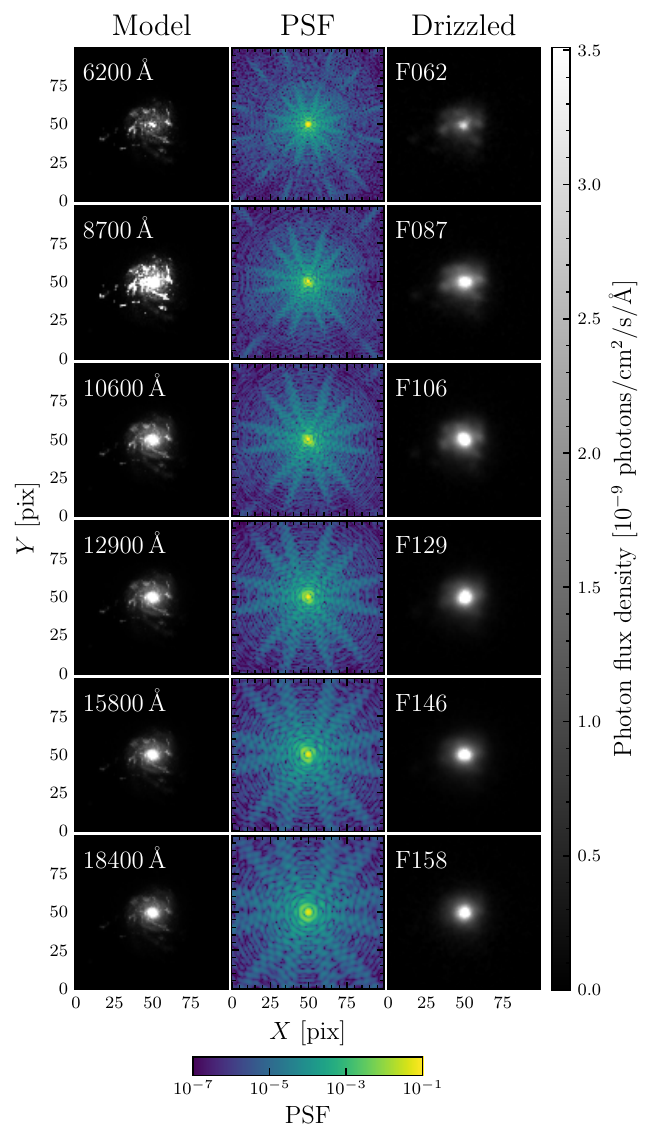}
   \caption{The ``true'' monochromatic images represented by the high-resolution models on the left column are convolved with the corresponding PSF on the middle column as part of the image formation process. The resulting images at various roll angles are then coadded using Drizzle and the results are shown on the right column. The drizzled and true images are shown in linear scale, while the PSFs are shown in logarithmic scale (see the corresponding color bars at the right and bottom of the plot).}
   \label{fig:drizzled}
\end{figure}

\begin{figure}
   \centering
   \includegraphics[width=\columnwidth]{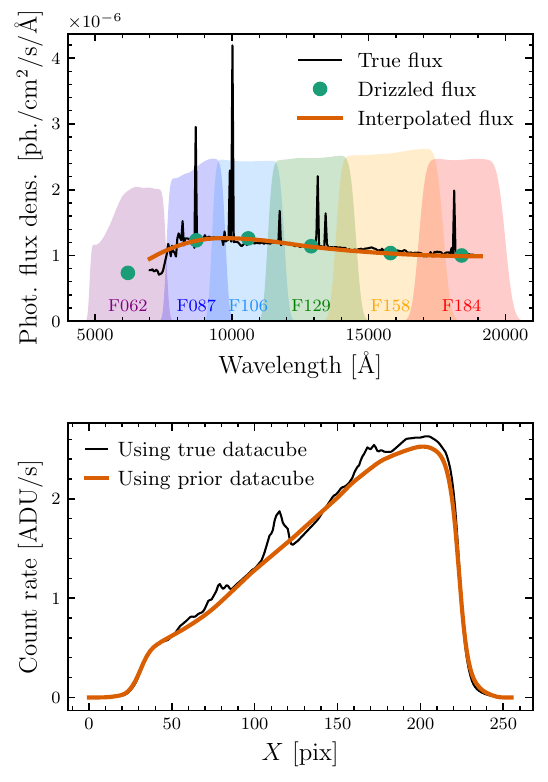}
   \caption{The top panel shows comparisons of the integrated flux density estimated from drizzled images of the scene in 6 \Roman{} filters (green dots) with the interpolated flux (brown line) and the true flux (black line). Also shown for reference are the passbands of the 6 filters, scaled to arbitrary value. The bottom panel shows a comparison of noise-free 1d spectrum generated using the true datacube shown on the top panel and those generated using the prior datacube. Some of the emission lines seen in the true spectrum on the top panel are still apparent in the true spectrum. However, because the prior datacube is constructed using WFI images, it is unable to detect the emission lines because they are too narrow compared with the filter bandwidth. Consequently the resulting spectrum generated from the prior datacube (brown curve) reproduces only the continuum but misses the emission lines. This is a prior mismatch and will be discussed in Section~\ref{sec:alpha}.}
   \label{fig:integratedflux_priorspec}
\end{figure}

\begin{figure}
    \centering
    \includegraphics[width=\columnwidth]{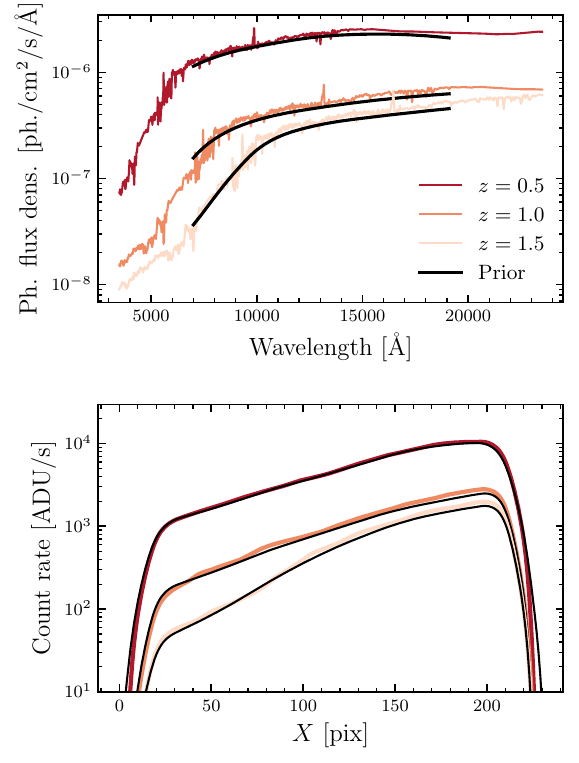}
    \caption{The upper panel shows the true integrated flux densities of the three simple galaxies used in our tests. The galaxies have uniform SED at all points in the galaxy, but with intensities scaled according to the S\'{e}rsic power law with index $n = 1$. The spectrum of IC~4553 from the \cite{bro14} atlas is used as templates and redshifted accordingly. Black lines show the corresponding prior fluxes estimated from drizzled images in 6 \Roman{} filters (cf. Figure~\ref{fig:integratedflux_priorspec}). The bottom panel shows the resulting noise-free \Roman{} 1d spectra simulated from the true datacube (shown in similar colors as the those in the upper panel), and the noise-free \Roman{} 1d spectra simulated from the corresponding prior datacubes (shown in black lines).}
    \label{fig:galaxyspec}
\end{figure}

\begin{figure}
    \centering
    \includegraphics[width=\columnwidth]{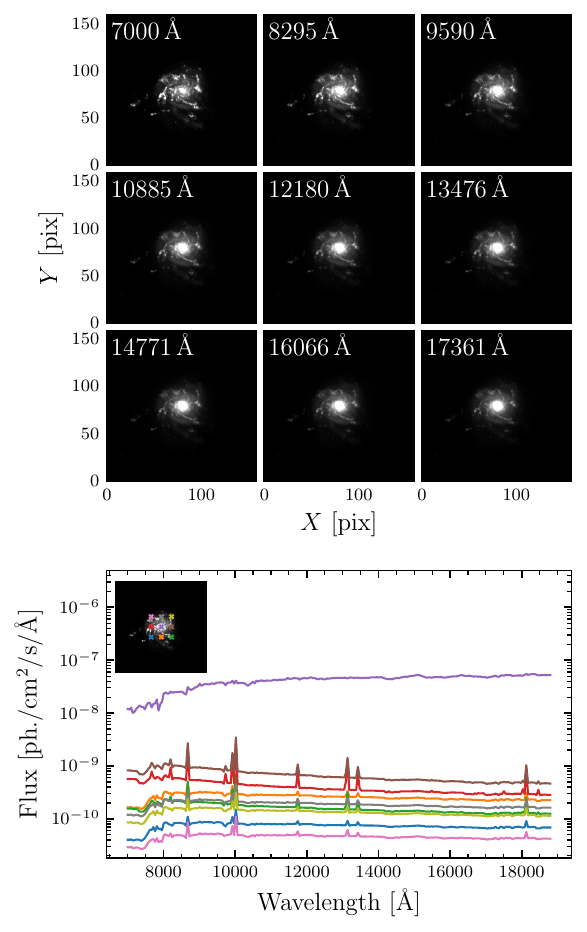}\\
    \caption{The VELA01 datacube at selected wavelengths are shown in the upper 9 panels. At the bottom panel, the spectra of the VELA01 datacube at selected points in the galaxy are shown on the main panel. The corresponding points are shown on the inset image with the same color. The integrated spectrum for this galaxy is shown on the top panel of Figure~\ref{fig:integratedflux_priorspec}.}
    \label{fig:vela01_imspec}
\end{figure}

\begin{figure*}
    \centering
    \includegraphics[width=\textwidth]{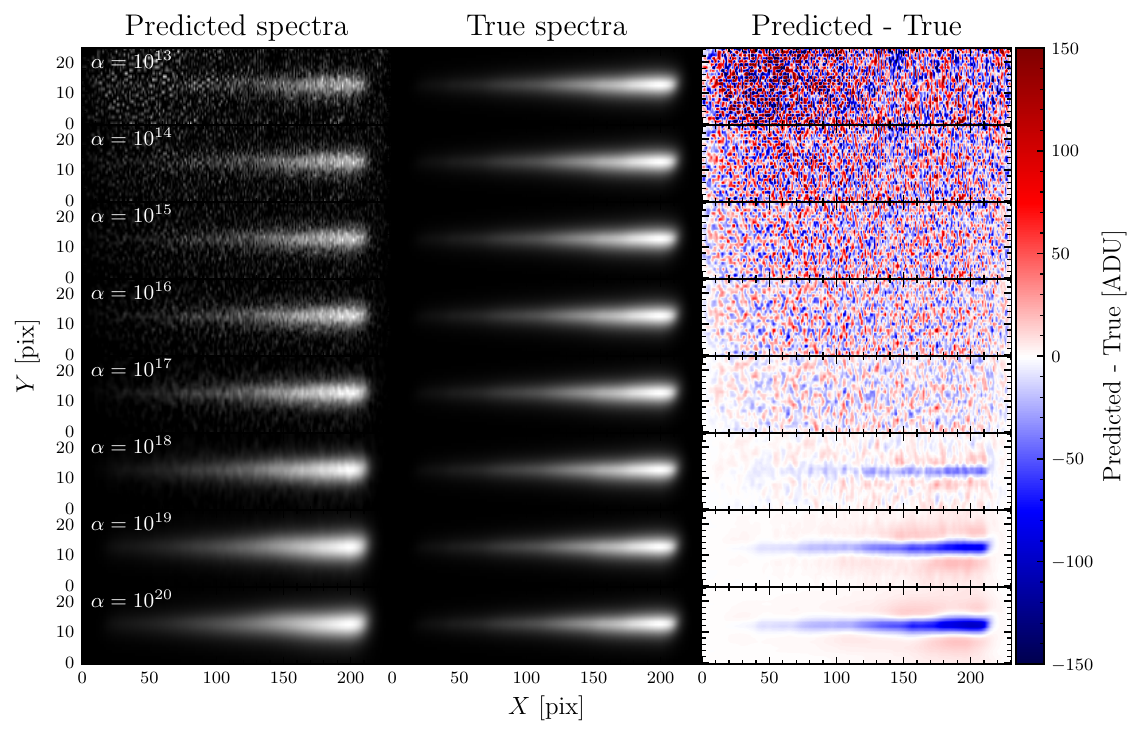}
    \caption{Comparisons between noise-free galaxy-only spectra (left column) predicted using the datacube reconstructed using the value of $\alpha$ indicated on the top left corner of each row (left column), and the corresponding noise-free spectra generated from the true scene (center column). This galaxy is at $z = 1.0$ and the reconstruction uses 24 roll angles (uniformly spaced at $15^\circ$). The right column shows the residuals between predicted and true spectra. As can be seen in the residual images, a spectrum generated using the datacubes inferred using high values of $\alpha$---shown in the bottom rows---will still have systematic residuals, shown as the blue underestimation streak in the middle of the image. On the other hand, low values of $\alpha$ will give noisy inferences of the scene, leading to very noisy predicted spectra, as shown on the top three rows.}
    \label{fig:predalpha}
\end{figure*}

\begin{figure*}
    \centering
    \includegraphics[width=\textwidth]{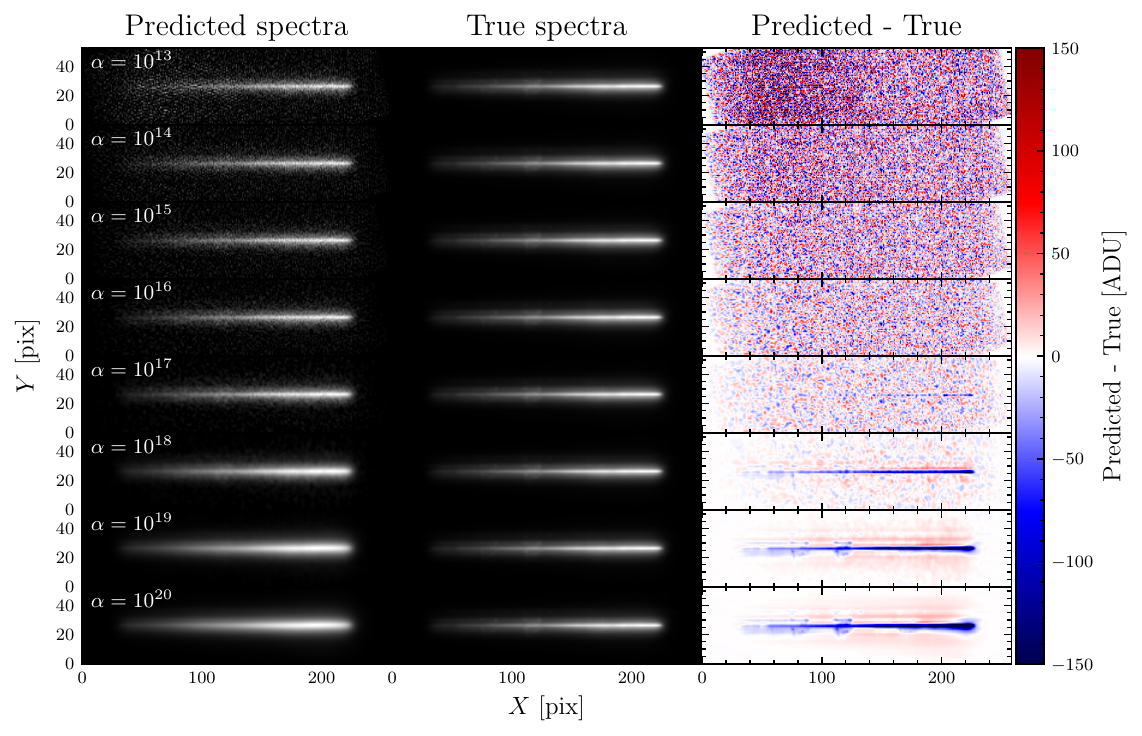}
    \caption{As Figure~\ref{fig:predalpha}, but for the VELA01 datacube. The galaxy is also at $z=1.0$, but with different SEDs at each pixel on the galaxy datacube. Because of this more complex structure of the true datacube and the simpler prior datacube employed in the inferences, prior mismatch is more severe here than in the simpler galaxy. Nevertheless, at $\alpha=10^{16}$, the subtraction is very good, as traces of the galaxy are no longer discernible by eye, and the residual image resembles white noise image.}
    \label{fig:predalpha_vela01}
\end{figure*}

\subsection{Using images to construct a prior datacube}
\label{sec:prior}

It is tempting to set the prior datacube $\cube_\text{\rm p}$ as zero everywhere\footnote{Which makes $Q$ the objective function for Ridge regression. See Appendix~\ref{app:bayesridge}.}. However, \Roman{} will have additional information that can be used to construct a prior: images of the host galaxy in 6 filters.  These images will be taken in conjunction with the spectra, and will also be taken at various roll angles and sub-pixel dithers.

We construct a prior datacube $\cube_\text{\rm p}$ based on these images by first using Drizzle to co-add all images with the same filter into a single image with the target datacube's oversample rate and orientation. Then we estimate the spectra on all oversampled pixels by interpolating the photon flux densities at the central wavelength of each filter into the wavelength grid $\wave$ of the target datacube. 

For each pixel $(p,q)$, the photon flux density at the central wavelength of filter $k$ is estimated by
\begin{equation}
    f_{kpq} = \frac{I_{kpq}}{\tau_\text{\rm tot}\int A_{\text{\rm eff},k}\mathrm{d}\lambda},
\end{equation}
where $I_{kpq}$ is the observed total brightness on pixel $(p,q)$ of the co-added image in filter $k$, $\tau_\text{\rm tot}$ is the total exposure time of the drizzled image, and $A_{\text{\rm eff},k}$ is the effective collecting area of the filter $k$, here integrated over wavelength.

Examples of the resulting co-added galaxy-only images in six \Roman{} filters are shown on the right column of Figure~\ref{fig:drizzled}, while the estimated integrated photon flux densities of the scene at the central wavelengths of the filters is shown on the panel of Figure~\ref{fig:integratedflux_priorspec} as the green dots. Using these six data points we then interpolate the photon flux densities at the wavelength grid $\wave$, overplotted as the purple line in the Figure. These then become our prior datacube $\cube_\text{p}$. Compared to the true integrated spectrum, shown as the black line on the bottom panel of Figure~\ref{fig:integratedflux_priorspec}, the prior integrated spectrum (made using the prior datacube $\cube_\text{p}$) matches well with the continuum. Here we are unable to reproduce the emission lines as they are too narrow compared with the bandwidth of the filters. As we can see in Figure~\ref{fig:integratedflux_priorspec}, this results in a spectrum that is missing the emission lines (some of them are still apparent in the true spectrum). This is a prior mismatch problem, in which the prior does not match well with the truth. We will discuss this further in Section~\ref{sec:alpha}.

\begin{figure}
    \centering
    \includegraphics[width=\columnwidth]{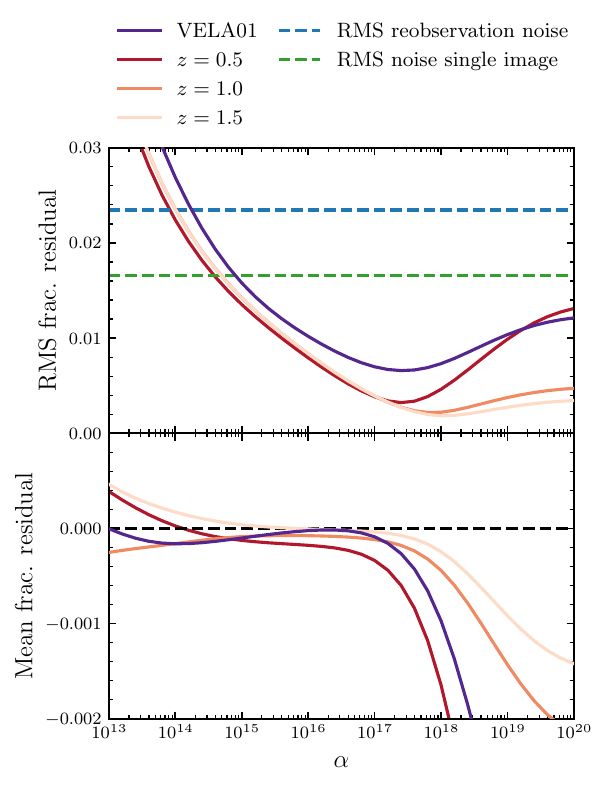}
    \caption{Performance of the reconstruction of the VELA01 galaxy (solid purple line) and galaxies with redshifts $z~=~0.5, 1.0$, and 1.5 (red-hued lines), as a function of $\alpha$, and for 1000 new roll angles not in the dataset used for scene reconstruction. As in previous Figures, the reconstruction uses 24 roll angles from $0^\circ$ to $360^\circ$ (exclusive) and uniformly spaced by $15^\circ$. The performance is shown in terms of the standard deviation of the fractional residuals (top panel) and the mean fractional residuals (bottom). For each residual image, only the central 11 rows along the cross-dispersion axis are considered. Also shown on the top panel for comparisons are the expected fractional RMS noise of host-galaxy subtraction by reobserving the scene (dashed blue line) and the expected fractional RMS noise of a single image (dashed green line). Note the different range of the vertical axes of both panels.} 
    \label{fig:rms}
\end{figure}
\begin{figure*}
    \centering
    \includegraphics[width=\textwidth]{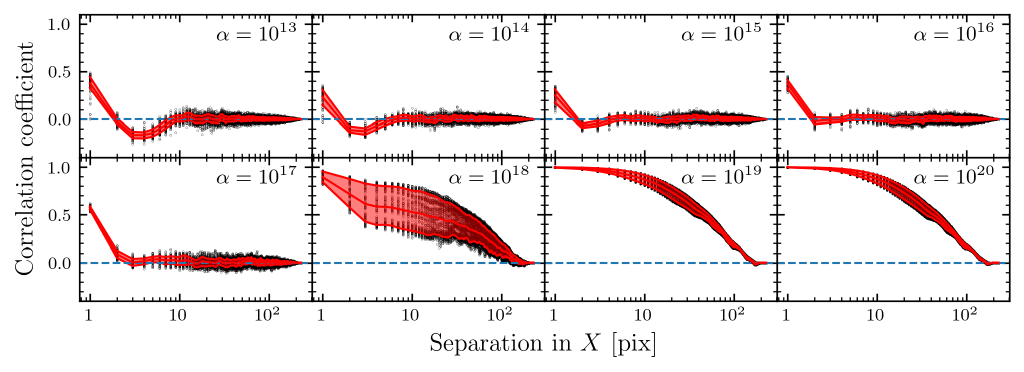}
    \caption{The autocorrelation of 1000 residual images between noise-free predicted spectra and true noise-free spectra as a function of pixel separation (black points), for spectra predicted using scenes reconstructed with various choices of $\alpha$ as indicated at the top right corner of each panel. This is the simple galaxy at redshift $z=1.0$. Red lines indicate the mean correlation coefficient as a function of pixel separation, and the red area shows the 68\% range of the correlation coefficients.}
    \label{fig:acf}
\end{figure*}
\begin{figure*}
\centering
    \includegraphics[width=\textwidth]{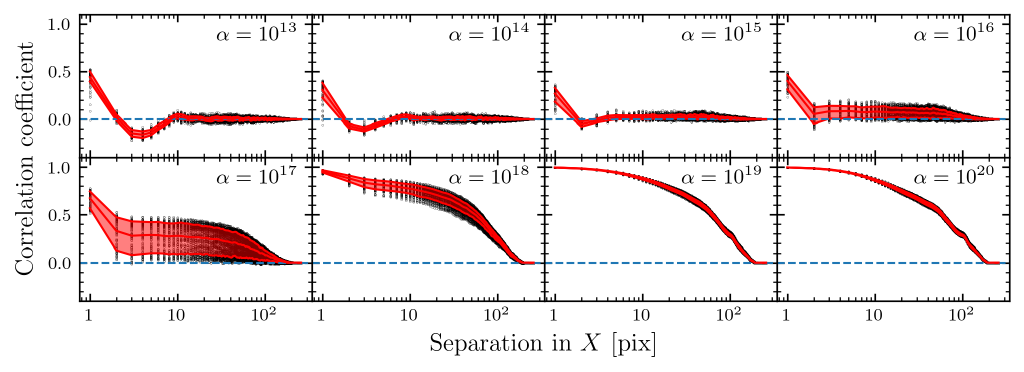}
    \caption{As Figure~\ref{fig:acf}, but for the VELA01 galaxy. Because of the more complex structure of this galaxy's datacube, stronger correlations are still present at $\alpha = 10^{17}$.}
    \label{fig:acf_vela01}
\end{figure*}
\begin{figure}
    \centering
    \includegraphics[width=\columnwidth]{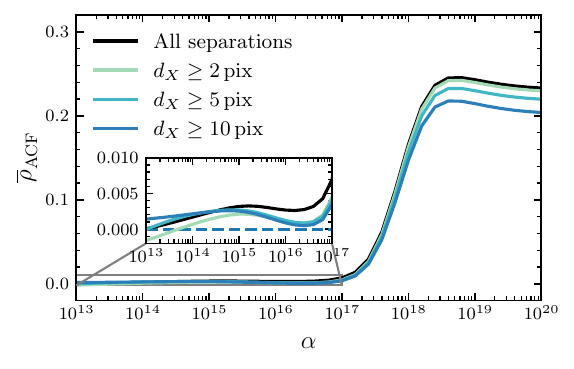}
    \caption{The mean autocorrelation coefficients of 1000 residual images between noise-free spectra predicted using scenes with various values of $\alpha$ and the corresponding true spectra. Three cuts in pixel separation $d$ are also shown.}
    \label{fig:rho}
\end{figure}
\begin{figure*}
    \centering
    \includegraphics[width=\textwidth]{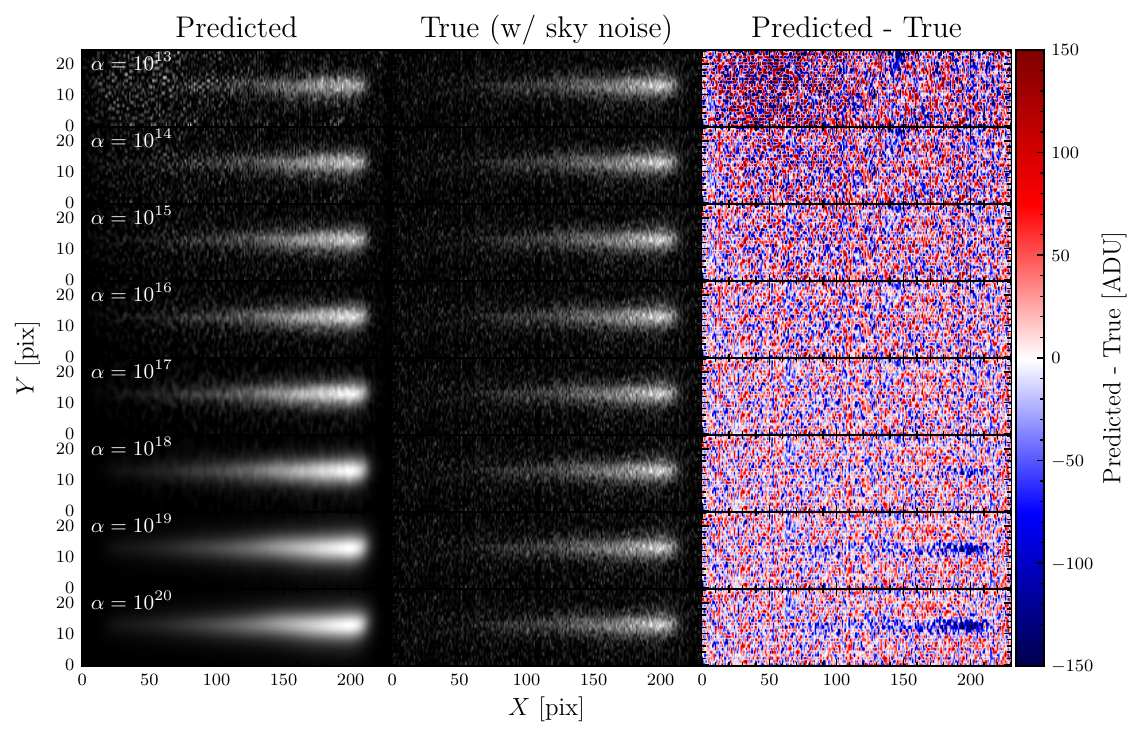}
    \caption{As Figure~\ref{fig:predalpha}, but the comparisons are now between predicted spectra and the corresponding \textit{noisy} spectra generated from the true scene (center column). The residual images are dominated by read and sky noise, but nevertheless at extreme values of $\alpha$, bias and noise from the reconstruction are still apparent. However, at $\alpha = 10^{17}$, the small amount of systematic residuals are now drowned out by the noise (cf. Figure~\ref{fig:predalpha}).}
    \label{fig:prednoisy1}
\end{figure*}
\begin{figure*}
    \centering
    \includegraphics[width=\textwidth]{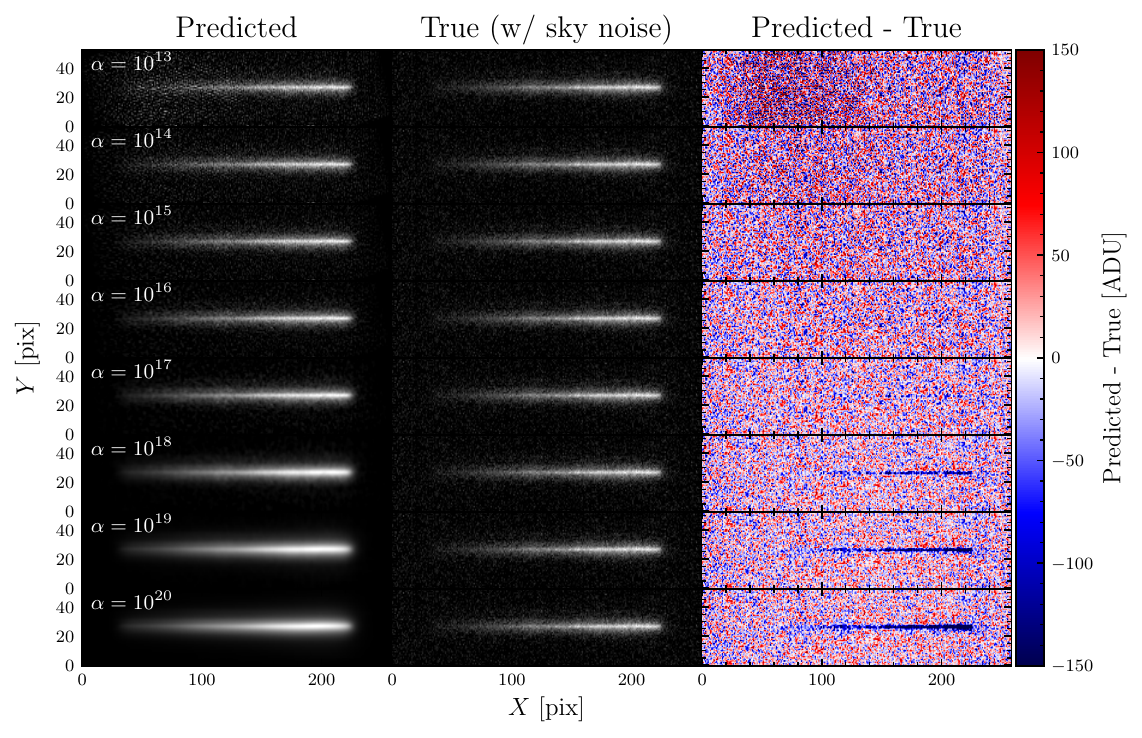}
    \caption{As Figure~\ref{fig:prednoisy1}, but for the VELA01 galaxy. Here at $\alpha = 10^{17}$ the systematic residuals are no longer apparent amid the noise.}
    \label{fig:prednoisy_vela01}
\end{figure*}
\begin{figure}
    \centering
    \includegraphics[width=\columnwidth]{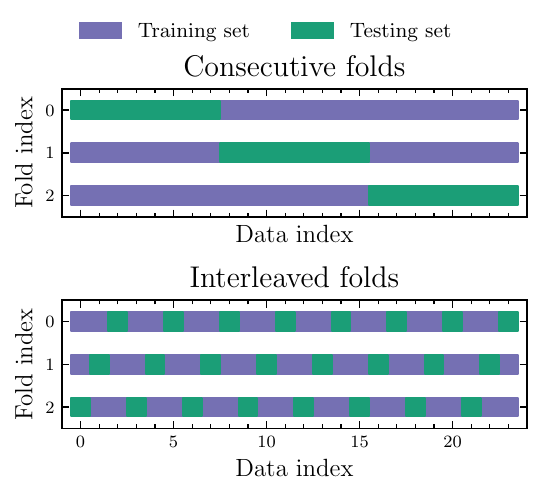}
    \caption{Two different schemes of arranging the training and testing set for 3-fold cross-validation. On the top panel, the data are partitioned into three equal-sized subsets. For each fold, we form a testing set using 1 subset and form a training set using all the other subsets. On the bottom panel, the testing set of each fold still has the same size as those on the top panel, but they are \textit{interleaved} with the training set. This ensures that each fold has equal coverage of training data.}
    \label{fig:cvscheme}
\end{figure}
\begin{figure}
    \centering
    \includegraphics[width=\columnwidth]{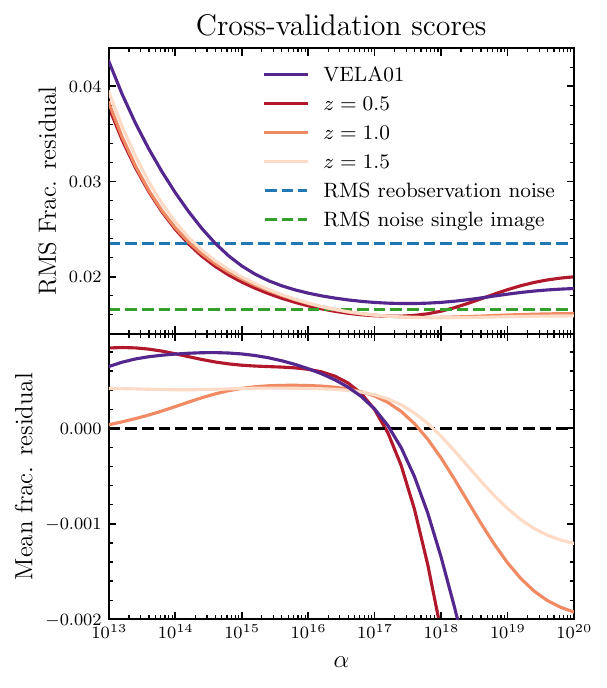}
    \caption{The 2-fold cross-validation RMS fractional residuals and mean fractional errors of the training as a function of the regularization parameter $\alpha$, for the VELA01 galaxy (solid purple line) and galaxies with redshifts $z = 0.5, 1.0,$ and $1.5$, each reconstructed using 24 spectra.}
    \label{fig:cv}
\end{figure}

\begin{figure*}
    \centering
    \includegraphics[width=\textwidth]{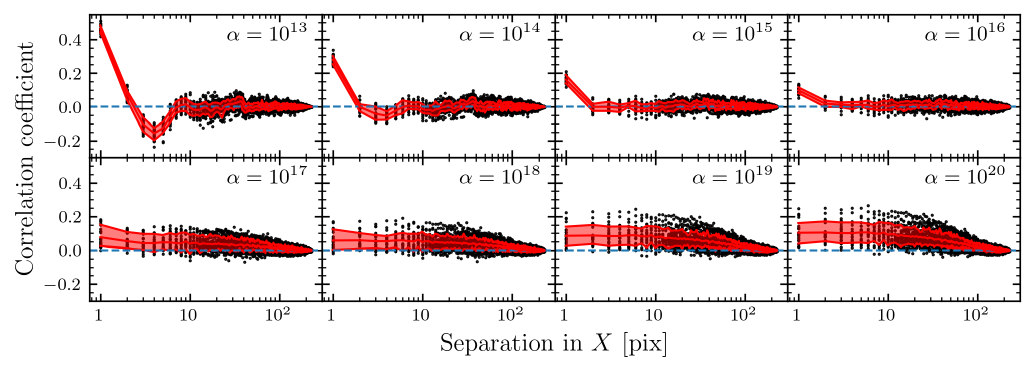}
    \caption{As Figure~\ref{fig:acf}, but now using the cross-validation test sets described in Section~\ref{sec:alphacv}. Note the narrower range of vertical axis. Because of the added instrument and sky noise, here the mean autocorrelation coefficient goes down to zero faster with decreasing regularization parameter $\alpha$ than those shown in Figure~\ref{fig:acf}.}
    \label{fig:acfcv}
\end{figure*}

\begin{figure}
    \centering
    \includegraphics[width=\columnwidth]{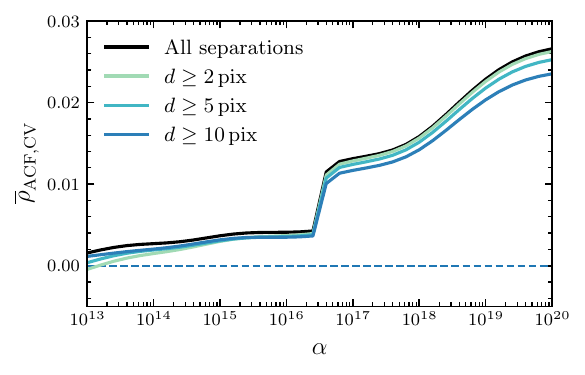}
    \caption{As Figure~\ref{fig:rho}, but using the cross-validation test sets as described in Section~\ref{sec:alphacv}. Note the much narrower range of vertical axis. The added instrument and sky noise flatten out the patterns we see in Figure~\ref{fig:rho}, but the mean correlation coefficient never actually reaches zero.} 
    \label{fig:acfstat}
\end{figure}

\begin{figure*}
    \centering
    \includegraphics[width=\textwidth]{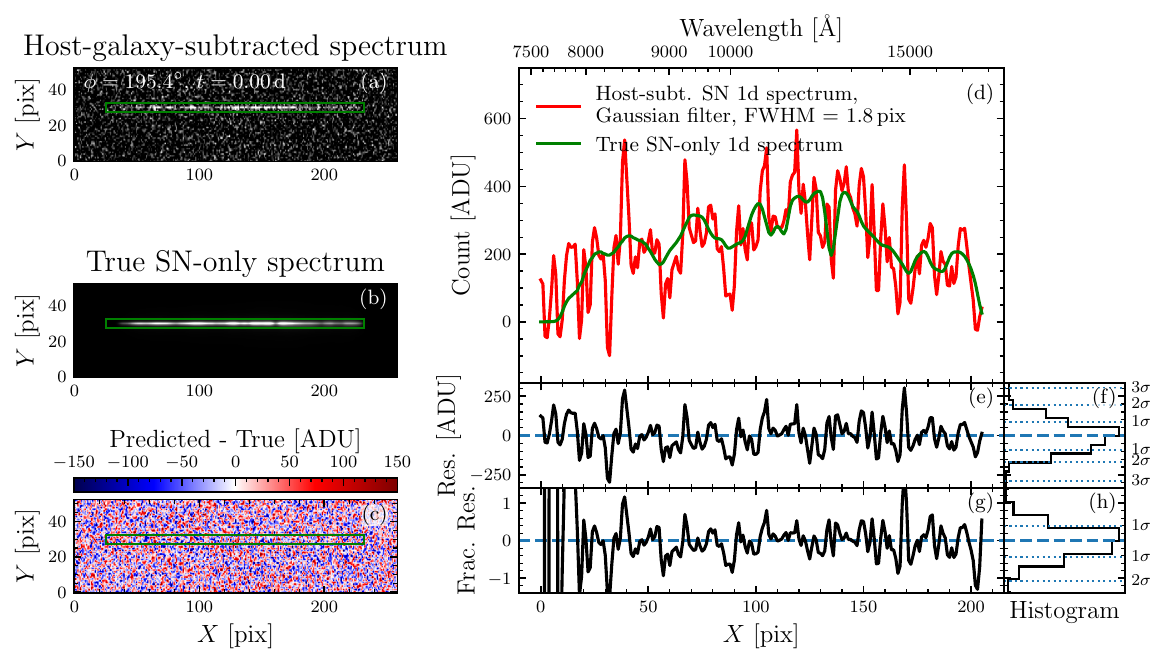}
    \caption{Host-galaxy subtracted, supernova-only spectrum at $z=1.0$ obtained using the scene reconstructed with $\alpha = 10^{16}$. The host galaxy is the VELA01 galaxy. The left column shows the host-subtracted SN-only 2d spectrum (panel a), the true SN-only 2d spectrum (panel b), and the residuals between the two spectra (panel c). Panel (d) shows the extracted 1d spectrum, obtained by summing the 2d spectrum over 5 pixels in the cross-dispersion direction, and to reduce noise it is smoothed by a Gaussian kernel with FWHM~=~1.8\,pix. In panel (e) the residuals of the 1d spectrum is plotted as a function of pixel coordinates, while panel (f) shows the corresponding histogram. Panel (g) is the fractional residuals of the 1d spectrum as a function of pixel coordinates, and panel (h) is the corresponding histogram. In panels (g) and (h), dotted lines show the $1\sigma$, $2\sigma$, and $3\sigma$ confidence interval respectively from the inner to outer part of the plot, and are marked as such. Similar plots for simple galaxies at redshifts $z=0.5$, $z=1.0$, and $z=1.5$ are shown in Figures~\ref{fig:snonly1}--\ref{fig:snonly3}, with the statistics of the residuals of the 1d spectra for all galaxies summarized in Table~\ref{tab:stat_snonly}.}
    \label{fig:snonly_vela01}
\end{figure*}

\begin{figure}
    \centering
    \includegraphics[width=\columnwidth]{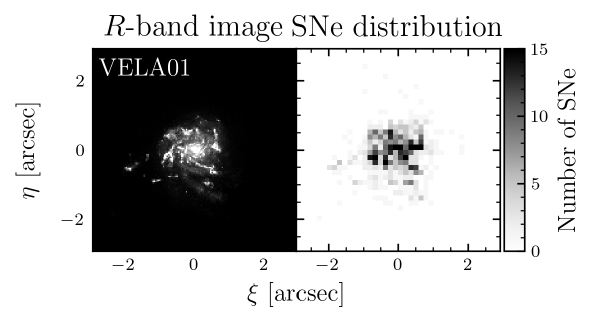}
    \caption{The $R$-band image of the VELA01 host-galaxy (left column) used for randomly drawing SN locations within the galaxy, and the resulting distribution of supernova locations (right column).}
    \label{fig:sndist_vela01}
\end{figure}

\begin{figure}
    \centering
    \includegraphics[width=\columnwidth]{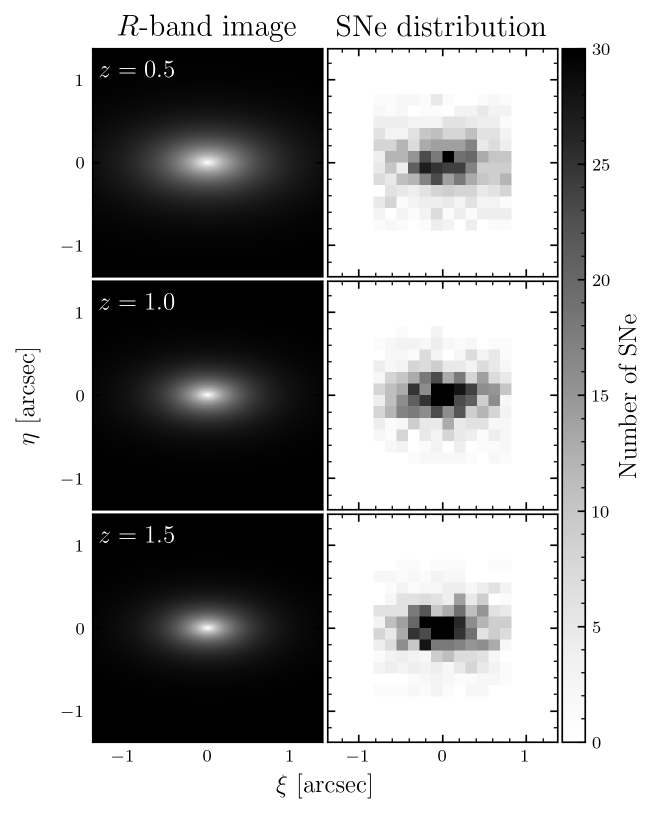}
    \caption{As Figure~\ref{fig:sndist_vela01}, but for the simple galaxies at redshifts $z=0.5$ (top row), $z=1.0$ (middle row), and $z=1.5$ (bottom row). Note the smaller field-of-view of the scene compared to VELA01.}
    \label{fig:sndist}
\end{figure}

\begin{figure}
    \centering
    \includegraphics[width=\columnwidth]{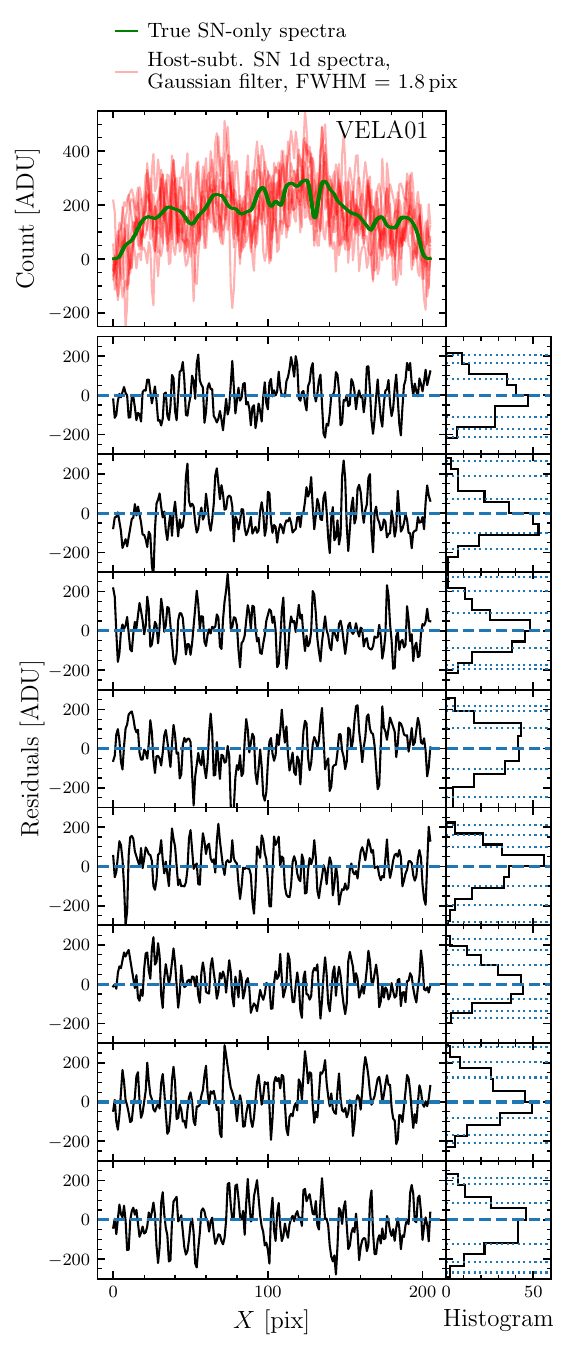}
    \caption{Eight randomly selected host-subtracted supernovae-only 1d spectra, for the VELA01 galaxy as host. Each panel in the second from top row down to the bottom row displays individual residuals and their corresponding 1d histograms. On the top panel we overplot the predicted 1d spectra and the corresponding true spectrum. Dotted lines in the histogram plot indicates the $1\sigma$, $2\sigma$, and $3\sigma$ confidence intervals in similar manner as Figure~\ref{fig:sndist}.}
    \label{fig:snonlyrandomselected1}
\end{figure}


\begin{figure}
    \centering
    \includegraphics[width=\columnwidth]{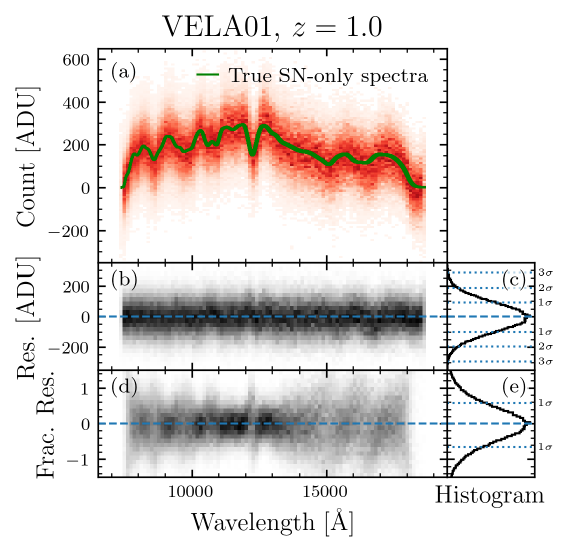}
    \caption{One thousand supernova-only 1d spectra obtained by subtracting the host-galaxy spectra using the datacube reconstructed with $\alpha = 10^{16}$, for the VELA01 galaxy. The 1d spectra are not individually shown here, but rather their distribution of flux versus wavelength is shown as a 2d histogram in panel~(a). For comparison we show the true SN-only 1d spectrum as a green line. Similarly, in panel~(b) and panel~(d) we show the 2d histograms of respectively the absolute and fractional residuals as a function of wavelength. The corresponding 1d histograms of these residuals are shown respectively in panel~(c) and panel~(e). Dotted lines indicate the $1\sigma$, $2\sigma$, and $3\sigma$ confidence intervals. The overall statistics of the residuals of the 1d spectra are summarized in Table~\ref{tab:stat_snrandom}.}
    \label{fig:snrandom_vela01}
\end{figure}

\subsection{The effect of the regularization parameter $\alpha$ on the predicted spectra}
\label{sec:alpha}
From a Bayesian point of view, the regularization parameter $\alpha$ is interpreted as the precision (viz. inverse variance) of a multivariate Gaussian centered around the prior datacube $\cube_\text{p}$ (see Appendix~\ref{app:bayesridge} for details). It determines how close to the given prior do we want the inferences to be. Larger $\alpha$ values will emphasize the prior $\cube_\text{p}$ more, making the inference prior-dominated, while conversely lower $\alpha$ values will put more emphasis on the data and thus the inference will be data-dominated.


We demonstrate how the choice of $\alpha$ will affect the predictions by reconstructing the datacube using various $\alpha$, using four different galaxy scenes. 

The first three scenes, which we call \textit{simple galaxy model}, are galaxies with uniform SED at all points in the galaxy, but with their intensities scaled with radius according to the S\'{e}rsic power law \citep{ser63} with index $n=1$, and redshifted to $z=0.5$, $z=1.0$, and $z=1.5$. Thus they are actually the same galaxy but at different distances. The galaxy has a semimajor axis of 3\,kpc, semiminor axis of 1.5\,kpc, and integrated absolute luminosity of $-21$\,mag in the SDSS $z$-band. We use the spectrum of IC~4553 from the \cite{bro14} spectral atlas as a template. We use \texttt{Pandeia} \citep{pon16} to generate the datacube specified as such and redshifted accordingly. The resulting datacube consists of monochromatic images uniformly spaced by 10\,\AA{} in the restframe between 3500\,\AA{} to 23500\,\AA. In spatial axes the size is $25\,\text{\rm pix}\times 25\,\text{\rm pix}$ in the \Roman{} pixel scale, oversampled by a factor of 5 (thus making it $125\,\text{\rm sub-pix}\times 125\,\text{\rm sub-pix}$). We define these datacubes as the \textit{true datacube}. In the top panel of Figure~\ref{fig:galaxyspec} we show the spatially-integrated spectra (the 1d spectra) of the true datacube of these simple galaxies.

The second model, the \textit{VELA01} galaxy, is based on an SED fitting of a galaxy from the VELA simulation \citep{sim19}, using the CIGALE \citep{boq19} code. The galaxy is at redshift $z=1.0$. The size of this scene is $53\,\text{\rm pix}\times 53\,\text{\rm pix}$ in the \Roman{} pixel scale and oversampled by a factor of 3 (thus $159\,\text{\rm sub-pix}\times 159\,\text{\rm sub-pix}$). Selected monochromatic images of this datacube and its spectra at selected spatial points are shown in Figure~\ref{fig:vela01_imspec}. As we can see this is a more complex scene of a galaxy, with different spectra at various points in the galaxy reflecting different stellar populations. The noise-free 1d \Roman{} spectra, simulated from both the true and prior datacubes are shown on the bottom panel of Figure~\ref{fig:integratedflux_priorspec}.

These four datacubes---the simple galaxies at redshifts $z = \{0.5, 1.0, 1.5\}$ and the VELA01 galaxy---are the \textit{true} datacube and become our input datacube to simulate \textit{noiseless} \Roman{} spectra, which we then use to simulate \textit{noisy} \Roman{} spectra by adding noise according to Equations~\ref{eq:noise1}--\ref{eq:noise2}. For each scene we simulate observations at 24 roll angles from $0^\circ$ to $360^\circ$ (exclusive), uniformly spaced by $15^\circ$. In addition to rotations, sub-pixel dithers are also included. Six of these 24 galaxy-only spectra have been shown back in Figure~\ref{fig:obs}. We do not use the observation marked in red in Figure~\ref{fig:obs}, as this has a supernova on it. We also simulate images in the 6 \Roman{} filter shown in Figure~\ref{fig:area}, each with the same number of roll angles as the spectroscopic observation. For all spectroscopic observations we set an exposure time of $\tau = 1800$\,s, whereas for imaging we set $\tau = 900$\,s.

For each galaxy, we use the images to construct a \textit{prior datacube} in the manner as discussed in Section~\ref{sec:prior}. For each simple galaxy, the 1d spectra from their corresponding prior datacubes are overplotted in black in the top panel of Figure~\ref{fig:galaxyspec}. Because of the simpler structure of the galaxies, the 1d spectra of the prior datacubes have closer resemblance to the corresponding true spectra than what is shown earlier in Figure~\ref{fig:integratedflux_priorspec}, which is more complex as it has different spectrum at each pixel on the galaxy. 

For each scene, we use the 24 spectroscopic observations and the corresponding prior datacube to recover the underlying datacube, which we define as the \textit{reconstructed datacube}. The reconstructed datacubes are oversampled by a factor of 3 (thus making it $75\,\text{\rm sub-pix}\times 75\,\text{\rm sub-pix}$ in spatial dimension for the simple galaxy and $159\,\text{\rm sub-pix}\times 159\,\text{\rm sub-pix}$ for the VELA01 galaxy). In the spectral dimension there are 221 wavelength elements between 7000\,\AA--19100\,\AA{} with uniform spacing. The choice of oversampling by a factor of 3 is a compromise between accuracy of predicted spectra and computing time.

The reconstructed datacubes are then projected back into spectra at new roll angles that were not in the dataset used for reconstruction. We define these as \textit{predicted spectra}, which are then compared with the corresponding \textit{true spectra}, i.e. spectra simulated using the \textit{true datacube}\footnote{Depending on the context, the \textit{true} spectra used for comparisons can be those without noise, or those with added noise.} . As previously discussed, the true datacube---which was also used to generate the simulated data set described above and with added observation noise---has a higher resolution than the reconstructed datacube.

In the interest of brevity, in the following examples we only present and discuss the results for the simple galaxy at $z = 1.0$ and for the VELA01 galaxy. Results for galaxies with redshifts $z=0.5$ and $z=1.5$ are qualitatively similar and discussed in Appendix~\ref{app:results}, or presented here when comparisons are necessary.

Figures~\ref{fig:predalpha}--\ref{fig:predalpha_vela01} show respectively the simple galaxy at $z=1.0$ and VELA01 galaxy-only spectra predicted using their respective datacubes reconstructed using various $\alpha$ (left column) compared with the corresponding true spectrum (middle column). In the right column we show the residuals between the two spectra. In both Figures, in the bottom three rows, we see underestimation streaks (colored blue) that get more prominent with larger $\alpha$. This is a \textit{prior mismatch}, which happens when inferences are prior-dominated and the prior datacube is not extremely well-matched with the true datacube (which is bound to occur in reality). 

Prior mismatches are systematic errors we have to watch out for and mitigate. On the other hand, when we use low values of $\alpha$, the inferences will be data-dominated. This gives more flexibility but will lead to very noisy predictions. Between these extreme values occurs a transition between prior-dominated and data-dominated posterior. We can expect that the appropriate value of $\alpha$ will vary from galaxy to galaxy.

Figures~\ref{fig:predalpha}--\ref{fig:predalpha_vela01} show the predictions \textit{only for a single new dither and roll angle}. To evaluate the average performance of the predictions over many new rolls and dithers as a function of $\alpha$, we generate 1000 random dither patterns and roll angles, then predict the spectra using the datacubes reconstructed with various values of $\alpha$. Again these predictions are then compared with the corresponding true spectra, and their performance is evaluated using summary statistics. For each pixel index $j$ in the $i$-th spectrum image predicted using the scene reconstructed with regularization parameter $\alpha$, we calculate the fractional residual
\begin{equation}
    r(\alpha)_{f, ij} = \frac{s(\alpha)_{ij} - s_{\text{\rm true}, ij}}{ s_{\text{\rm true}, ij}},
    \label{eq:res}
\end{equation}
where $s(\alpha)_{ij}$ is the brightness in the predicted 2d spectra image and $s_{\text{\rm true},ij}$ is the corresponding true brightness. Using these residuals, for a given $\alpha$ we then calculate the \textit{fractional bias}, $b_f(\alpha) = \overline{r}_f(\alpha)$:
\begin{equation}
    \overline{r}_f(\alpha) = \frac{1}{N_\text{\rm s}\times N_\text{\rm pix}}\sum_{i}^{N_\text{\rm s}}\sum_{j}^{N_\text{\rm pix}} r(\alpha)_{f, ij},
    \label{eq:bias}
\end{equation}
which is an average of the residuals over all spectra and pixels. We also calculate the root-mean-square (RMS) fractional residual
\begin{equation}
    \overline{r^2_f}^{1/2}(\alpha) = \sqrt{\frac{1}{N_\text{\rm s}\times N_\text{\rm pix}}\sum_{i}^{N_\text{\rm s}}\sum_{j}^{N_\text{\rm pix}} r(\alpha)^2_{f,ij}},
    \label{eq:rms}
\end{equation}
which measures the entire error, i.e. the bias and the standard deviation (the standard deviation measures the scatter about the bias).

We calculate the fractional bias and RMS fractional residuals of the subtraction as a function of $\alpha$ for the four galaxies, and plot them in Figure~\ref{fig:rms}. At optimum $\alpha$---we preliminarily define it here as the $\alpha$ that gives the smallest RMS fractional residuals on the test data (later we will re-discuss this subject about what is \textit{optimum} for the purpose of host-galaxy subtraction)---$\alpha\sim 10^{18}$ for the simple galaxy at $z = 1.0$ and $\alpha\sim 3\times 10^{17}$ for the VELA01 galaxy, the RMS fractional residuals is very much below the noise level expected from host-galaxy subtraction by reobserving the galaxy\footnote{One way to subtract the host galaxy spectrum from the galaxy+supernova spectrum is to reobserve the host galaxy well after the light of the supernova has significantly dimmed, at \textit{the exact dither and roll angle} at which we observed the supernova. Due to the survey design this might not necessarily happen.}---shown as the dashed blue line. It is also lower than the expected noise level of a single empty image (the noise that came only from the sky background and instrumental noise), shown as the dashed green line. For low values of $\alpha$ (i.e. $\alpha\lesssim 10^{14}$), where much more emphasis is put on the data rather than the prior, the noise of the predictions are worse than even the reobservation noise. 

The prediction noise at optimum $\alpha$ is very much lower than the expected noise of a single image because we are effectively combining the signals from multiple spectra in making the predictions. As shown by the dashed green line in Figure~\ref{fig:rms}, the expected RMS noise of a single image is $\sim$0.017 for an exposure time of 1800\,s, thus for $N_\text{\rm s} = 24$ spectra, the expected prediction noise is $0.017/\sqrt{24} = \sim$0.003. This is comparable to what we see in Figure~\ref{fig:rms}, where at the optimum $\alpha$ the RMS residuals is $\sim$0.003 for the simple galaxy and $\sim$0.007.

The bias, which can be used to see how much of the host galaxy is left after it has been subtracted, suggests that very little of the host-galaxy remains. At optimum $\alpha$ it is very close to zero.

Up to this point we define the optimum $\alpha$ to be that with the smallest RMS fractional residuals. The RMS residuals measure both the bias and the standard deviation\footnote{It can be shown that the RMS error is a quadrature summation of the bias and the standard deviation.}, however one of our goals for host galaxy subtraction is to remove possible (hidden) systematic errors from the supernova spectra, since for cosmological measurements the bias is as, or even more, important than the RMS. 
For this purpose the best choice of $\alpha$ is then that for which the bias is closest to zero rather than that which minimizes the RMS residuals. 

\subsection{Autocorrelations of residual images}
\label{sec:autocorr}
We present another way to evaluate the cleanliness of the host galaxy subtraction, which is by evaluating the closeness of the residual image to a white noise image, in which the noise is uncorrelated. We quantify this by measuring the correlation coefficients between two points on the residual images separated by a distance $d$.

As we have seen in Figures~\ref{fig:predalpha}--\ref{fig:predalpha_vela01}, inaccurate predictions appear as underestimation streaks. These are systematic biases that spatially correlate with each other, and we can expect them to have high correlation coefficients.

To measure the autocorrelation coefficients (correlation between an image with itself), we use the same set of 1000 random spectra used previously to calculate the RMS and mean residuals. For each residual image $\res{}$ we calculate the autocorrelation coefficient $\rho_{\res}$ for all two points in the image separated by a distance $d$:
\begin{equation}
    \rho_{\res}(d) = \frac{\mathbb{E}[(\res_{\vecbold{x}+d} - \mu_{\res})\overline{(\res_{\vecbold{x}} - \mu_{\res})}]}{\sigma_{\res}^2},
    \label{eq:autocorr}
\end{equation}
where $\mathbb{E}$ is the expected value operator for the quantities inside the square bracket, overline is complex conjugation, $\mu_{\res}$ is the mean of the residual image, $\sigma^2_{\res}$ is the variance of the residual image, and $\vecbold{x}$ is the image coordinate. In the following analysis we will always ignore correlations at $d = 0$, viz. correlation of a pixel with itself, because they will always be unity. We will also consider only correlations along the dispersion direction and ignore those along the cross-dispersion direction, as imperfections in the reconstructed datacube will be more amplified along the dispersion direction. 

Figure~\ref{fig:acf} shows the autocorrelation function for all 1000 random spectra and for different choices of $\alpha$, again for the simple host galaxy at $z=0.5$. For $\alpha\gtrsim 10^{18}$, on average any two points in the residual image separated at any distances along the dispersion direction are highly correlated. This is a measure of the systematic bias due to the prior mismatch we observed in Figure~\ref{fig:predalpha} and discussed in Section~\ref{sec:alpha}, this time quantified as the autocorrelation coefficient.

As $\alpha$ gets lower and we put more emphasis on the data rather than the prior, the average correlation coefficient gets closer to zero for distances $d_{X}\gtrsim 2$\,pix. Again we see here the manifestation of the noisy predictions, which tend to hide the bias under the prediction noise, due to the increased flexibility of data-dominated inferences. The trade-off is the higher variance in the predictions because even small changes in the data (due to noise) will produce significantly different fits. 


At short distances, $d_X\lesssim 2$\,pix, some correlations are still present at intermediate and low values of $\alpha$ (i.e. $\alpha\lesssim 10^{16}$). This is due to the smearing of the reconstruction errors by the PSF, which has a FWHM of $\sim$1.6\,pix. For very low values of $\alpha$, i.e. $\alpha\lesssim 10^{14}$, the reconstruction errors are so high that the PSF smearing in fact increases correlations between points separated by intermediate distances, e.g. $d_X\lesssim 10$\,pix. This is something we should bear in mind when evaluating the summary statistics of the correlation coefficients. 

Figure~\ref{fig:acf_vela01} shows the autocorrelation function for the VELA01 galaxy. We observe the same behaviour as the simple galaxy, but with more pronounced correlations at $\alpha = 10^{17}$. This is because of more complex structure of this galaxy's datacube, which has different spectra for each point in the galaxy. 

We show the mean correlation coefficients $\overline{\rho}_\text{\rm ACF}$ over all distances as a function of $\alpha$ in Figure~\ref{fig:rho} (black line) for the simple galaxy at $z = 1.0$. As we see in Figures~\ref{fig:acf}--\ref{fig:acf_vela01}, at lower values of $\alpha$, points separated at distances shorter than $\sim$10\,pix are highly-correlated and will surely dominate $\overline{\rho}_\text{\rm ACF}$. Thus in Figure~\ref{fig:rho} we also show $\overline{\rho}_\text{\rm ACF}$ if several shorter distances are excluded.


The highly-correlated residuals due to prior mismatch at $\alpha\gtrsim 10^{18}$ dominate the mean autocorrelation coefficients $\overline{\rho}_\text{\rm ACF}$ at this range of $\alpha$. As $\alpha$ decreases, $\overline{\rho}_\text{\rm ACF}$ goes toward zero, indicating that for $\alpha\lesssim 10^{17}$ the average residuals between true and predicted 2d spectra look like white noise with very little systematic noise present. Nevertheless, as we see in the inset image in Figure~\ref{fig:rho}, very small correlations remains if we include all distances, and that can be reduced by excluding correlations between distances shorter than $\sim$$5$\,pix. We attribute this short-distances correlations between residuals to PSF smearing.


\subsection{Using cross-validation to fine-tune the regularization parameter $\alpha$}
\label{sec:alphacv}

In the previous section we showed the performance of the predicted spectra for various $\alpha$ by comparing them with the true spectra. This is useful for predicting the expected performance, but \textit{in real observations we will never be able to know what the true spectra would look like}. We will therefore never be able to determine the optimum $\alpha$ by comparing the predicted with the true spectra as discussed in the previous Section. Rather, we will make comparisons between the predicted and the (noisy) observed spectra.

Figures~\ref{fig:prednoisy1}--\ref{fig:prednoisy_vela01} show the comparison between spectra predicted using various $\alpha$, with the corresponding ``observed'' spectra, generated using the true scene added with the expected read and sky noise as prescribed in Section~\ref{sec:noise}, respectively for the simple galaxy at $z=1.0$ and the VELA01 galaxy. We see in the residual plots (rightmost columns) that any systematic residuals we saw in Figures~\ref{fig:predalpha}--\ref{fig:predalpha_vela01} are now drowned out by the noise. For example, whereas in Figure~\ref{fig:predalpha}, for $\alpha = 10^{17}$ we can still see by eye that small amounts of systematic residuals are still left, for the same $\alpha$ in Figure~\ref{fig:prednoisy1} these systematic residuals are buried under the read and sky noise. We observe the same behaviour as well for the VELA01 galaxy in Figure~\ref{fig:prednoisy_vela01}, where the slight systematic residuals observed in $\alpha = 10^{17}$ is now completely hidden under the read and sky noise.

To address this we evaluate the predictive performance of the reconstruction as a function of $\alpha$ using $k$-fold cross-validation, i.e. splitting the data into $k$ training and validation sets. The $N_\text{\rm s}$ spectra we are working with are partitioned into $k$ (more-or-less) equal-sized subsets. For each subset\footnote{Also called \textit{fold}, hence the name $k$-fold cross-validation.}, we form a \textit{training set} $\dset_{\text{\rm train},k}$ containing all subsets \textit{except} the $k$-th subset, which form the \textit{test set} $\dset_{\text{\rm test},k}$ (an illustration is shown in the top panel of Figure~\ref{fig:cvscheme}). 

For a fixed value of $\alpha$ and for each training set, we use the $k$-th training set $\dset_{\text{\rm train},k}$ to infer $\cube_k$, then we use $\cube_k$ to predict the corresponding test set $\dset_{\text{\rm test},k}$. We then evaluate the quality of the reconstruction by calculating the sum of squared fractional residuals
\begin{equation}
    SSR_\text{\rm CV}(\alpha) = \sum_k\sum_{i\in\dset_{\text{\rm test},k}}\sum_{j}\left[\frac{y_{ij}- \left(\model_i\cube_{k}\right)_j}{y_{ij}}\right]^2,
\end{equation}
which is then used to calculate the cross-validation RMS residuals:
\begin{equation}
    \overline{r^2}^{1/2}_\text{\rm CV}(\alpha) = \sqrt{\frac{1}{N_\text{\rm s}\times N_\text{\rm pix}}SSR_\text{\rm CV}(\alpha)}
\end{equation}
and similarly to Equation~\ref{eq:bias} we calculate the cross-validation bias
\begin{equation}
    \overline{r}_\text{\rm CV}(\alpha) = \frac{1}{N_\text{\rm s}\times N_\text{\rm pix}}\sum_k\sum_{i\in\dset_{\text{\rm test},k}}\sum_j \frac{y_{ij} - \left(\model_i\cube_{k}\right)_j}{y_{ij}}.
\end{equation}

There are various schemes to assign the data into training and testing set. Figure~\ref{fig:cvscheme} shows two different schemes. In this Figure, the horizontal axis correspond to the indices of the spectra, sorted from small to large roll angles. On the top panel is the simple arrangement discussed previously. This arrangement, however, does not equally cover the range of roll angles in the training set. For example, in fold index 0, the training set spans indices 8 to 23, while the test is spans indices 0 to 7, which are outside the span of the training set. Similarly, the test set of fold index 2 (indices 16 to 23) is also outside the span of its training set (indices 0 to 15). Only the test set of fold index 1 is within the boundaries of its training set. The test beyond the bounds of the training set is an extrapolation and can result in biased predictions.

To minimize extrapolating beyond the span of the training set and ensuring that the test set in each fold is within the span of its training set, we interleave the testing set in-between the training set, such that the test set of each fold is within the boundaries of available training data. This scheme is shown on the bottom panel of Figure~\ref{fig:cvscheme}.

Figure~\ref{fig:cv} plots the 2-fold cross-validation RMS residuals and bias as a function of $\alpha$, for the same four galaxies analyzed in Section~\ref{sec:alpha}. Naturally the cross-validation RMS errors are higher than what is shown in Figure~\ref{fig:rms} because of the additional noise, but it is only slightly above what is expected from a single image (shown by the green dashed line). Additionally, the optimum $\alpha$ is still consistent with what we see in Figure~\ref{fig:rms}, which is at $\alpha\sim 2\times 10^{17}$.

Using the test sets, we repeat the autocorrelation analysis performed in Section~\ref{sec:autocorr}. This time we use the residuals between predicted and observed 2d spectra images in the test set. Because we now work with noisy data, we concentrate only on regions where we expect signals and exclude the outer parts of the residual images. We define a box $11\,\text{\rm pix}\times 216\,\text{\rm pix}$ centered on each spectral image (the images are $25\,\text{\rm pix}\times 256\,\text{\rm pix}$ in dimension) and only consider residuals within this box. The autocorrelation function for various choices of $\alpha$ is shown in Figure~\ref{fig:acfcv}. Unlike the correlation coefficients in residual images between predicted and noise-free data shown in Figure~\ref{fig:acf}, which has broader range of correlation coefficient values, here it is immediately apparent that the correlation coefficients are closer to zero even at high $\alpha$ values (note that the range of the vertical axis is narrower in this Figure). The added instrument and sky noise makes the residuals look more like white noise. On the opposite end, at lower $\alpha$ values, the highly-correlated residuals at short distances---which we attribute to reconstruction noise smeared by the PSF---can still be seen as the major source of systematic errors. 

Because of the presence of instrument and sky noise, we expect that the residual images will have very low mean cross-validation correlation coefficients, $\overline{\rho}_\text{\rm CV, ACF}$. We plot $\overline{\rho}_\text{\rm CV, ACF}$ in Figure~\ref{fig:acfstat} as a function of $\alpha$, which is similar to Figure~\ref{fig:rho} but now with much narrower range of vertical axis. Despite the much lower mean correlation coefficients, similar structure observed in Figure~\ref{fig:rho} can still be seen, in which $\overline{\rho}_\text{\rm CV, ACF}$ starts to flatten out for $\alpha\lesssim 3\times 10^{16}$. 

We conclude that analyzing the cross-validation autocorrelation of the residuals is a powerful method to evaluate whether clean subtractions that leave only white noise have been achieved, provided that we exclude the areas where no signal is expected. When working with noisy spectra, the correlations of the residuals are more subdued due to the presence of noise, yet we obtained similar results as working with noise-free data.

The choice of cross-validation score to fine-tune the hyperparameters\footnote{Parameters of the prior distribution. In our case $\alpha$ is the sole hyperparameter.} is usually motivated by the need to find those that minimize the overall errors, for example the root-mean-square residuals, which measures both the bias and the standard deviation. Because the overall goal of this project is to obtain more precise cosmological measurements free of systematic errors, here we choose to fine-tune the regularization parameter $\alpha$ such that the subtraction produces a residual image that leaves no systematic structure and resembles white noise. To evaluate this we can calculate the average cross-validation autocorrelation coefficient in the residual image. The optimum $\alpha$ is then that where $\overline{\rho}_\text{\rm CV,\,ACF}(\alpha_\text{\rm opt})\simeq 0$. This can be obtained using root-finding algorithms such as Brent's method \citep{bre73}, which does not require derivatives. 

\section{The host-subtracted supernova-only spectra}
\label{sec:subtraction}

We now return to the supernova+galaxy spectrum shown in Figure~\ref{fig:obs}, observed at roll angle $\phi = 195.4^\circ$, and use the datacube $\D$ reconstructed using 24 2d spectral images and with $\alpha = 10^{16}$ as discussed in the previous Section, to subtract the host galaxy from the supernova+galaxy spectrum. For the VELA01 galaxy, the resulting SN-only spectrum is shown in Figure~\ref{fig:snonly_vela01}. We compare the host-subtracted supernova only spectrum (panel a) with the true supernova-only spectrum (panel b) by creating the residual image (panel c). We see no indications that there are systematic residuals. From the 2d supernova-only spectrum in panel~a, we extract the 1d spectrum by summing the fluxes inside the green rectangle along the spatial direction (the rectangle is 5\,pix in height), then to reduce the noise we convolve it with a Gaussian kernel with FWHM 1.8\,pix. The resulting 1d spectrum is shown in red in panel~(d). Comparing this with the true 1d spectrum (shown in green in panel~d), we calculate the residuals between the 1d spectra. In panel~(e) we plot the residuals as a function of pixel coordinate along the dispersion axis, with the corresponding histogram shown in panel~(f). In panel~(g) we plot the fractional residuals, i.e.
\begin{equation}
    r_{f,j} = \frac{s_j - s_{\text{\rm true}, j}}{s_{\text{\rm true}, j}},
\end{equation}
for each pixel index $j$. As in Equation~\ref{eq:res}, here $s_j$ is the predicted brightness at pixel index $j$ in the predicted 1d spectra, and $s_{\text{\rm true}, j}$ is the corresponding true brightness. In panel~(h) the corresponding histogram for the fractional residuals is plotted.

We repeat this for the simple galaxies, with the results plotted in Figures~\ref{fig:snonly1}--\ref{fig:snonly3} in Appendix~\ref{app:results}. In Table~\ref{tab:stat_snonly}, for all galaxies we summarize the statistics of the residuals both in absolute (columns~3--6) and relative (columns~7--10) terms. Overall the matches betweeen the host-subtracted SN-only 1d spectra with the corresponding true 1d spectra is very good, except for the galaxy with $z=1.5$, which are dominated by sky background and instrumental noise (hence lower SNR), it is difficult to make an unbiased predicted spectrum using only 24 rotation angles.

\begin{deluxetable*}{crrrrrrrrr}
\tablecaption{Summary statistics of the residuals of the supernova-only 1d spectra shown in Figures~\ref{fig:snonly_vela01}, \ref{fig:snonly1}--\ref{fig:snonly3}. Shown here is the bias (column 3), median (column 4), standard deviation (column 5), and RMS residuals (column 6). Columns 7--10 are the corresponding statistics in terms of fractional residuals. Note that this is for a single roll angle and that the host-galaxy spectra were subtracted by making use of the datacube reconstructed using 24 observations and $\alpha = 10^{16}$.}
\label{tab:stat_snonly}
\tablehead{
\colhead{Model} & \colhead{$z$} & \colhead{$\overline{r_\text{\rm 1d}}$} & \colhead{${\rm Md}(r_\text{\rm 1d})$} & \colhead{$\sigma_{r_\text{\rm 1d}}$} & \colhead{$\overline{r_\text{\rm 1d}^2}^{1/2}$} & \colhead{$\overline{r_{f,\text{\rm 1d}}}$} & \colhead{${\rm Md}(r_{f,\text{\rm 1d}})$} & \colhead{$\sigma_{r_{f,\text{\rm 1d}}}$} & \colhead{$\overline{r_{f,\text{\rm 1d}}^2}^{1/2}$} \\
\colhead{} & \colhead{} & \colhead{} & \colhead{(ADU)} & \colhead{(ADU)} & \colhead{(ADU)} & \colhead{(ADU)}
}
\decimalcolnumbers
\startdata
VELA01 & $1.0$ & $-0.51$ & $-1.10$ & $43.21$ & $43.21$ & $0.006$ & $-0.007$ & $0.256$ & $0.256$\\
\hline
\multirow{3}{3em}{Simple Galaxy} & $0.5$ & $  1.61$ & $  1.42$ & $ 46.07$ & $ 46.08$ & $ 0.005$ & $ 0.019$ & $ 0.118$ & $ 0.118$\\
 & $1.0$ & $ -3.11$ & $ -2.69$ & $ 41.99$ & $ 42.02$ & $-0.032$ & $-0.007$ & $ 0.290$ & $ 0.290$\\
 & $1.5$ & $ 16.88$ & $ 18.89$ & $ 40.90$ & $ 41.59$ & $ 0.088$ & $ 0.221$ & $ 1.337$ & $ 1.337$\\ 
\enddata
\tablecomments{Here the statistics are summarized for the 1d spectra, which sum over 5 pixels vertically as mentioned in the text. The standard deviation and RMS displayed here however is normalized to a single pixel (such that they are comparable to the statistics shown in Figures~\ref{fig:rms} and \ref{fig:cv}) by dividing them with $\sqrt{5}$.}
\end{deluxetable*}



For $z=0.5$, the single-pixel RMS residuals is $46.08$\,ADU. This is comparable to the expected noise of a single image, which is $\sim$55\,ADU (shown as the dashed green line in Figures~\ref{fig:rms} and \ref{fig:cv}), but it is somewhat lower because of the Gaussian smoothing applied to the 1d spectra. Using an optimum spectrum extraction technique \citep[][q.v.]{hor86} will ultimately reduce the noise further.

The results above are for a single roll angle and supernova location in its host galaxy. Naturally we are also interested to see the expected performance of the subtraction over various new roll angles and supernovae locations in its host galaxy. To assess this, we generate various supernovae locations based on the $R$-band light distribution of the host galaxy. For the VELA01 galaxy and for the simple galaxies, on each galaxy we generate 1000 supernovae locations, with the 2d histograms of the injected supernovae locations shown in Figures~\ref{fig:sndist_vela01}--\ref{fig:sndist}. The corresponding observatory roll angles and sub-pixel dithers are also randomly generated.

For each of these 1000 spectroscopic observations of supernova+galaxy scene, we subtract the host-galaxy from the spectra again using the datacube inferred using $\alpha = 10^{16}$, and extract the SN-only 1d spectrum in the same manner as discussed previously. In Figure~\ref{fig:snonlyrandomselected1} we show the 1d spectra residuals for eight random selections from our pool of 1000 SN-only spectra. As we can see from the histograms of the residuals, their distribution resemble a normal distribution.

Panel~(a) of Figure~\ref{fig:snrandom_vela01} shows all 1000 of the extracted SN-only 1d spectra as the 2d histogram as a function of fluxes and wavelengths, for the VELA01 host-galaxy. For comparison, the corresponding true SN-only 1d spectra are shown individually in this panel as the green lines. Although difficult to see in the Figure, the true spectra are slightly different from each other and form a very thin band. This is due to the different values of sub-pixel dithers and the limited extraction window (5 pixels over the spatial direction), which shifts a small part of the flux out of the extraction area. Panel~(b) in Figure~\ref{fig:snrandom_vela01} shows the absolute residuals of the 1000 1d SN-only spectra again as a 2d histogram, with the corresponding 1d histogram shown in panel~(c). Similarly, panel~(d) plots the fractional residuals, with the corresponding histogram shown in panel~(e). The statistics of these residuals is summarized in Table~\ref{tab:stat_snrandom} for all four galaxies, again in terms of the absolute residuals as well as fractional residuals.

Looking at the summary statistics in Table~\ref{tab:stat_snrandom}, the skewness and kurtosis of the 1d distribution of the absolute residuals (respectively columns (7) and (8)) are close to zero, suggesting that overall the shape of the 1d distributions are normally distributed. The same cannot be said for the fractional residuals, however. Column (14) show the kurtosis of the 1d distribution of the fractional residuals, showing considerably higher value of kurtosis for all galaxies. This indicates heavier tail and a higher probability of outliers in the 1d distribution of the fractional residuals. We can expect heaver tail as the source grows fainter and dominated more by the sky background and instrumental noise.

The mean of this distribution, which measures the bias, is shown in Column~(3) of Table~\ref{tab:stat_snrandom}, while the scatter about this mean is shown in Column~(5) of the same Table. Comparing these summary statistics with those for a single roll angle in Table~\ref{tab:stat_snonly}, the overall noise---represented by the RMS residual $\overline{r_\text{\rm 1d}^2}^{1/2}$ in column~(5)---is consistent with a single roll angle observation. This means on average the noise does not change with SN locations or roll angles. What is different is the bias $\overline{r_\text{\rm 1d}}$, which is worse than those for a single roll angle. This is because the single supernova location shown in Figure~\ref{fig:snonly1} is far from the core of the galaxy (see Figure~\ref{fig:obs}), while the majority of the 1000 locations we have generated are concentrated around the galactic core (see Figure~\ref{fig:sndist}). This is naturally the brightest part of the galaxy and we can expect this to be a problematic area. Nevertheless, looking at column~8 of Table~\ref{tab:stat_snrandom}, from the host-galaxy subtraction, on average we can expect a median fractional residuals of $1$\%--$3$\% of the true flux. 

\begin{deluxetable*}{crrrrrrrrrrrrr}
\tablecaption{As Table~\ref{tab:stat_snonly}, but for the summary statistics of the residuals of supernovae-only 1d spectra shown in Figures~\ref{fig:snrandom_vela01} and \ref{fig:snrandom123}, which are 1000 randomly generated supernovae spread over the corresponding host galaxy. Skewness and kurtosis of the distribution of the residuals are respectively shown in columns (7) and (8) for the absolute residuals and in columns (13) and (14) for the fractional residuals. As throughout the paper, the host-galaxy spectra were predicted by making use of the datacube reconstructed using 24 observations and $\alpha = 10^{16}$.}
\label{tab:stat_snrandom}
\tablehead{
\colhead{Model} & \colhead{$z$} & \colhead{$\overline{r_\text{\rm 1d}}$} & \colhead{${\rm Md}(r_\text{\rm 1d})$} & \colhead{$\sigma_{r_\text{\rm 1d}}$} & \colhead{$\overline{r_\text{\rm 1d}^2}^{1/2}$} & \colhead{$\beta_{1,{r_\text{\rm 1d}}}$} & \colhead{$\beta_{2,{r_\text{\rm 1d}}}$} & \colhead{$\overline{r_{f,\text{\rm 1d}}}$} & \colhead{${\rm Md}(r_{f,\text{\rm 1d}})$} & \colhead{$\sigma_{r_{f,\text{\rm 1d}}}$} & \colhead{$\overline{r_{f,\text{\rm 1d}}^2}^{1/2}$} & \colhead{$\beta_{1,{r_{f,\text{\rm 1d}}}}$} & \colhead{$\beta_{2,{r_{f,\text{\rm 1d}}}}$}\\
\colhead{} & \colhead{} & \colhead{(ADU)} & \colhead{(ADU)} & \colhead{(ADU)} & \colhead{(ADU)}
}
\decimalcolnumbers
\startdata
VELA01          & $1.0$ & $ -5.18$ & $ -4.99$ & $ 42.91$ & $ 42.97$ & $-0.003$ & $ 0.021$ & $-0.031$ & $-0.026$ & $ 0.296$ & $ 0.296$ & $-0.056$ & $ 2.666$\\
\hline
\multirow{3}{3em}{Simple Galaxy} & $0.5$ & $ -6.41$ & $ -5.81$ & $ 44.62$ & $ 44.71$ & $-0.021$ & $ 0.005$ & $-0.018$ & $-0.011$ & $ 0.117$ & $ 0.117$ & $-0.199$ & $ 3.154$\\
 & $1.0$ & $ -1.52$ & $ -1.21$ & $ 42.08$ & $ 42.08$ & $-0.003$ & $ 0.015$ & $-0.012$ & $-0.005$ & $ 0.292$ & $ 0.292$ & $-0.074$ & $ 2.789$\\
 & $1.5$ & $ -2.73$ & $ -2.83$ & $ 42.07$ & $ 42.09$ & $-0.003$ & $ 0.018$ & $-0.037$ & $-0.030$ & $ 1.137$ & $ 1.137$ & $-0.086$ & $ 5.695$\\
\enddata
\end{deluxetable*}

\begin{figure}
    \centering
    \includegraphics[width=\columnwidth]{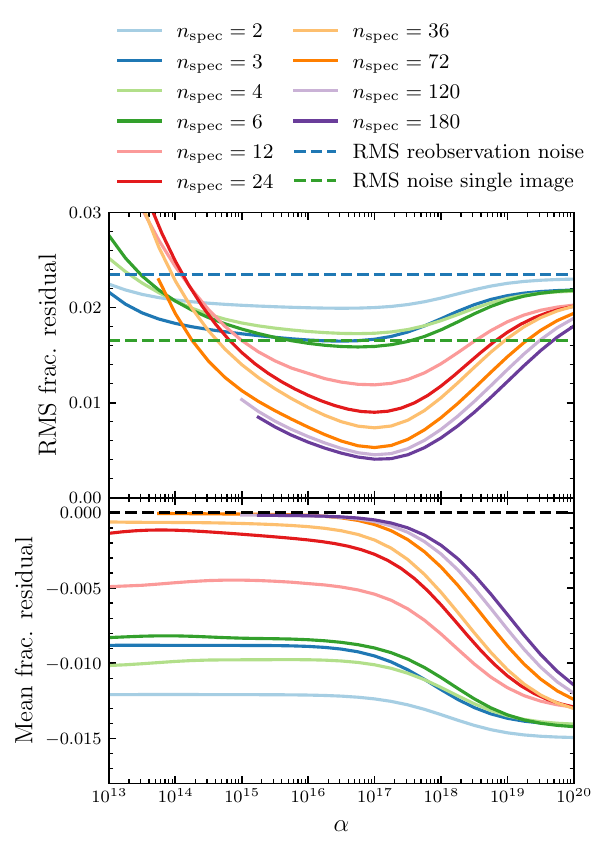}
    \caption{As Figure~\ref{fig:rms}, but now for the performance of reconstructing the VELA01 galaxy using various number of available roll angles $n_{\rm spec}$. Here the performance for a given $n_{\rm spec}$ is plotted as a function the regularization parameter $\alpha$, and for 1000 new roll angles not in the dataset used for reconstruction. A general trend is that the fractional RMS fractional residuals is decreasing with increasing number of roll angles used for reconstruction, and that the mean fractional residuals gets closer to zero.}
    \label{fig:rms_nspec}
\end{figure}

\begin{figure}
    \centering
    \includegraphics[width=\columnwidth]{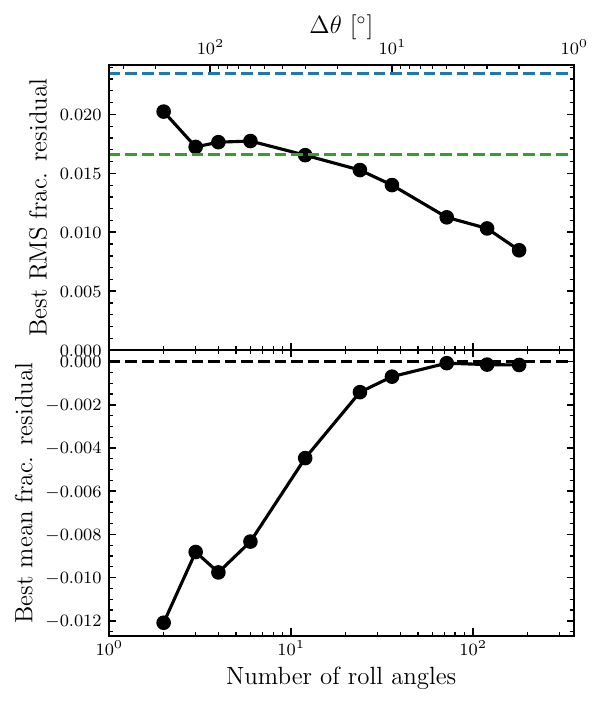}
    \caption{The reduction of the RMS fractional reconstruction residuals error with increasing number of available roll angles (top panel), and the corresponding plot for the mean fractional residuals (bottom panel).}
    \label{fig:rms_alpha_nspec}
\end{figure}

\begin{figure*}
    \centering
    \includegraphics[width=\textwidth]{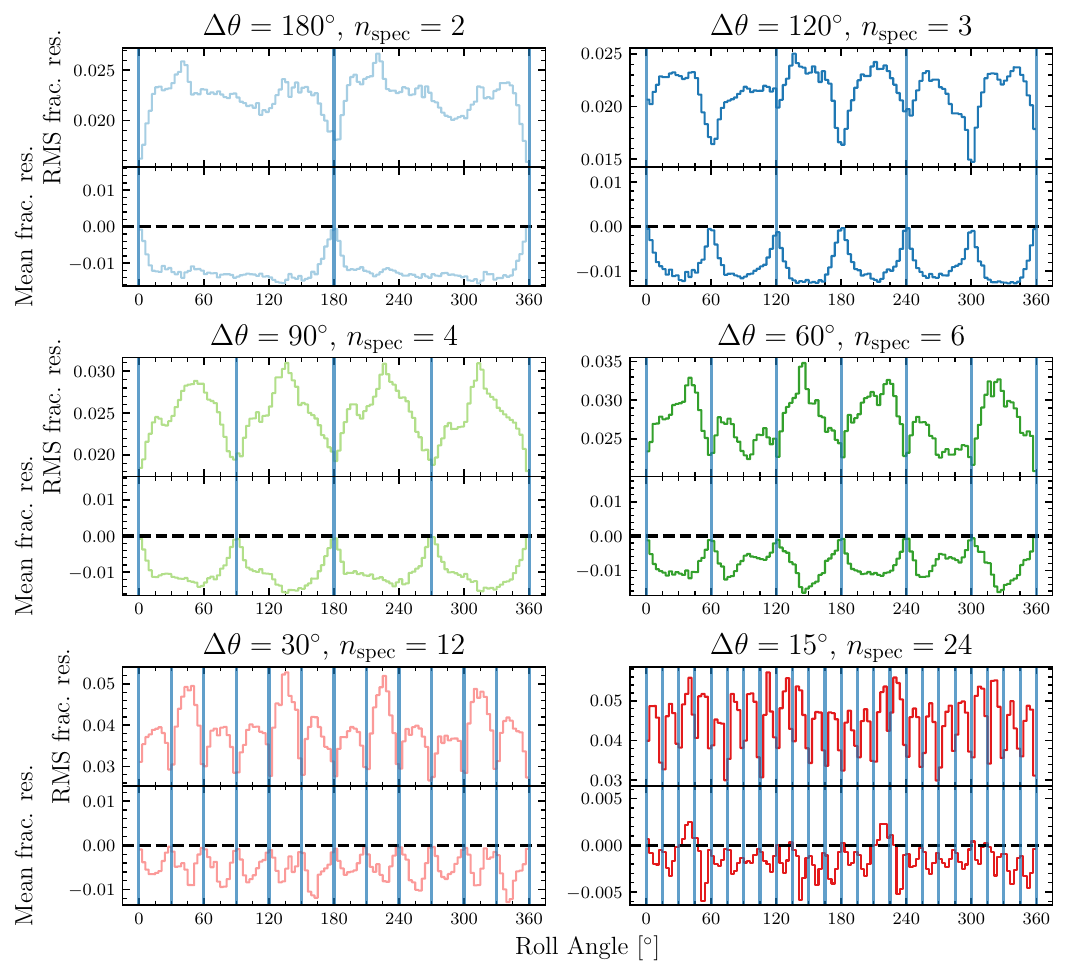}
    \caption{Each panel plot the RMS fractional reconstruction residual errors (upper sub-panel) and the mean fractional residuals (lower sub-panel) as a function of new roll angle bin used as input to predict new spectra. For each panel the predictions of new spectra make use of datacube reconstruction with the given number of available roll angles $n_{\rm spec}$ as shown on the top of each panel. In each panel, vertical lines mark the available roll angles used for reconstruction. A general trend is that the minimum RMS and mean fractional residuals between predicted and true spectra are when the new roll angles are close to the available roll angles.}
    \label{fig:rms_rollangles}
\end{figure*}


\section{Discussion and Conclusion}

We have developed a datacube reconstruction algorithm that makes use of multiple slitless spectroscopic observations of a scene with various observatory roll angles and sub-pixel dithers. We do not make any assumption about the nature of the scene, making our approach applicable for arbitrary scenes. The reconstructed datacube can be used to predict the spectra at new roll angles that were not in the dataset used for reconstruction, with very little systematic residuals left. 

Because the predictions are made by a joint analysis of multiple spectra, the noise of the predictions is lower than the noise of a single 2d spectrum image, and decreases as more roll angles are included in the analysis. To see how the prediction noise decreases with increasing the number of roll angles, we reconstruct the VELA01 scene using various number of available roll angles $n_{\rm spec}$ and again for each $n_{\rm spec}$ we use various choices of regularization parameter $\alpha$. In Figure~\ref{fig:rms_nspec} we show the summary statistics of the resulting residuals in predicting 1000 new roll angles not in the dataset used for reconstruction, again in terms of the RMS fractional residuals and the mean fractional residuals.

The top panel of Figure~\ref{fig:rms_nspec} shows a general trend of decreasing fractional RMS residuals as we use more roll angles for scene reconstruction. The bottom panel, which shows the corresponding mean fractional residuals, shows that the mean residuals also get ever closer to zero with increasing number of roll angles. In Figure~\ref{fig:rms_alpha_nspec} we show the reduction of the RMS fractional residuals as we increase the number of data for reconstruction, as well as the best possible achievable mean fractional residuals. As we can see here, at optimum regularization parameter $\alpha$, the general trend is that the RMS and mean fractional residuals improves with increasing data.




Going to the other end, when we do not have too much roll angle to work with, for $n_{\rm spec}\lesssim 10$, the reconstruction puts more emphasis on the data. Reconstruction is still possible albeit noisy, so much so it is noisier than the RMS expected from a single image and similar to reobservation noise. We can see that host-galaxy subtraction using scene reconstruction can do better predictions over reobservation when the number of available roll angle is more than 10.



So far our analysis are averaged over all possible new roll angles. It would also be interesting, however, to see if there are deviations from the overall trend we have seen, as a function of new roll angles. We bin the 1000 new, predicted, spectra into bins of roll angles, and calculate the RMS and mean fractional residuals between predicted and true spectra for each bin. In Figure~\ref{fig:rms_rollangles} we plot the results for selected datacubes reconstructed with $n_{\rm spec} = 2, 3, 4, 6, 12,$~and $15$. We see here that the best RMS and mean fractional residuals tend to concentrate around the new roll angles closest to those where we have data. Gaps in roll angle coverage tend to worsen the predictions, but these predictions are improved as these gaps are covered by new data.

Our method is however computationally intensive and this limits us to testing the algorithm using scenes with small sizes. There are two reasons for this slow computing time. First, although $\model$ is sparse, the number of non-zero elements in the matrix is still very high. We show in Appendix~\ref{app:psf} that a huge reduction in the sparsity of $\model$ can be made by reducing the size of the PSF kernel used in convolution. A substantial reduction in the time to compute the predicted spectra $\vect{\spec} = \model\vect{\D}$ can be achieved by reducing the PSF kernel size. The minimum kernel size that can still provide adequate accuracy will be addressed in a future publication.

Second, the size of the reconstructed scene will determine the number of parameters $P$ we want to infer. As we have discussed, the reconstructed datacube must be oversampled to accurately map all the various transformations described in Section~\ref{sec:forward}. We find that an oversampling factor of 3 is sufficiently accurate. Oversampling the scene by this factor (with no increase in the size of the wavelength grid $\wave$) will increase $P$ ninefold. A potential avenue for decreasing the number of parameters is to use B-splines \citep{deb78}, where a sub-pixel is no longer represented by a rectangular function but with piecewise polynomials. This additional flexibility can reduce the oversampling factor without significant reduction in accuracy. This will be the subject of a future study.

Our method has been tested not only on simple galaxies with uniform SEDs at every point on the galaxies, but also on a galaxy with varying SEDs. The reconstructed scene performed equally well in both cases. However, because in real applications galaxies will be more complex---varying not only in its SED at every point, but also in morphology---we need to expand the library of realistic scenes that \Roman{} will encounter over the course of the survey. Galaxies come in different morphologies and physical properties such as age, metallicity, star-formation rate, and velocities that depend on the location within the galaxy.  While these features are generally spatially smooth, dust attenuation and local star-forming regions can produce discontinuous features in the SED. We are currently working on generating realistic scenes of galaxies using multi-band images of nearby galaxies and integral-field-unit spectra from surveys like MaNGA\footnote{Mapping Nearby Galaxies at APO, \href{https://www.sdss4.org/surveys/manga}{https://www.sdss4.org/surveys/manga}} \citep{bun15}. This added realism will improve the performance assessment. A particularly important question to answer is whether SED gradients and sharp features will result in biased host-subtracted spectrum. 

Finally we remark that our method is predicated on the assumption that all parameters used to form the operator $\model$ are accurate and their uncertainties are negligible. This means that the PSF $\PSF$, the dispersion curve, and the passband should all be well measured, and that the uncertainties of the sub-pixel dithers and roll angles measurements are negligible. How well we can still reconstruct the datacube and predict new spectra with inaccurate $\model$ is another subject of a future study.

\begin{acknowledgments}
This research has made use of NASA’s Astrophysics Data System (ADS).

The authors thank Steven W. Brown and Bhavin Joshi for suggestions that improved this paper. This research is carried out with support of subcontract Nos. 7323250 and 7636013 between the Space Telescope Science Institute (STScI) and the Lawrence Berkeley National Laboratory (LBNL), awarded by the National Aeronautics and Space Administration (NASA). T.L.A. acknowledges the support from the STScI Discretionary Research Fund (DRF) D0001.82514, as well as the often-overlooked labor of the custodial, facilities, and security staff of STScI, without which this research would not be possible.

L.G. acknowledges financial support from the Spanish Ministerio de Ciencia e Innovaci\'on (MCIN) and the Agencia Estatal de Investigaci\'on (AEI) 10.13039/501100011033 under the PID2020-115253GA-I00 HOSTFLOWS project, from Centro Superior de Investigaciones Cient\'ificas (CSIC) under the PIE project 20215AT016 and the program Unidad de Excelencia Mar\'ia de Maeztu CEX2020-001058-M, and from the Departament de Recerca i Universitats de la Generalitat de Catalunya through the 2021-SGR-01270 grant.

R.A.H states that this material is based upon work supported by NASA under award number 80GSFC21M0002.

Funding for the \Roman{} Supernova Project Infrastructure Team has been provided by NASA under contract number 80NSSC24M0023.

Part of the research described in this paper was carried out at the Space Telescope Science Institute, which occupies the traditional, ancestral, and unceded territory of the Piscataway-Conoy and Susquehannock peoples. We honor their original stewardship of the land we live and work on, and acknowledge the deep root of colonialism, racism, and inequality in this country. We support the redress of centuries of genocide, ethnic cleansing, and forced removal against Native nations, including---but not limited to---land tax that benefits Indigenous communities, and returning land back under Indigenous stewardship.
\end{acknowledgments}

\software{Astropy \citep{astropy:2013, astropy:2018, astropy:2022}, IPython \citep{ipython}, Matplotlib \citep{Hunter:2007}, NumPy \citep{numpy}, Pandeia \citep{pon16}, PGF \& TikZ \citep{tan23}, SciPy \citep{scipy}, SNCosmo \citep{sncosmo}, synphot \citep{synphot}, STPSF (formerly called WebbPSF) \citep{per11,per12,per14}.}

\appendix

\section{Source codes and notebook}
The source codes implementing the algorithms discussed in this paper---written in \texttt{python}---as well as a \texttt{Jupyter} notebook described the installation procedure and a demonstration of the functionalities of the package, are available at \href{https://gitlab.com/astraatmadja/Ilia}{https://gitlab.com/astraatmadja/Ilia}.

\section{Vectorization of matrices and matricization of vectors}
\label{app:vec}

Let $\A$ be an $M\times N$ matrix. The \textit{vectorization} of $\A$, denoted $\vectorize{\A}$, is a linear transformation that rearranges the elements of $\A$ into an $MN\times 1$ \textit{column vector}:
\begin{equation}
    \vect{\A} = \vectorize{\A} = \left(A_{11}, A_{12}, \cdots, A_{1N}, A_{21}, A_{22}, \cdots, A_{2N}, \cdots, A_{M1}, A_{M2}, \cdots, A_{MN}\right)\transpose.
    \label{eq:vectorization}
\end{equation}
As an example, given a matrix $\A = \begin{pmatrix}A_{11} & A_{12}\\ A_{21} & A_{22}\end{pmatrix}$, then $\vectorize{\A} = \left(A_{11}, A_{12}, A_{21}, A_{22}\right)\transpose$. The vectorized matrix is denoted in lower case to reflect the conversion into vectors.

By default we vectorize a matrix in \textit{row-major order}, viz. we start from the top left element ($A_{11}$) then to the right going through the first row, etc. This is a convention we adopt from the way some programming languages store arrays, whereas mathematicians vectorize matrices in column-major order by default (starting from the top left element then downward going through the first column etc.), e.g.~\cite{zha17}.

A given matrix element index $(m,n)$ of an $M\times N$ matrix can be converted into the corresponding vector index $p$ by the transformation 
\begin{equation}
    p = (m - 1) N + n,\quad 1\leq p \leq MN.
    \label{eq:mnp}
\end{equation}
In \texttt{NumPy} this can be done using the \texttt{ravel\_multi\_index} routine (here the indices start from zero). Vectorization can be done using the \texttt{flatten} method.

We vectorize a three-dimensional matrix (such as a datacube) by starting from the first slice, then for each slice it is as Equation~\ref{eq:vectorization}. Suppose $\A$ is an $I\times J\times K$ array, then
\begin{equation}
    \vectorize{\A} = \left(A_{111}, \cdots, A_{11K}, \cdots, A_{1J1}, \cdots, A_{1JK}, A_{211}, \cdots, A_{2J1}, \cdots, A_{2J1}, \cdots, A_{2JK}, \cdots, A_{IJ1}, \cdots, A_{IJK}\right)\transpose,
\end{equation}
with a given element index $(i,j,k)$ can be converted into vector index $p$ by
\begin{equation}
    p = K\left[(i-1)J + (j-1)\right] + k,\quad 1\leq p \leq IJK.
    \label{eq:ijkp}
\end{equation}

The inverse operation of vectorization is \textit{matricization}, in which an $MN\times 1$ column vector $\vect{\A} = \left(a_{1}, a_{2}, \cdots, a_{MN}\right)\transpose$ is transformed into an $M\times N$ matrix $\A$, denoted as $\matricize{\A}{M}{N}$ and is defined as
\begin{equation}
    \A = \matricize{\A}{M}{N} =
    \begin{pmatrix}
        a_{1}    & a_{2}    & a_{3}     & \cdots & a_{N}\\
        a_{N+1}  & a_{N+2}  & a_{N+3}   & \cdots & a_{2N}\\
        a_{2N+1} & a_{2N+2} & a_{2N+3}  & \cdots & a_{3N}\\
        \vdots   & \vdots   & \vdots    & \ddots & \vdots\\
        a_{MN}   & a_{MN+1} & a_{MN+2} & \cdots  & a_{MN}
    \end{pmatrix},
\end{equation}
where a given vector index $p$ is converted into matrix element index $(m,n)$ by
\begin{equation}
    m = \floor{\frac{(p-1)}{N}} + 1,\qquad n = \left[(p - 1)\;\text{\rm mod}\;N\right] + 1,\qquad 1\leq p \leq MN,
    \label{eq:pmn}
\end{equation}
which can be done in \texttt{NumPy} using the \texttt{unravel\_index} routine (here the indices start from zero). Matricization can be done using the \texttt{reshape} method. As we can see we matricize $\vect{\A}$ by making the first $N$ elements of $\vect{\A}$ as the first row of $\A$, the next $N$ elements of $\vect{\A}$ as the second row of $\A$, etc.

When we are matricizing an $IJK\times 1$ vector $\vect{\A}$ into an $I\times J \times K$ three-dimensional matrix $\A$, denoted as $\A = \matricizend{\A}{I}{J}{K}$, the first $JK$ elements of the vector is matricized as $\A_1$ (the first slice of $\A$), the next $JK$ elements as the $\A_2$, etc. A given vector index $p$ can be converted into matrix element $(i,j,k)$ by
\begin{equation}
    i = \floor{\frac{(p-1)}{JK}} + 1,\quad j = \floor{\frac{(p-1) - (i-1)JK}{K}}+1,\quad k = \left[\left((p-1)-(i-1)JK\right)\;\text{\rm mod}\;K\right] + 1,\quad 1\leq p \leq IJK.
    \label{eq:pijk}
\end{equation}

\section{Discrete 2d convolution as a matrix operation}
\label{app:conv}
Suppose an $m\times n$ matrix $\A$ is convolved with a $u\times v$ matrix $\krl$ to produce $\vecbold{B}$ with the same size as $\A$. The value of $\vecbold{B}$ at $(i,j)$ is then \citep{ber22}
\begin{equation}
    B_{ij} = (\krl\star\A)_{ij} = \sum_{p,q}K_{(i-p+u_0)(j-q+v_0)}A_{pq},
    \label{eq:conv2}
\end{equation}
where $(u_0,v_0)$ are indices of the central element of $\krl$:
\begin{equation}
    u_0 = \lfloor\tfrac{1}{2}u\rfloor + 1,\qquad
    v_0 = \lfloor\tfrac{1}{2}v\rfloor + 1,
\end{equation}
and the summation is performed over the range
\begin{align}
    \max(1,i-u_0+1)\leq p \leq\min(u,i+u_0-1),\qquad
    \max(1,j-v_0+1)\leq q \leq\min(v,j+v_0-1).
\end{align}

If $\A$ and $\vecbold{B}$ are vectorized into $N\times 1$ vectors, i.e.
\begin{equation}
\vect{\A} = \vectorize{\A},\qquad
\vect{\vecbold{B}} = \vectorize{\vecbold{B}},
\end{equation}
it is possible to express the 2d convolution operation as a matrix multiplication $\vect{\vecbold{B}} = \C\vect{\A}$, where $\C$ is an $N\times N$ matrix that performs the convolution $K\star A$, and the output $\vect{\vecbold{B}}$ is an $N\times 1$ vector, which can then be converted back row-wise into an $m\times n$ matrix to recover the convolved matrix.

To illustrate on the form of $\C$, let us set an example that a $3\times 3$ matrix $\krl$ is convolved with a $3\times 3$ matrix $\A$, producing a $3\times 3$ matrix $\vecbold{B}$:
\begin{equation}
    \begin{pmatrix}
        B_{11} & B_{12} & B_{13}\\
        B_{21} & B_{22} & B_{23}\\
        B_{31} & B_{32} & B_{33}\\
    \end{pmatrix}
    = 
    \begin{pmatrix}
        K_{11} & K_{12} & K_{13}\\
        K_{21} & K_{22} & K_{23}\\
        K_{31} & K_{32} & K_{33}\\
    \end{pmatrix}
    \star
    \begin{pmatrix}
        A_{11} & A_{12} & A_{13}\\
        A_{21} & A_{22} & A_{23}\\
        A_{31} & A_{32} & A_{33}\\
    \end{pmatrix}.
    \label{eq:conv3}
\end{equation}

With $\A$ and $\vecbold{B}$ converted into $9\times 1$ vectors $\vect{\A}$ and $\vect{\vecbold{B}}$, the form of the convolution matrix $\C$ that can be operated on $\vect{\A}$ and is equivalent to the expression in Equation~\ref{eq:conv2} and \ref{eq:conv3} is
\begin{equation}
    \label{eq:bca}
    \begin{pmatrix}
        B_{11}\\
        B_{12}\\
        B_{13}\\
        B_{21}\\
        B_{22}\\
        B_{23}\\
        B_{31}\\
        B_{32}\\
        B_{33}\\
    \end{pmatrix}
    =
    \begin{pmatrix}
    K_{22} & K_{21} & 0      & K_{12} & K_{11} & 0      & 0      & 0      & 0     \\
    K_{23} & K_{22} & K_{21} & K_{13} & K_{12} & K_{11} & 0      & 0      & 0     \\
    0      & K_{23} & K_{22} & 0      & K_{13} & K_{12} & 0      & 0      & 0     \\
    K_{32} & K_{31} & 0      & K_{22} & K_{21} & 0      & K_{12} & K_{11} & 0     \\
    K_{33} & K_{32} & K_{31} & K_{23} & K_{22} & K_{21} & K_{13} & K_{12} & K_{11}\\
    0      & K_{33} & K_{32} & 0      & K_{23} & K_{22} & 0      & K_{13} & K_{12}\\
    0      & 0      & 0      & K_{32} & K_{31} & 0      & K_{22} & K_{21} & 0     \\
    0      & 0      & 0      & K_{33} & K_{32} & K_{31} & K_{23} & K_{22} & K_{21}\\
    0      & 0      & 0      & 0      & K_{33} & K_{32} & 0      & K_{23} & K_{22}\\
    \end{pmatrix}
    \times
    \begin{pmatrix}
        A_{11}\\
        A_{12}\\
        A_{13}\\
        A_{21}\\
        A_{22}\\
        A_{23}\\
        A_{31}\\
        A_{32}\\
        A_{33}\\
    \end{pmatrix}.
\end{equation}

\newcommand{\tikzmark}[1]{\tikz[overlay,remember picture] \node (#1) {};}
\newcommand{\tikzdrawbox}[2][]{
    \tikz[overlay,remember picture]{
    \fill[#1,opacity=0.3]
      ($(left#2)+(-0.3em,1.1em)$) rectangle
      ($(right#2)+(0.3em,-0.4em)$);}
}
\newcommand{\tikzdrawboxzero}[2][]{
    \tikz[overlay,remember picture]{
    \fill[#1,opacity=0.3]
      ($(left#2)+(-0.90em,1.1em)$) rectangle
      ($(right#2)+(0.90em,-0.4em)$);}
}

We can see from Equation~\ref{eq:bca} that convolution can be seen as a linear summation of all elements of $\vect{\A}$, weighted by certain elements of $\krl$ that form the operator $\C$. The convolution matrix $\C$ also has special properties that can be easier seen if we divide $\C$ into nine $3\times 3$ block matrices:
\begin{equation}
\C = 
\left(
\begin{array}{ccc|ccc|ccc}
    \tikzmark{left1}K_{22} & K_{21} & 0      & \tikzmark{left4}K_{12} & K_{11} & 0      & \tikzmark{left8}0      & 0      & 0     \\
    K_{23} & K_{22} & K_{21} & K_{13} & K_{12} & K_{11} & 0      & 0      & 0     \\
    0      & K_{23} & K_{22}\tikzmark{right1} & 0      & K_{13} & K_{12}\tikzmark{right4} & 0      & 0      & 0\tikzmark{right8}     \\
    \hline
    \tikzmark{left6}K_{32} & K_{31} & 0      & \tikzmark{left2}K_{22} & K_{21} & 0      & \tikzmark{left5}K_{12} & K_{11} & 0     \\
    K_{33} & K_{32} & K_{31} & K_{23} & K_{22} & K_{21} & K_{13} & K_{12} & K_{11}\\
    0      & K_{33} & K_{32}\tikzmark{right6} & 0      & K_{23} & K_{22}\tikzmark{right2} & 0      & K_{13} & K_{12}\tikzmark{right5}\\
    \hline
    \tikzmark{left9}0      & 0      & 0      & \tikzmark{left7}K_{32} & K_{31} & 0      & \tikzmark{left3}K_{22} & K_{21} & 0     \\
    0      & 0      & 0      & K_{33} & K_{32} & K_{31} & K_{23} & K_{22} & K_{21}\\
    0      & 0      & 0\tikzmark{right9}      & 0      & K_{33} & K_{32}\tikzmark{right7} & 0      & K_{23} & K_{22}\tikzmark{right3}\\
\end{array}
\right).
\tikzdrawbox[yellow]{1}
\tikzdrawbox[yellow]{2}
\tikzdrawbox[yellow]{3}
\tikzdrawbox[red]{4}
\tikzdrawbox[red]{5}
\tikzdrawbox[green]{6}
\tikzdrawbox[green]{7}
\tikzdrawboxzero[blue]{8}
\tikzdrawboxzero[blue]{9}
\end{equation}
The block matrices highlighted in the same color are the same matrices. Looking at one of these block matrices, we can see that they have constant elements in their main diagonals, as do their sub-diagonals. This kind of matrix, that has constant elements both in its main diagonal and off-diagonals, is called a Toeplitz matrix. Furthermore, if we look at the block matrices, we can see that the block diagonals are also constant (viz. $\C$ is composed of Toeplitz blocks). This is more discernible if we write $\C$ in terms of its block matrices:
\begin{equation}
    \C = 
    \begin{pmatrix}
    \C_1  & \C_2 & \zero\\
    \C_3  & \C_1 & \C_2\\
    \zero & \C_3 & \C_1\\
    \end{pmatrix}.
\end{equation}
This kind of matrix, where a matrix consists of Toeplitz blocks and the block matrices themselves are in turn Toeplitz matrices, is called a \textit{doubly blocked Toeplitz matrix} or a \textit{Toeplitz-block Toeplitz matrix}.

\section{PSF kernel size and the sparsity of the convolution operator}
\label{app:psf}

By reducing the PSF kernel size, the sparsity of the convolution operator $\C_l$ can be increased and substantial computing time can be reduced, with convolution accuracy as the tradeoff.

\begin{figure*}
    \centering
    \includegraphics[width=\textwidth]{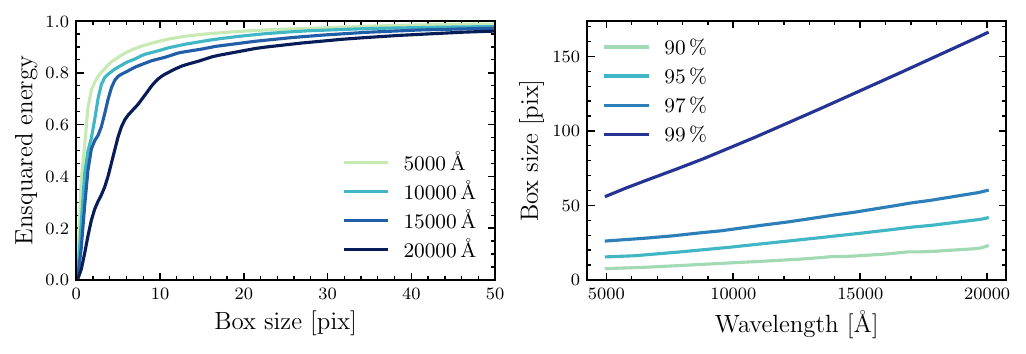}
    \caption{The left panel shows the ensquared energy of several monochromatic \Roman{} PSFs as a function of the box size. The PSFs of bluer wavelengths are more compact than those of redder wavelengths. The right panel shows the size of the box we have to make if we want to contain 90\%, 95\%, 97\%, or 99\% of the total flux within the box centered on the centroid of the PSF, here plotted as a function of the PSF wavelength.}
    \label{fig:width}
\end{figure*}

In choosing the appropriate PSF kernel size, we can consider the fraction of light concentrated within a box centered on the PSF centroid and with width $w$. This quantity is called the \textit{ensquared energy}, similar to encircled energy which measures the concentration of light within a circle with radius $r$. On the left panel of Figure~\ref{fig:width} we plot the ensquared energy as a function of the enclosing box size, for selected wavelengths between 5000\,\AA--20000\,\AA. As we can see the PSFs of redder wavelengths are less compact than those of bluer wavelengths, which causes more blurring of images in redder wavelengths we previously saw in Figure~\ref{fig:convolution}. We use these ensquared energy curves to calculate the box size we have to make if we want to contain 90\%--99\% of the light within the box, as a function of wavelength. These are plotted on the right panel of Figure~\ref{fig:width}. Again we see the same trend previously seen, that for redder wavelengths we have to enlarge the box to contain the same concentration of light as bluer wavelengths. As expected, the size of the box increases linearly with wavelength. However, what is particularly striking is the large size of the required kernel if we want to enclose 99\% of the light within the box.  

\begin{figure*}
    \centering
    \includegraphics[width=\textwidth]{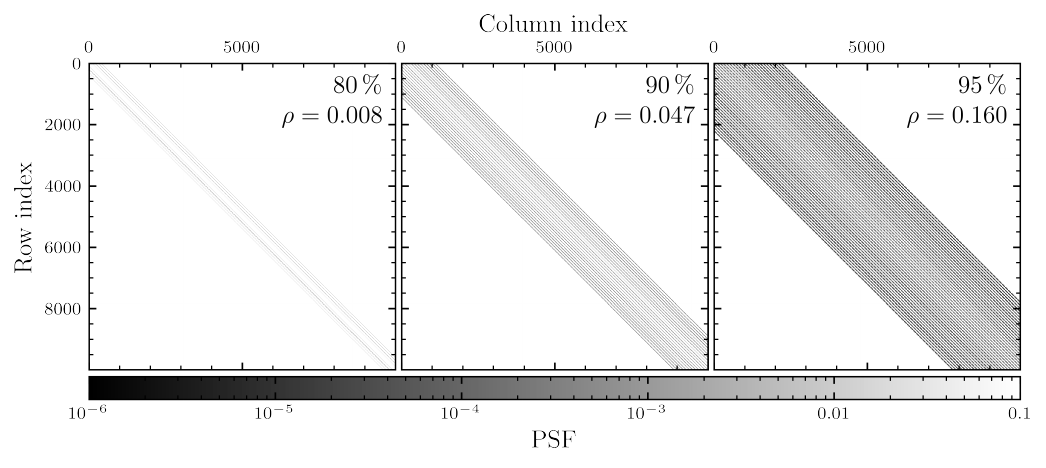}
    \caption{Examples of convolution operator $\C_l$ (which convolves a vectorized image $\cube_l$ with the corresponding PSF $\PSF_l$) for three different PSF kernel widths, such that the flux enclosed within the box are 80\% (left panel), 90\% (middle panel), or 95\% (right panel) of the total flux. The kernel widths are then respectively 9\,pixels, 23\,pixels, and 45\,pixels (these are oversampled by a factor of 2). Also shown in each panel is the density of the matrix, $\rho$, the ratio between the non-zero elements and total number of elements in the matrix.}
    \label{fig:cvar}
\end{figure*}

In Figure~\ref{fig:cvar} we show the reduction in $\C_l$ density that can be achieved by reducing the requirement on the light enclosed within the box. For an image size of $100\times 100$ (this is oversampled by 2), the density of the convolution matrix $\C_l$ can be reduced to 4.7\% if we enclose only 90\% of the light (middle panel). The sparsity can further be increased if we relax the enclosure of light to (for example) 80\%, as shown in the left panel of Figure~\ref{fig:cvar}. Further study to determine the minimum kernel width that can still give adequate accuracy in modelling and predicting the spectra is warranted and will be addressed in our future publications.


\section{The probabilistic interpretation of Ridge regression}
\label{app:bayesridge}
Ridge regression is a form of regularized $\chi^2$ regression. It minimizes the objective function
\begin{equation}
    Q_R(\cube) = \res\transpose\weight\res + \alpha\cube\transpose\cube,
    \label{eq:QR}
\end{equation}
in which 
\begin{equation}
    \res\transpose\weight\res = \left(\data - \X\cube\right)\transpose\weight\left(\data - \X\cube\right)
\end{equation}
is the $\chi^2$ term and $\alpha\cube\transpose\cube$ is the regularization term, here defined as the squared norm of $\cube$ multiplied by the regularization parameter $\alpha$.

We can use Bayes theorem to interpret the meaning of $Q_R$ in a probabilistic manner. We start from Bayes Theorem:
\begin{equation}
    \posterior = \frac{\likelihood\prior}{p(\data|\X)},
\end{equation}

where $\cube$ is the vector of the flattened datacube, and $\X$ is the forward model that projects $\cube$ into observed counts $\data$.

The likelihood $\likelihood$ of observing $\data$ given $\X$, $\cube$, and covariance matrix $\cov$ is an $N$-dimensional Gaussian distribution:
\begin{equation}
    \likelihood = (2\pi)^{-N/2}|\cov|^{-1/2}\exp\left[-\frac{1}{2}(\data-\X\cube)^\intercal\cov^{-1}(\data-\X\cube)\right],
\end{equation}
in which $N$ is the number of data, i.e. the total number of pixels in the set of 2d spectra images.

For the prior $\prior$, we assume that the datacube $\cube$ is normally distributed around the prior datacube $\cube_\text{\rm p}$ with single scale length $\scalelength$:
\begin{equation}
    \prior = (2\pi)^{-P/2}\scalelength^{-P}\exp\left[-\frac{1}{2\scalelength^2}(\cube - \cube_\text{\rm p})\transpose(\cube - \cube_\text{\rm p})\right],
\end{equation}
where $P$ is the size of the datacube.

The natural logarithm of the posterior then becomes
\begin{equation}
    \ln\posterior\propto -\frac{1}{2}\res^\intercal\weight\res - \frac{1}{2}\alpha(\cube - \cube_\text{\rm p})^\intercal(\cube - \cube_\text{\rm p}),
    \label{eq:lnpost}
\end{equation}
where $\alpha = 1/\scalelength^2$, $\weight = \cov^{-1}$, and $\res = \data - \X\cube$. 

Here we make another simplification about the covariance matrix $\cov$: We assume no covariance between pixels. $\cov$ is a diagonal matrix and its inverse is denoted as $\cov^{-1} = \weight$, with $W_{ii} = 1/\sigma^2_{i}$ and zero otherwise. This simplifies the calculation of $\res^\intercal\weight\res$ to just the summation of the residuals weighted by the inverse variance, i.e.
\begin{equation}
    \res^\intercal\weight\res = \sum^{N}_{i}\frac{(y_i - \X_i \cube)^2}{\sigma^2_i}.
    \label{eq:simplesummation}
\end{equation}
In real applications the pixel variance will correlate with each other due to inter-pixel capacitance (IPC), resulting in non-zero off-diagonals. In this case it is non-trivial to invert the covariance matrix $\cov$ to form $\weight$ (but we only have to do this once because they are static), and $\res^\intercal\weight\res$ will no longer be a simple summation of the weighted residuals as in Equation~\ref{eq:simplesummation}.

In Equation~\ref{eq:lnpost}, we also remove all terms that do not depend on $\cube$. We now take the inverse of $\ln\posterior$ to obtain the objective function 
\begin{equation}
\begin{aligned}
    Q(\cube) &= -2\ln\posterior\\
             &= \res^\intercal\weight\res + \alpha(\cube - \cube_\text{\rm p})^\intercal(\cube - \cube_\text{\rm p}).
    \label{eq:Q}
\end{aligned}
\end{equation}
If we set the prior $\cube_\text{\rm p}$ to be zero everywhere, we then arrive back at Equation~\ref{eq:QR}. This is then the probabilistic interpretation of Ridge regression. Just like ordinary least square regression (OLS) it adopts a normally distributed likelihood, but whereas OLS adopts a uniform distribution with no boundaries for its prior (a non-informative prior), Ridge regression uses a weakly-informative prior, i.e. a normal distribution centered around zero, with no covariance between each other, and uniform width $\sigma$ (c.f. \citealt[Sect.~12.2.3]{cbj17}). In this work we adopt the same prior as Ridge regression but centered around a non-zero prior.

Since the objective function $Q(\cube)$ is the inverse natural logarithm of the posterior, minimizing $Q(\cube)$ is thus equivalent to finding the \textit{maximum a posteriori} (MAP) estimate of $\cube$.

\section{Results of regularization parameter optimization and host-galaxy subtraction for the simple galaxies with redshifts $\MakeLowercase{z}=0.5$, $\MakeLowercase{z}=1.0$, and $\MakeLowercase{z}=1.5$}
\label{app:results}

\begin{figure*}
    \centering
    \includegraphics[width=\textwidth]{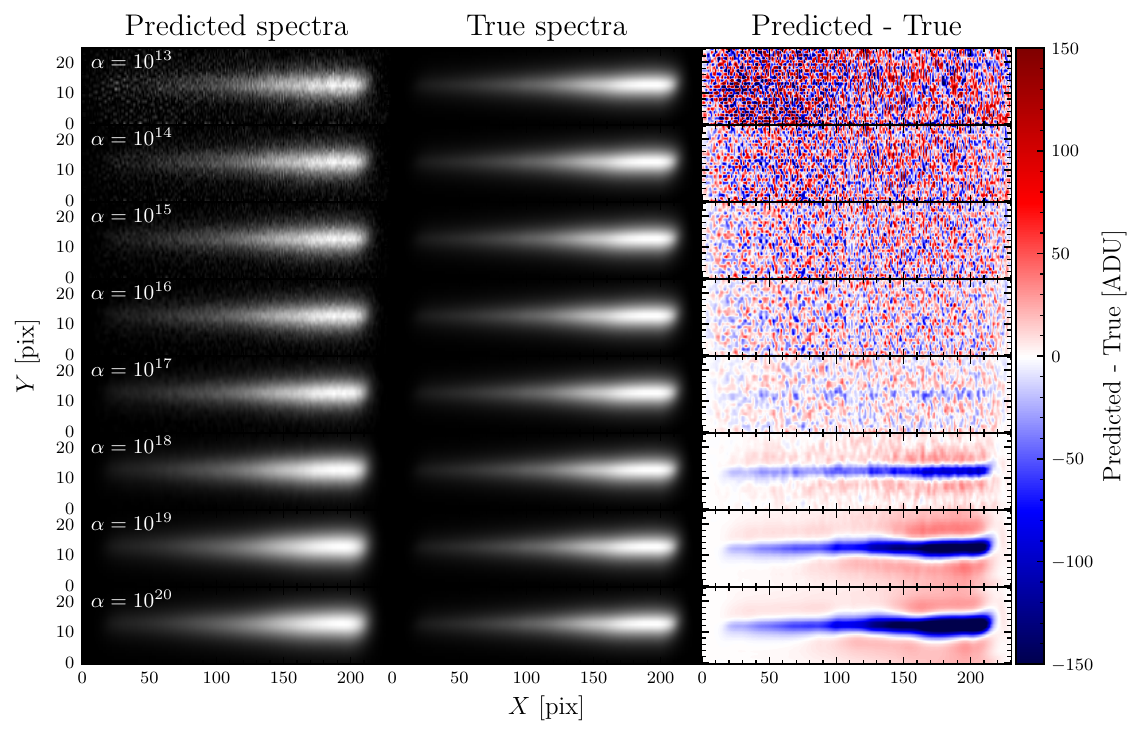}
    \caption{Comparisons between noise-free galaxy-only spectra (left column) generated from the scenes reconstructed using the value of $\alpha$ indicated on the top left corner of each row (left column), and the corresponding noise-free spectra generated from the true scene (center column). This galaxy is at redshift $z = 0.5$ and the reconstruction uses 24 roll angles (uniformly spaced at $15^\circ$). The right column shows the residuals between predicted and true spectra.}
    \label{fig:predalpha2}
\end{figure*}

\begin{figure*}
    \centering
    \includegraphics[width=\textwidth]{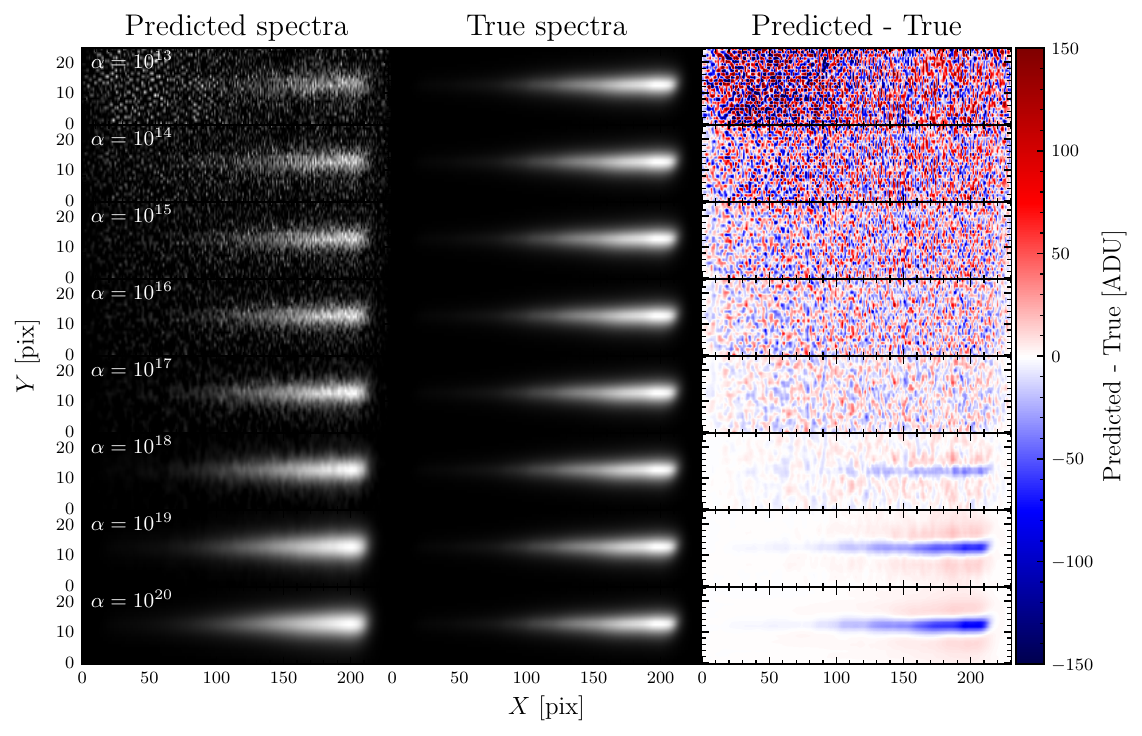}
    \caption{As Figure~\ref{fig:predalpha2}, but for a galaxy with $z=1.5$.}
    \label{fig:predalpha3}
\end{figure*}

\begin{figure*}
    \centering
    \includegraphics[width=\textwidth]{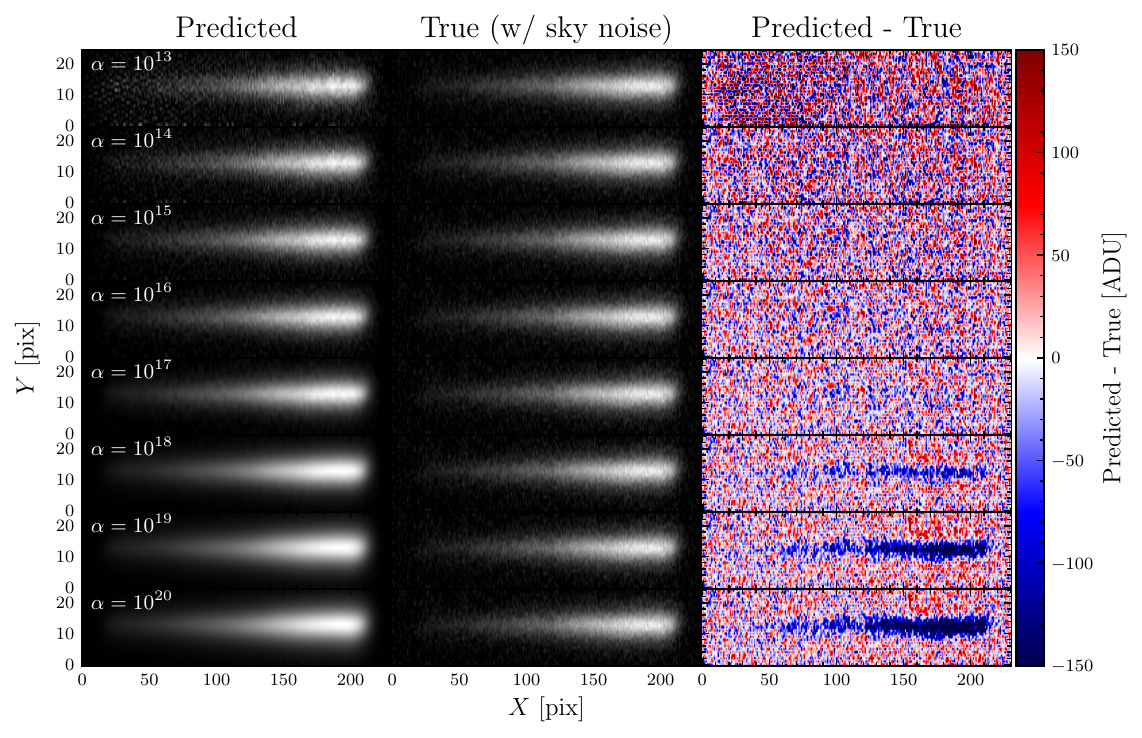}
    \caption{As Figure~\ref{fig:prednoisy1}, which compares predicted spectra and the corresponding \textit{noisy} spectra generated from the true scene (center column), but for a galaxy $z=0.5$. Similar to Figure~\ref{fig:prednoisy1}, which is for $z=1.0$, the residual images are dominated by read and sky noise.}
    \label{fig:prednoisy2}
\end{figure*}
\begin{figure*}
    \centering
    \includegraphics[width=\textwidth]{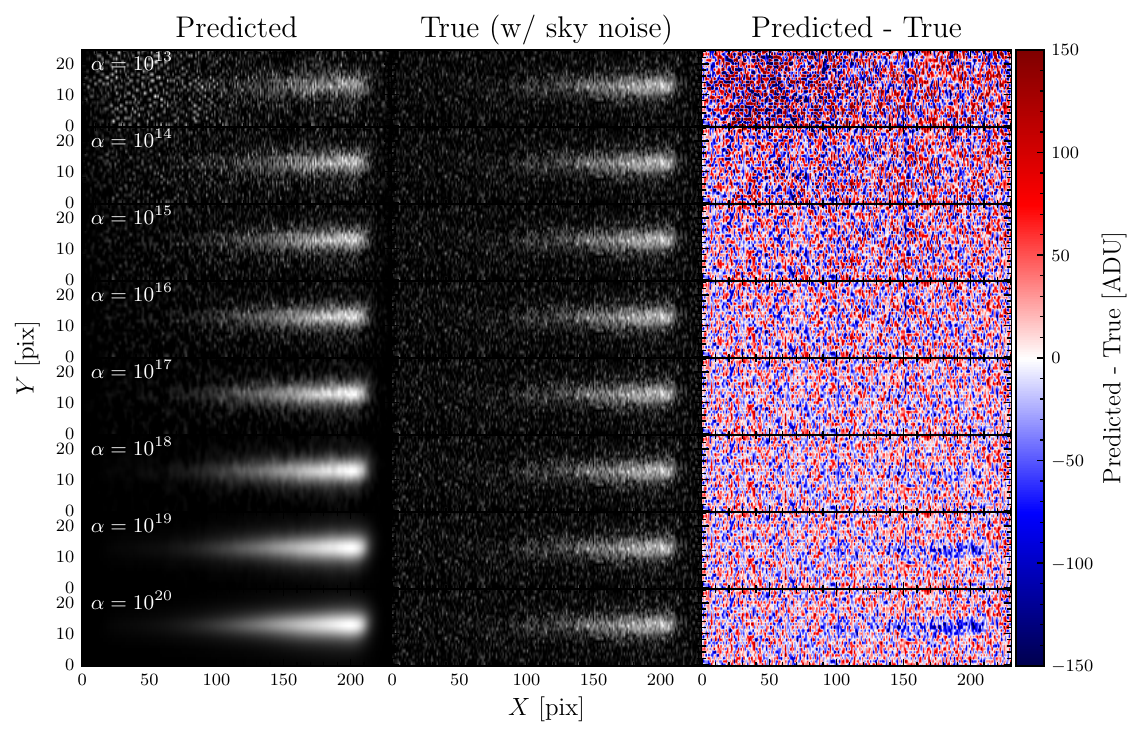}
    \caption{As Figure~\ref{fig:prednoisy2}, but for a galaxy with $z=1.5$. Here the systematic residuals are weaker because of the the fainter galaxy}
    \label{fig:prednoisy3}
\end{figure*}

\begin{figure*}
    \centering
    \includegraphics[width=\textwidth]{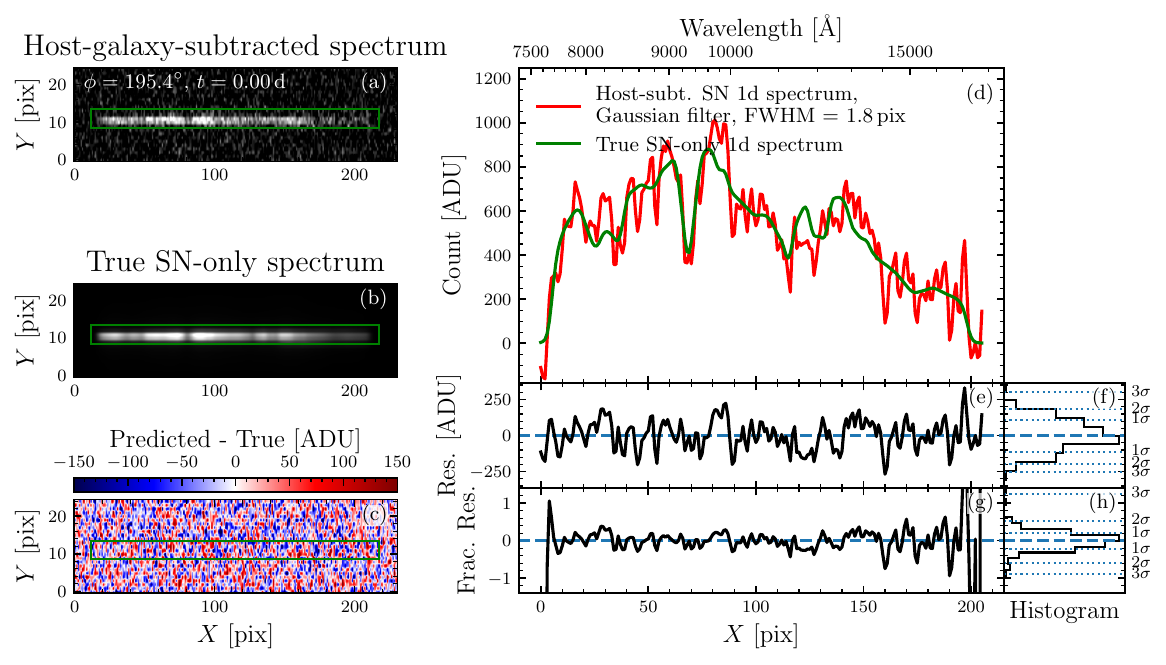}
    \caption{As Figure~\ref{fig:snonly_vela01}, but for a supernova with $z=0.5$. Note the different range of vertical axis in the 1d spectrum plot (panel d).}
    \label{fig:snonly1}
\end{figure*}

\begin{figure*}
    \centering
    \includegraphics[width=\textwidth]{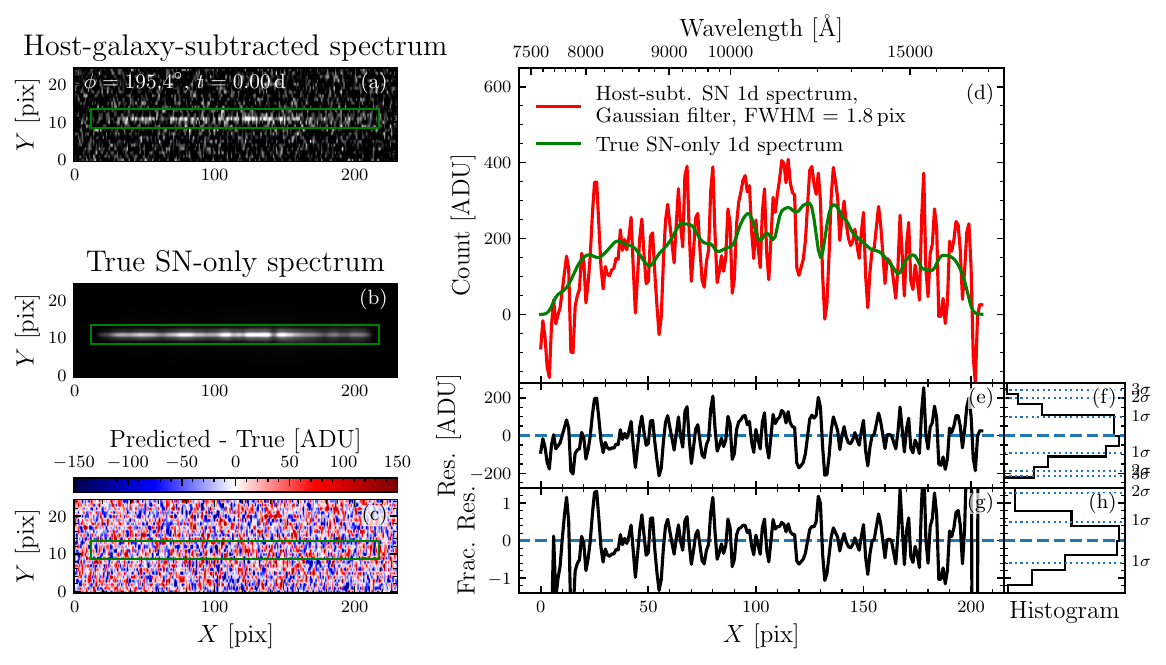}
    \caption{As Figure~\ref{fig:snonly1}, but for a supernova with $z=1.0$. Note the different range of vertical axis in the 1d spectrum plot (panel d).}
    \label{fig:snonly2}
\end{figure*}

\begin{figure*}
    \centering
    \includegraphics[width=\textwidth]{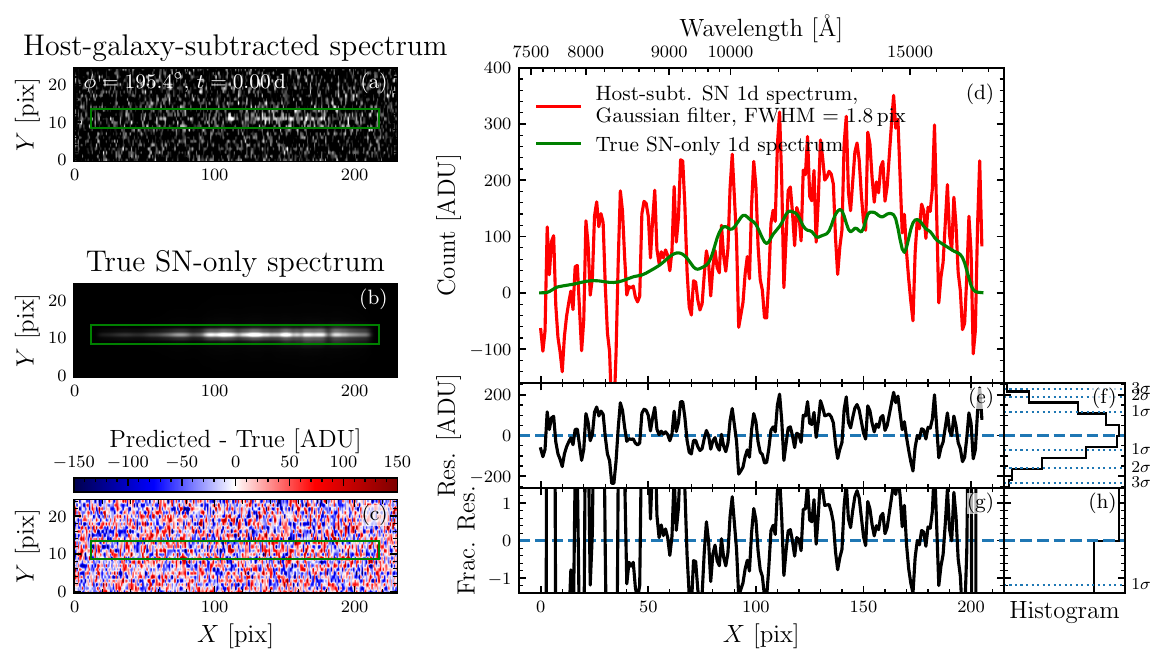}
    \caption{As Figure~\ref{fig:snonly1}, but for a supernova with $z=1.5$.}
    \label{fig:snonly3}
\end{figure*}

\begin{figure*}
    \centering
    \includegraphics[width=\textwidth]{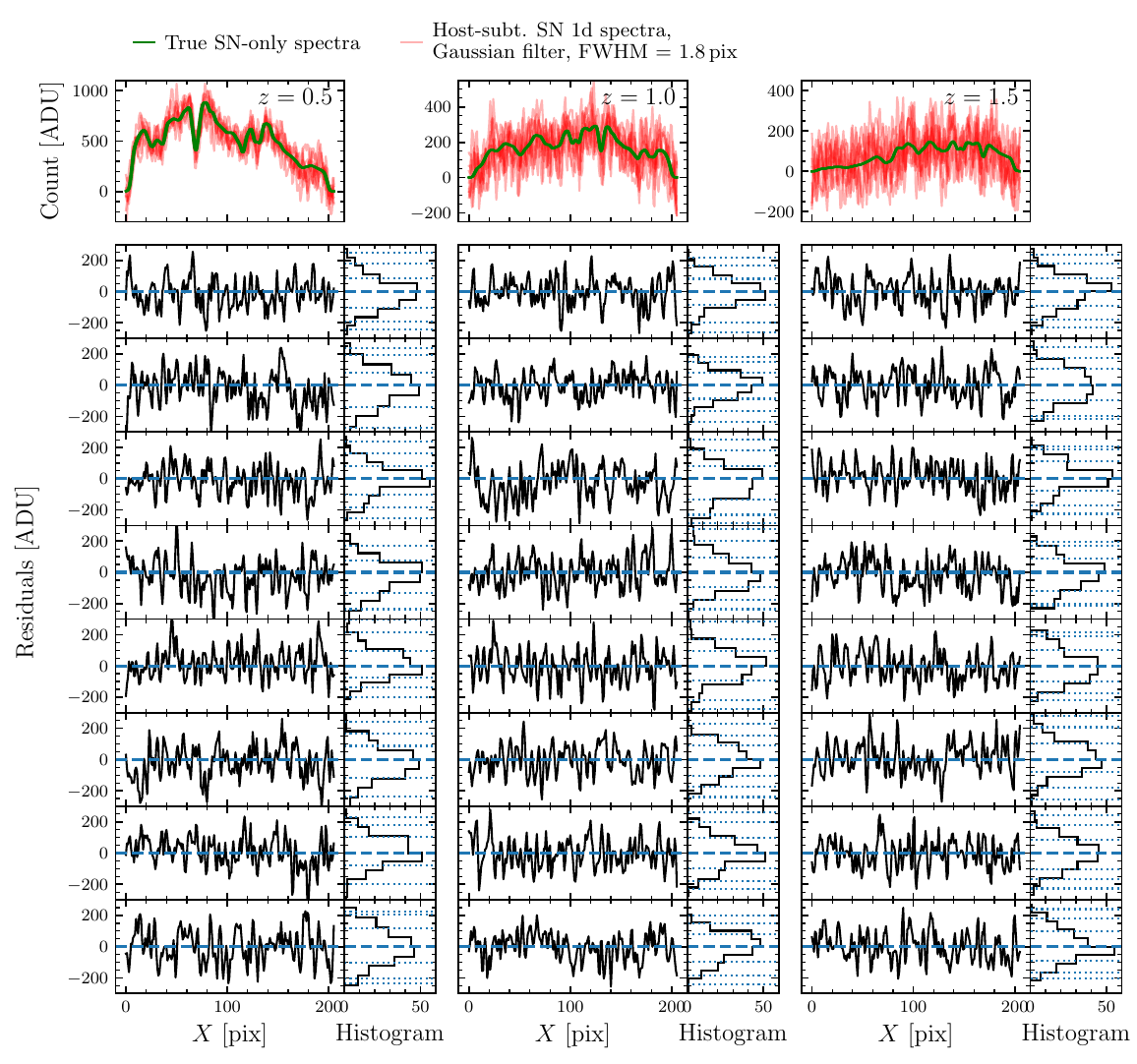}
    \caption{As Figure~\ref{fig:snonlyrandomselected1}, but now for the simple galaxies with redshifts $z=0.5$, $z=1.0$, and $z=1.5$. For each panel in the second from top row down to the bottom row displays individual residuals and their corresponding 1d histograms. On the top panel of each column we overplot the predicted 1d spectra and compare them with the corresponding true spectra. Dotted lines in the histogram plot indicates the $1\sigma$, $2\sigma$, and $3\sigma$ confidence intervals in similar manner as Figures~\ref{fig:snonly1}--\ref{fig:snonly3}.}
    \label{fig:snonlyrandomselected123}
\end{figure*}

\begin{figure*}
    \centering
    \includegraphics[width=0.495\columnwidth]{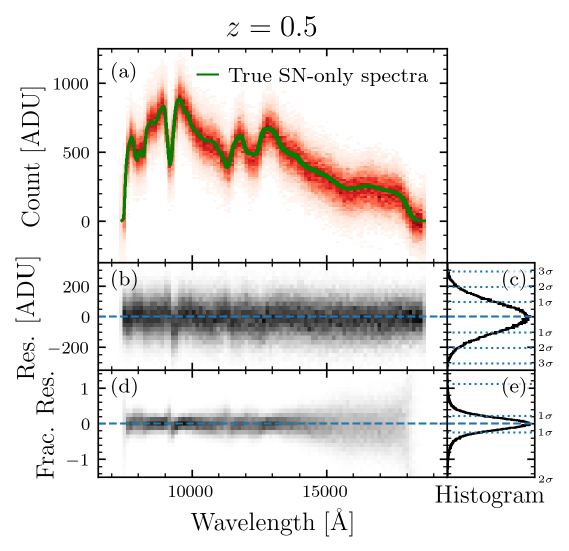}
    \includegraphics[width=0.495\columnwidth]{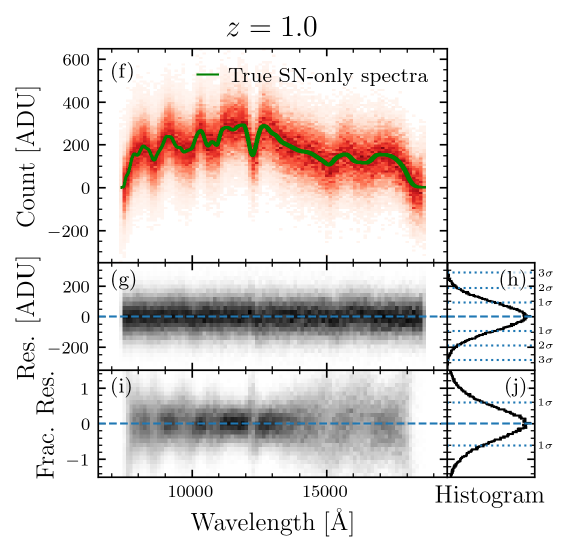}
    \includegraphics[width=0.495\columnwidth]{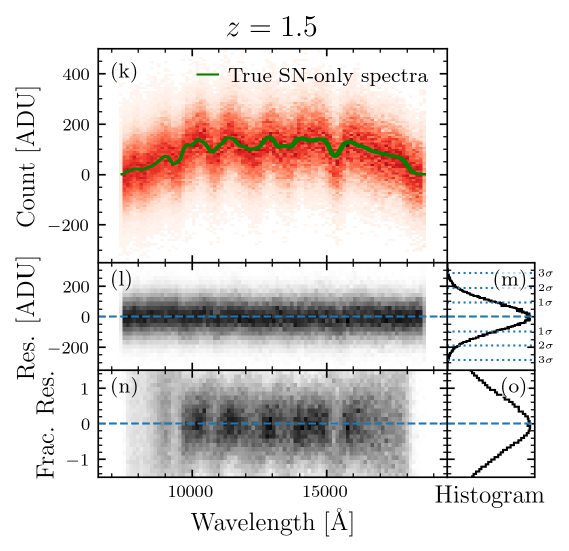}
    \caption{As Figure~\ref{fig:snrandom_vela01}, but for the simple host-galaxies with redshifts $z=0.5$, $z=1.0$, and $z=1.5$.}
    \label{fig:snrandom123}
\end{figure*}

We repeat the same datacube reconstruction process described in Section~\ref{sec:alpha}, which uses various $\alpha$, for two other galaxies at redshifts $z=0.5$ and $z=1.5$. Figures~\ref{fig:predalpha2}--\ref{fig:predalpha3} shows the same comparisons as Figure~\ref{fig:predalpha} for these galaxies. We observe the same behaviour discussed in Section~\ref{sec:alpha}, although at these redshifts, the underestimation streaks at high $\alpha$ are relatively weaker compared to those at $z=0.5$ due to the lower fluxes of these galaxies.

For the three redshifts, we also use the datacube reconstructed using $\alpha = 10^{16}$ to subtract the host-galaxy in which we observe a supernova, and similar to Figure~\ref{fig:snonly_vela01}, in Figures~\ref{fig:snonly1}--\ref{fig:snonly3} we show the host-galaxy subtracted, supernova-only 1d spectra for a single observatory roll angle. The residuals as a function of pixel coordinates as well as the one-dimensional histograms of the residuals, shown respectively in panel~(e) and panel~(f) of these Figures, suggest that at these redshifts there also no systematic residuals and that they are normally distributed.

We also generate 1000 random supernovae locations, observatory roll angles, and sub-pixel dithers for these galaxies, also using the $R$-band images of the host-galaxies. The resulting residuals of the host-subtracted supernova-only 1d spectrum is shown in Figure~\ref{fig:snonlyrandomselected123} for 8 supernovae randomly selected from our pool of 1000 supernovae, and all 1000 of the spectra shown in Figure~\ref{fig:snrandom123} as 2d histograms in similar manner as Figure~\ref{fig:snrandom_vela01}. Again we see that averaged over locations, roll angles, and dithers, the residuals---shown in panel~(c) and panel~(f) of Figure~\ref{fig:snrandom_vela01}---are normally distributed with the mean very close to zero (see Table~\ref{tab:stat_snrandom}).

\clearpage

\newpage

\section{Glossary, list of acronyms, and list of symbols}
\label{app:acronyms}

Table~\ref{tab:glossary} lists the description of some key concepts used in this paper, Table~\ref{tab:acronyms} lists the acronyms used in this paper, and Table~\ref{tab:variables} lists the most important mathematical symbols used in this paper along with a short description and where they are introduced or explained.

\startlongtable
\begin{deluxetable}{>{\bfseries\RaggedRight\arraybackslash}p{0.15\columnwidth}p{0.80\columnwidth}}
\tablecaption{Glossary}
\label{tab:glossary}
\tablehead{\colhead{Terminology} & \colhead{Description}}
\startdata
1d spectrum           & A spectrum extracted from the 2d spectrum using a certain technique (e.g. optimal spectrum extraction, \citealt{hor86}), providing a mapping of fluxes as a function of wavelength. See \textbf{2d spectrum}.\\
2d spectrum           & An image of the dispersed light of an astronomical scene, that formed on the detector. Along the dispersion direction, the image maps the fluxes as a function of wavelength, whereas the fluxes along the cross-dispersion direction are mapped along the spatial dimension.\\
block diagonal matrix & A special kind of block matrix which consists of only block matrices along its main diagonal, and zero matrices elsewhere. See \textbf{block matrix}.\\
block matrix          & A matrix that is partitioned into smaller matrices called \textit{blocks}. These blocks can be square or rectangular, and arranged in grid-like structure to form the larger block matrix.\\
cross-validation      & One technique for hyperparameter optimization by using the data itself to evaluate the predictive performance of the model. The data is split into training set and validation set. The model parameters are inferred using the training set, and the performance of the model in predicting new values (unseen in the training set) is evaluated using the validation set. The performance evaluation commonly uses summary statistics (e.g. root-mean-square residuals between observed and predicted values). The performance evaluation is then repeated for different values of the hyperparameter. The appropriate value of the hyperparameter is those that results in a model with the best performance in predicting new values. See Sect.~\ref{sec:alphacv} and \textbf{hyperparameter}.\\ 
datacube              & A representation of an astronomical scene in the form of a mapping of flux densities in three-dimensions: 2d spatial and 1d spectral. In real applications the datacube is always digitized, viz. binned into discrete three-dimensional grid of flux densities and indexed by a predefined, discrete, and uniform 2d grid of celestial coordinates and 1d grid of wavelengths.\\
fine-tuning           & The process of optimizing the hyperparameters of a prior distribution. See \textbf{hyperparameter} and \textbf{cross-validation}.\\
forward model         & In statistical inferences, a forward model is a mathematical model that transforms the model parameters (that we want to infer) into the observables.\\
hyperparameter        & In Bayesian inferences, hyperparameters are the parameters of the prior distribution. They are not a part of the model parameters that we want to infer, and are usually set at fixed values before we start the inference process. See \textbf{fine-tuning}.\\
ill-posed problem     & A problem that violates at least one of the three properties of a well-posed problem. Per \cite{had02}, the three properties are that a solution to the problem exists; the solution is unique for any data; and the solution changes continuously with the data. Scene reconstruction is an ill-posed problem if there are more parameters to solve then there are data points available. One way to address this is to include \textit{a priori} information and find solutions that not only can reproduce the observables but also physically plausible.\\
inverse problem       & The process of finding the underlying cause of an event given the observed consequences, as opposed to \textit{direct problems} in which we predict the consequence of a known or assumed underlying cause.\\
matricization         & The rearrangement of the elements of a $K \times 1$ vector into an $I \times J$ matrix of equivalent shape, i.e. $I \times J = K$. See \textbf{vectorization} and Appendix~\ref{app:vec}.\\
noise-free spectrum   & A synthetic spectrum without the addition of simulated random noise.\\
noisy spectrum        & A synthetic spectrum with added random noise generated from the adopted noise model, or a real spectrum observed with a real instrument.\\
objective function    & In statistics, an objective function is a mathematical function that measures the closeness of the predicted values with the observed values, as a function of the model parameters. From a Bayesian standpoint, finding the model parameters that minimize the objective function is finding the mode of the posterior distribution, i.e. the maximum a posteriori (MAP) estimate, if the objective function can be derived using Bayes' Theorem. See Appendix~\ref{app:bayesridge}.\\
regularization        & In regression, regularization is the addition of a term in the objective function that penalize complex models and prevent overfitting. See \textbf{objective function}. From a Bayesian standpoint we can interpret the regularization term as the assumed prior distribution of the model parameters, with the parameters of the prior distribution---called hyperparameters---either set at fixed values or fine-tuned. See \textbf{hyperparameter}, \textbf{fine-tuning}, \textbf{cross-validation}, and Appendix~\ref{app:bayesridge}.\\
slitless spectroscopy & A technique in spectroscopy that does not use a slit to block light sources falling outside the slit. Commonly employed in spectroscopic surveys to rapidly obtain spectra for all observed sources in the field-of-view  of the telescope, at the expense of elevated sky background noise and possible overlapping spectra of nearby sources.\\
sparse matrix         & A matrix in which most of its elements are zero. The sparsity of the matrix is often defined as the ratio between the number of zero-valued elements and the total number of elements in the matrix.\\
support               & The set of points in the domain of a function, in which the function give non-zero values.\\
Toeplitz-block Toeplitz matrix & A block matrix with constant block diagonals, along both its main diagonal and the off-diagonals, and the blocks themselves also have constant elements on both the main diagonal and off-diagonals. Also called doubly blocked Toeplitz matrix. See \textbf{Toeplitz matrix} and Appendix~\ref{app:conv}.\\
Toeplitz matrix       & A matrix with constant elements along both its main diagonal and off-diagonals. See Appendix~\ref{app:conv}.\\
trace                 & A curve on the focal plane of a spectrograph system, that maps where a beam of light with a particular wavelength will fall along the curve. A trace does not necessarily have to be a straight line, it can be (slightly) curved depending on the properties of the disperser.\\
vectorization        & The rearrangement of the elements of an $M \times N$ matrix into an $MN\times 1$ vector. See \textbf{matricization} and Appendix~\ref{app:vec}.\\
\enddata
\end{deluxetable}

\clearpage
\newpage

\begin{deluxetable}{l>{\RaggedRight\arraybackslash}p{0.8\columnwidth}}
\tablecaption{List of abbreviations}
\label{tab:acronyms}
\tablehead{\colhead{Acronym} & \colhead{Description}}
\startdata
ADC          & Analog-to-digital converter\\
ADU          & Analog-to-digital unit\\
CIGALE       & Code Investigating GALaxy Emission\\
CV           & Cross validation\\
FoV          & Field of View\\
FITS         & Flexible Image Transport System\\
FWHM         & Full width at half maximum\\
GSFC         & Goddard Space Flight Center\\
HLTDS        & High-Latitude Time Domain Survey\\
HST          & \textit{Hubble} Space Telescope\\
IC           & Index Catalogue, supplement to the New General Catalogue of Nebulae and Clusters of Stars (NGC), published in two parts by \cite{dre95, dre10}\\
$\Lambda$CDM & Lambda Cold Dark Matter, a concordant cosmological model\\
H4RG         & Hawaii-4RG, a photodiode arrays developed by Teledyne Imaging\\
HgCdTe       & Mercury Cadmium Tellurite, also called MCT, a chemical compound used for infrared detectors\\
L-BFGS       & Limited memory Broyden–Fletcher–Goldfarb–Shanno, an optimization algorithm\\
MaNGA        & Mapping Nearby Galaxies at APO (Apache Point Observatory)\\
MAP          & Maximum a posteriori\\
NASA         & National Aeronautics and Space Administration\\
PSF          & Point spread function\\
RMS          & Root mean square\\
RON          & Readout noise\\
SALT         & Spectral Adaptive Light curve Template\\
SCA          & Sensor chip assembly\\
SDSS         & Sloan Digital Sky Survey\\
SED          & Spectral energy distribution\\
SN           & Supernova\\
SN Ia        & Supernova Type Ia (read: "type One-A")\\
SNe          & Supernovae, i.e. plural of SN\\
SNR          & Signal-to-noise ratio\\
SSR          & Sum of squared residuals\\
WFC3         & Wide Field Camera 3, an instrument on-board the \Hubble{} Space Telescope\\
WFI          & Wide Field Instrument\\
WFIRST       & Wide-Field Infrared Survey Telescope, former name of the \Romanc{} Space Telescope\\
\enddata
\end{deluxetable}

\clearpage
\newpage

\startlongtable
\begin{deluxetable}{>{\RaggedRight\arraybackslash}p{0.10\columnwidth}>{\RaggedRight\arraybackslash}p{0.73\columnwidth}>{\RaggedRight\arraybackslash}p{0.12\columnwidth}}
\tablecaption{List of mathematical symbols.}
\label{tab:variables}
\tablehead{\colhead{Notation} & \colhead{Description} & \colhead{Ref.}}
\startdata
$\zero$         & Zero matrix, i.e. a matrix in which all of the entries are zero & Sect.~\ref{sec:interp}\\
$A_{\rm eff}$   & Telescope effective area & Eq.~\ref{eq:spectra}, Tab.~\ref{tab:im_params}, Fig.~\ref{fig:area}\\
$a_{pqij}$      & Fractional overlap of origin pixel $(i,j)$ in destination pixel $(p,q)$, used in Drizzle & Sect.~\ref{sec:Dp}\\
$\C$            & Convolution operator, i.e. a matrix that convolves the datacube $\D$ with the PSF $\vecbold{P}$ & Sect.~\ref{sec:C}\\
$\datacube$     & Datacube in continuous spatial coordinates $(\xi, \eta)$ and wavelength $\lambda$ & Sect.~\ref{sec:specform}\\
$\D$            & Digitized datacube, viz. binned into discrete 3d grid of flux densities, indexed by a predefined, discrete, and uniform 2d grid of celestial coordinates $\vecbold{\xi}$ and 1d grid of wavelengths $\wave$ & Sect.~\ref{sec:forward}\\
$\tilde{\D}$    & Interpolated $\D$. The fluxes in the interpolated datacube $\tilde{\D}$ are interpolations at a new wavelength grid $\wavetilde$, using the original datacube $\D$ with wavelength grid $\wave$ as input & Sect.~\ref{sec:interp}\\
$\cube$         & Vectorized (digitized) datacube & Eq.~\ref{eq:projection}, Sect.~\ref{sec:forward}\\
$\tilde{\cube}$ & Interpolated, vectorized, datacube & Sect.~\ref{sec:interp}\\
$\hat{\cube}$   & Datacube inferred from observed data & Eq.~\ref{eq:objfunc}, Sect.~\ref{sec:rec}\\
$\cube_{\rm p}$ & Prior datacube & Eq.~\ref{eq:objfunc}, Sect.~\ref{sec:rec}, Sect.~\ref{sec:prior}, App.~\ref{app:bayesridge}\\
$\model$        & Forward model that transforms the model parameter---that we want to infer---into observables. In this paper the forward model projects the datacube into spectrum & Eq.~\ref{eq:projection}, Sect.~\ref{sec:forward}\\
$\eye$          & Identity matrix & Sect.~\ref{sec:interp}\\
$(k,m)$         & Discrete coordinates in detector space & Sect.~\ref{sec:specform}\\
$\L$            & Linear interpolation operator, i.e. a matrix that acts upon the datacube $\D$ to perform a linear interpolation of flux densities at new wavelength grid $\wavetilde$ & Sect.~\ref{sec:interp}\\
$N$             & Total number of pixels in the cumulative set of 2d spectral images used to infer the datacube & Sect.~\ref{sec:rec}\\
$N_{\rm s}$     & Number of 2d spectra images used to infer the datacube & Sect.~\ref{sec:rec}\\
$N_\lambda$     & Number of elements in the wavelength grid $\vecbold{\lambda}$ & Sect.~\ref{sec:forward}\\
$P$             & Total number of elements in the datacube $\D$ & Sect.~\ref{sec:forward}\\
$\mathcal{P}$   & Point spread function (PSF) in continuous detector space & Sect.~\ref{sec:specform}\\
$\vecbold{P}$   & Digitized PSF & Sect.~\ref{sec:forward}\\
$p(A)$          & Probability of event $A$ to occur & Sect.~\ref{sec:rec}, App.~\ref{app:bayesridge}\\
$p(A|B)$        & Probability of event $A$ to occur given that event $B$ is known to have occured. Also called conditional probability & Sect.~\ref{sec:rec}, App.~\ref{app:bayesridge}\\
$Q$             & Objective function in statistical inferences & Sect.~\ref{sec:rec}\\
$r_G(0|1)$      & A number randomly drawn from the standard normal distribution (zero mean and 1 unit of standard deviation) & Sect.~\ref{sec:noise}\\
$\vecbold{R}$   & Dispersion operator, i.e. a matrix that use the Drizzle algorithm to disperse the datacube along the trace. At the same time the operator also shifts and rotates the datacube as well as pixelizes it to the detector scale & Sect.~\ref{sec:Dp}\\
$\vecbold{r}$   & Vector of residuals between observed and inferred quantities & Sect.~\ref{sec:rec}\\
$\overline{r}$  & Mean of $r$ & Eq.~\ref{eq:bias}\\
$\overline{r^2}^{1/2}$ & Root-mean-square (RMS) of $r$ & Eq.~\ref{eq:rms}\\
$\mathcal{S}$   & 2d spectrum image in continuous detector space $(\kappa,\mu)$ & Sect.~\ref{sec:specform}\\
$\spec$         & 2d spectrum image in discrete detector space $(k,m)$, i.e. pixelated version of $\mathcal{S}$ & Sect.~\ref{sec:specform}\\
$\vect{\spec}$  & Vectorized 2d spectrum image & Eq.~\ref{eq:projection}, Sect.~\ref{sec:forward}\\
$\matricize{\A}{M}{N}$ & Matricization of $\vect{\A}$, viz. rearranging the elements of a $K\times 1$ vector $\vect{\A}$ into an $M \times N$ matrix $\A$, where $K = M \times N$ & Sect.~\ref{sec:math}, Sect.~\ref{sec:forward}, App.~\ref{app:vec}\\
$\vectorize{\A}$ & Vectorization of $\A$, viz. rearranging the elements of an $M\times N$ matrix $\A$ into an $MN\times 1$ vector $\vect{\A}$ & Sect.~\ref{sec:math}, Sect.~\ref{sec:forward}, App.~\ref{app:vec}\\
$\weight$       & Inverse covariance matrix, also called precision or weight matrix & Sect.~\ref{sec:rec}, App.~\ref{app:bayesridge}\\
$z$             & Redshift & Sect.~\ref{sec:intro}, Sect.~\ref{sec:rec}\\
$\alpha$        & Regularization parameter in statistical inferences & Sect.~\ref{sec:rec}, App.~\ref{app:bayesridge}\\
$\epsilon_S$    & Noise in the observed 2d spectrum image & Eq.~\ref{eq:specsamp1}\\
$(\kappa,\mu)$  & Continuous coordinates in detector space & Sect.~\ref{sec:specform}\\
$\kappa_\lambda(\lambda;\lambda_0,\kappa_0)$ & Prism dispersion curve as a function of wavelength $\lambda$ given the coordinate $\kappa_0$ of the reference wavelength $\lambda_0$ & Eq.~\ref{eq:traceline}, Eq.~\ref{eq:dispcurve}, Sect.~\ref{sec:specform}, Sect.~\ref{sec:prism}, Fig.~\ref{fig:disp}\\
$\lambda$       & Wavelength & Sect.~\ref{sec:specform}\\
$\wave$         & Grid of $N_\lambda$ wavelengths, i.e. $\wave = \left\{\lambda_l\right\}_{l=1}^{N_\lambda}$ & Sect.~\ref{sec:forward}\\
$\wavetilde$    & A new set of wavelength grid onto which we want to interpolate a datacube $\D$ with wavelength grid $\wave$. $\wavetilde = \left\{\tilde{\lambda}_{\tilde{l}}\right\}_{\tilde{l}=1}^{N_{\tilde{\lambda}}}$ & Sect.~\ref{sec:interp}\\
$(\xi, \eta)$   & Celestial coordinate in the gnomonic projection, i.e. on the tangent plane of the celestial sphere. Also called standard coordinates. Can be expressed in angular units (e.g. degree, arcsecond, milliarcsecond etc.) or in units of pixel scales. See \cite{vdk67} and \cite{alt13} for details & Sect.~\ref{sec:specform}\\
$\vecbold{\xi}$ & Grid of $I\times J$ celestial coordinates $(\xi, \eta)$, i.e. $\vecbold{\xi} = \left\{(\xi_i, \eta_j)\right\}_{i,j=1}^{I,J}$ & Sect.~\ref{sec:specform}\\
$\Pi$           & Pixel kernel function & Eqs.~\ref{eq:specsamp1}--\ref{eq:smearing}, Sect.~\ref{sec:specform}\\
$\rho(d)$       & Spatial autocorrelation function, i.e. correlation between a dataset and a spatially shifted copy of itself as a function of the shift magnitude $d$ & Eq.~\ref{eq:autocorr}, Sect.~\ref{sec:autocorr}\\
$\cov$          & Covariance matrix & Sect.~\ref{sec:rec}, App.~\ref{app:bayesridge}\\
$\sigma_S$      & Standard deviation of the count measurement of the 2d spectrum image & Sect.~\ref{sec:noise}\\
$\tau$          & Integration time & Eq.~\ref{eq:spectra}, Tab.~\ref{tab:im_params}\\
\enddata
\end{deluxetable}

\clearpage
\newpage

\bibliography{bibliography}
\bibliographystyle{aasjournalv7}

\end{document}